\documentclass[twocolumn,showpacs,preprintnumbers,amsmath,amssymb]{revtex4}
\usepackage{amsmath,amssymb,color,epsfig,latexsym,natbib,graphicx,dcolumn}
\usepackage{bm}% bold math
\bibpunct{[}{]}{,}{n}{,}{,}		% see natbib.sty
\def\DEL#1{{\textcolor{green}{}}}         % suggested deletion in text
\def\eg{{\it e.g. }}

\def\ie{{\it i.e. }}

\def\rms{{\it r.m.s. }}

\def\vs{{\it vs. }}

\newcommand{\bB}{\bf{B}}

\newcommand{\rem}[1]{}

\newcommand\vecp[1]{\vec{#1}}			% physical vector

\newcommand{\be}{\begin{equation}}
\newcommand{\ee}{\end{equation}}

\def\bB0{\vecp{B}_0} 
\def\kpe{k_{\perp}} 
 
\def\kpa{k_{\parallel}}

\begin{document}

\title{\bf Development of Anisotropy in Incompressible Magnetohydrodynamic Turbulence} 

\author{Barbara Bigot$^{1,2}$, S\'ebastien Galtier$^1$ and  H\'el\`ene Politano$^2$}
\affiliation{1. Institut d'Astrophysique Spatiale (IAS), B\^atiment 121,
F-91405 Orsay (France), Universit\'e Paris-Sud XI and CNRS (UMR 8617)}
\affiliation{2. Laboratoire Cassiop\'ee, UMR 6202, OCA, BP 42229, 06304 Nice Cedex 4, France}

\date{\today}

\begin{abstract}
We present a set of three-dimensional (3D) direct numerical simulations of incompressible decaying 
magnetohydrodynamic turbulence in which we investigate the influence of an external uniform 
magnetic field ${\bf B_0}$. A parametric study in terms of $B_0$ intensity is made where, in 
particular, we distinguish the shear- from the pseudo-Alfv\'en waves dynamics. The initial kinetic 
and magnetic energies are equal with a negligible cross-correlation. Both the 
temporal and spectral effects of ${\bf B_0}$ are discussed. A sub-critical balance is found 
between the Alfv\'en and nonlinear times with both a global and a spectral definition. 
The nonlinear dynamics of strongly magnetized flows is characterized by a different $\kpe$-spectrum 
(where ${\bf B_0}$ defines the parallel direction) if it is plotted at a fixed $\kpa$ (2D spectrum) or 
if it is integrated (averaged) over all $\kpa$ (1D spectrum). In the former case a much wider inertial 
range is found with a steep power law, closer to the wave turbulence prediction than the Kolmogorov 
one like in the latter case. It is believed that the averaging effect may be a source of difficulty to detect 
the transition towards wave turbulence in natural plasmas. For the first time, the formation of filaments
is reported within current and vorticity sheets in strongly magnetized flows which modifies our classical 
picture of dissipative sheets in conductive flows. 
\end{abstract}

\maketitle

%%%%%%%%%%%%%%%%%%%%%%%%%%%%%%%%%%%%%%%%%%%%%%%%%
\section{Introduction}

The magnetohydrodynamics (MHD) approximation has proved to be quite successful in the study 
of a variety of astrophysical plasmas, electrically conducting gas or fluids, like those found in the 
solar corona, the interplanetary medium or in the interstellar clouds. These media are characterized 
by extremely large Reynolds numbers (up to $10^{13}$) \cite{Tajima} with a range of available scales 
from $10^{18}$m to few meters. The isotropy assumption, usually used in hydrodynamic turbulence, 
is particularly difficult to justify when dealing with astrophysical flows since a large-scale magnetic
field is almost always present like in the inner interplanetary medium where the magnetic field lines
form an Archimedean spiral near the equatorial plane (see \eg  \cite{goldstein99,galtier06}). 
Thus, MHD turbulence is much more complex than Navier-Stokes turbulence with, in particular, a 
nonlinear transfer between structures of various sizes due to both nonlinear couplings and Alfv\'en 
wave propagation along the background magnetic field. 

In the mid sixties, Iroshnikov \cite{iro} and Kraichnan \cite{Kraichnan65} (hereafter IK) proposed a 
first description of incompressible MHD turbulence. In this approach {\it \`a la} Kolmogorov, the 
large-scale magnetic field is supposed to act on small-scales as a uniform magnetic field, leading 
to counterpropagating Alfv\'en waves whose interactions with turbulent motions produce a 
slowdown of the nonlinear energy cascade. The typical transfer time through the scales is then 
estimated as $\tau_{NL}^2/\tau_A$ (instead of $\tau_{NL}$ for Navier-Stokes turbulence), where 
$\tau_{NL} \sim \ell /u_{\ell}$ is the nonlinear eddy turnover time at characteristic length scale $\ell$ 
and $u_{\ell}$ is the associated velocity. The Alfv\'en time is $\tau_A \sim \ell/B_0$ where $B_0$ 
represents the large-scale magnetic field normalized to a velocity 
(${\bf B_0} \to {\bf B_0} \sqrt{\mu_0\rho_0}$, with $\mu_0$ the magnetic permeability of free space and 
$\rho_0$ the uniform plasma density). Note that we will use this renormalization in the rest of the paper. 
Hence, the IK energy spectrum in $k^{-3/2}$ unlike the $k^{-5/3}$ Kolmogorov one for neutral flows. 

The weakness of the IK's phenomenology is the apparent contradiction between the presence of 
Alfv\'en waves and the absence of an external uniform magnetic field. The external field is supposed 
to be played by the large-scale magnetic field but its main effect, \ie anisotropy, is not included 
in the description. The role of a uniform magnetic field has been widely discussed in the literature and, 
in particular, during the last two decades \cite{MontgoTurner,Shebalin,Higdon,Oughton94,GS95,ng96,
Kinney,Matthaeus98,Galtier2000,Galtier2002,milano2001,Muller03,verma,chandran,Muller,Boldyrev06,
Bigot07a}. At strong ${\bf B_0}$ intensity, one of the most clearly established results is the 
bi-dimensionalization of MHD turbulent flows with a strong reduction of nonlinear transfers along 
$\bf {B_0}$. In the early eighties, it was shown that a strong $B_0$ leads to anisotropic turbulence 
with an energy concentration near the plane ${\bf k} \cdot {\bf B_0} =0$ \cite{MontgoTurner}, a result 
confirmed later on by direct numerical simulations in two and three space dimensions 
\cite{Shebalin,Oughton94}. A linear dependence between anisotropy and ${\bf B_0}$ intensity was 
also suggested \cite{Matthaeus98}. From an observational point of view, we have also several 
evidences that astrophysical (and laboratory) plasmas are mostly in anisotropic states like in 
the solar wind (see \eg \cite{solarobs,Dasso05}) or in the interstellar medium (see \eg \cite{elmegreen}). 

The effects of a strong uniform magnetic field may be handled through an analysis of resonant triadic 
interactions \cite{Shebalin} between the wavevectors (${\bf k},{\bf p},{\bf q}$) which satisfy the 
relation ${\bf k} = {\bf p} + {\bf q}$, whereas the associated wave frequencies satisfy, for example, 
$\omega({\bf k}) = \omega({\bf p}) - \omega({\bf q})$. The Alfv\'en frequency is 
$\omega({\bf k})={\bf k} \cdot {\bf B_0}= k_{\parallel}B_0$, where $\parallel$ defines the direction 
along ${\bf B_0}$ ($\perp$ will be the perpendicular direction to ${\bf B_0}$). The solution of these 
three-wave resonant conditions directly gives, $q_{\parallel}=0$, which implies a spectral transfer 
only in the perpendicular direction. 
For a strength of $B_0$ well above the \rms level of the kinetic and magnetic fluctuations, the nonlinear 
interactions of Alfv\'en waves may dominate the dynamics of the MHD flow leading to the regime of 
(weak) wave turbulence where the energy transfer, stemming from three-wave resonant interactions,
can only increase the perpendicular component of the wavevectors, while the nonlinear transfers 
is completely inhibited along ${\bf B_0}$ \cite{Galtier2000,Galtier2002}. 

Another important issue discussed in the literature is the relationship between perpendicular and parallel 
scales in anisotropic MHD turbulence (see \cite{Higdon,GS95,Boldyrev06}). In order to take into account 
the anisotropy, Goldreich and Shridar \citep{GS95} proposed a heuristic model based on a critical 
balance between linear wave periods and nonlinear turnover time scales, respectively 
$\tau_A \sim \ell_\parallel/B_0$ and $\tau_{NL}\sim \ell_\perp/u_{\ell}$ (where $\ell_\parallel$ and 
$\ell_\perp$ are the typical length scales parallel and perpendicular to ${\bf B_0}$), with 
$\tau_A=\tau_{NL}$ at all inertial scales. Following the Kolmogorov arguments, one ends up with a 
$E(\kpe,\kpa)\sim \kpe^{-5/3}$ energy spectrum (where ${\bf k} \equiv$ (${\bf k}_\perp$,$\kpa$) and 
$\kpe \equiv |{\bf k}_\perp|$) with the anisotropic scaling law 
\be
\kpa \sim \kpe^{2/3} \, .
\label{aniso1}
\ee

A generalization of this result has been proposed recently \cite{Galtier2005} in an attempt to model MHD 
flows in both the weak and strong turbulent regimes, as well as in the transition between them. In this 
heuristic model, the time-scale ratio $\chi=\tau_A/\tau_{NL}$ is supposed to be constant at all 
scales but not necessarily equal to unity. The relaxation of this constraint enables to still recover the 
anisotropic scaling law (\ref{aniso1}) and to find a universal prediction for the total energy spectrum 
$E(\kpe,\kpa) \sim \kpe^{-\alpha} \kpa^{-\beta}$, with $3\alpha + 2 \beta =7$. 
According to direct numerical simulations (see, \eg \cite{Cho00,MaronGoldreich01,shaikh}), one of the 
most fundamental results seems to be the anisotropic scaling law between parallel and perpendicular 
scales (\ref{aniso1}) and an approximately constant ratio $\chi$, generally smaller than one, between 
the Alfv\'en and the nonlinear times. This sub-critical value of $\chi$ implies therefore a dynamics mainly 
driven by Alfv\'en waves interactions. 

In the weak turbulence limit, the time-scale separation, $\chi \ll 1$, leads to the destruction of some 
nonlinear terms, including the fourth-order cumulants, and only the resonance terms survive 
\cite{ZLF,Newell01,Galtier2000,Galtier2002} which allows to obtain a natural asymptotic closure for 
the wave kinetic equations. In absence of helicities and for $\kpe \gg \kpa$, the dynamics is then entirely 
governed by shear-Alfv\'en waves, the pseudo-Alfv\'en waves being passively advected by 
the previous one. In the case of an axisymmetric turbulence, and in the absence of cross-correlation
between velocity and magnetic field fluctuations, the exact power law solution is 
$E(\kpe,\kpa)\sim \kpe^{-2} f(\kpa)$, where $f$ is an arbitrary function taking into account the 
transfer inhibition along ${\bf B_0}$. The regime of wave turbulence is quiet difficult to reproduce by
direct numerical simulations since it requires a strong external magnetic field as well as a high spatial 
resolution. According to a recent theoretical analysis, it seems to be currently not possible to reach fully 
this regime \cite{Naza07} and only the transition towards such a regime is likely to be obtained 
\cite{Boldyrev08,Bigot07a}. 

In order to better understand the development of anisotropy in natural magnetized plasmas, we perform 
a set of tri-dimensionnal numerical simulations of incompressible MHD. In this work, the regime of freely 
decaying flows is chosen in an attempt to model the nonlinear evolution of outward and inward propagating 
Alfv\'en waves. We mainly focus our analysis on the development of anisotropy in flows at moderate 
Reynolds numbers which freely evolve under the influence of a uniform magnetic field whose strength 
will be taken as a parameter. 
The details of the numerical setup and simulations are given in the next Section. In Section~III, we 
investigate the temporal characteristics of the different flows, as well as different global quantities to 
measure the spectral anisotropy. Section~IV is devoted to the evolution of the energy spectra together 
with their fluxes. The flow spatial properties are examined in Section~V. Section VI discusses the second 
set of simulations. A summary and a conclusion are finally given in Section~VII.

%%%%%%%%%%%%%%%%%%%%%%%%%%%%%%%%%%%%%%%%%%%%%%%%%
\section{Numerical setup}
\subsection{Incompressible MHD equations}

The MHD equations that describe the large-scale and low frequency dynamics of magnetized plasmas 
are, in the incompressible case and in the presence of a uniform magnetic field ${\bf B_0}$, 
\be
\partial_t {\bf v} - B_0 \partial_{\parallel} {\bf b} + {\bf v} \cdot \nabla \, {\bf v} = 
- {\bf \nabla} P_* + {\bf b} \cdot \nabla \, {\bf b} + \nu \Delta {\bf v} \, ,
\label{mhd1}
\ee
\be
\partial_t {\bf b} - B_0 \partial_{\parallel} {\bf v} + {\bf v} \cdot \nabla \, {\bf b} = 
{\bf b} \cdot \nabla \, {\bf v} + \eta \Delta {\bf b} \, , 
\label{mhd2}
\ee
\be
\nabla \cdot {\bf v} = 0 \, ,
\label{mhd1b}
\ee
\be
\nabla \cdot {\bf b} = 0 \, ,
\label{mhd2b}
\ee
where ${\bf v}$ is the plasma flow velocity, ${\bf b}$ the magnetic field (normalized to a velocity), 
$P_*$ the total (magnetic plus kinetic) pressure, $\nu$ the viscosity and $\eta$ the magnetic diffusivity. 
It is convenient to introduce the Els\"asser fields ${\bf z}^\pm ={\bf u} \pm {\bf b}$ for the fluctuations; in 
this case and assuming a unit magnetic Prandtl number (\ie $\nu=\eta$), we get
\be
\partial_t {\bf z}^\pm + {\bf z}^\mp  \cdot \nabla {\bf z}^\pm \mp B_0 \partial_{\parallel} {\bf z}^\pm = 
-\nabla P_* + \nu \nabla^2 {\bf z}^\pm  \, ,
\label{mhd3}
\ee
\be
\nabla \cdot {\bf z}^\pm = 0 \, . 
\label{mhd4}
\ee
Note that the second term in the left hand side (LHS) of Eq.~(\ref{mhd3}) represents the nonlinear
interactions between the ${\bf z}^\pm$ fields, while the third term represents the linear Alfv\'enic 
wave propagation along the ${\bf B_0}$ field which will be assimilated to the z-direction in our 
numerical box. 
In the present analysis, a unit magnetic Prandtl number is taken in order to extend at maximum 
the inertial range for both the kinetic and magnetic energies. We believe that such analysis is the first step 
to understand turbulence in anisotropic media. The extension to other magnetic Prandtl numbers is the 
second step: this situation, more realistic for turbulence like in the interstellar medium, supposes to keep a 
high level of turbulence for both the kinetic and magnetic energies which necessitates a higher spatial 
resolution.

%%%%%%%%%%%%%%%%%%%%%%%%%%%%%%%%%%%%%%%%%%%%%%
\begin{table*}
\caption{\label{table1}
Computational parameters for runs {\bf Ia} to {\bf VIIa} with isotropic initial conditions, and for runs 
{\bf Ib} and {\bf IIb} with specific initial conditions (see text). Note that simulations {\bf VIIa} and {\bf IIb} 
use a hyperviscosity and a hyperdiffusivity (dissipation terms in $\nabla^4$). Spatial resolution, 
viscosity $\nu$ ($=\eta$) and applied magnetic field intensity $B_0$ are given, followed by initial integral 
length scales: isotropic $L=2\pi\int{(E^v(k)/k) dk}/\int{E^v(k) dk}$, perpendicular  
$L_\perp= 2\pi\int{(E^v(\kpe) /  \kpe) d\kpe}/\int{E^v(\kpe) d\kpe}$,
and parallel $L_\parallel= 2\pi\int{(E^v(\kpa)/\kpa)d\kpa}/\int{E^v(\kpa)d\kpa}$ scales.
Initial \rms velocity $u_{rms}=<{\bf v}^2>^{1/2}$ ($=b_{rms}=<{\bf b}^2>^{1/2}$) fluctuation 
is given together with the initial kinetic Reynolds number $\mathcal{R}_v=$$\ u_{rms}L/\nu$. 
Finally, we find typical times: isotropic eddy turnover time $\tau^i_{NL}=L/u_{rms}$ (based on the isotropic 
length-scale $L$), eddy turnover time $\tau_{NL}=L_\perp/u_{rms}$ (based on $L_\perp$), Alfv\'en time 
based on \rms magnetic fluctuations $\tau^i_{A}=L/b_{rms}$, Alfv\'en wave period $\tau_{A}=L_\parallel/B_0$ 
and the final time $t_{M}$ of the numerical simulation.}
\begin{ruledtabular}
\begin{tabular}{cccccccccccccc}
&  $  $ & $\nu$ & $B_0$ &  $L$ & $L_\perp$ & $L_\parallel$ & $u_{rms}$& $\mathcal{R}_v$ & $\tau^i_{NL}$  & $\tau_{NL}$ &$\tau^i_A$& $\tau_A$ & $t_{M}$\\
{\bf Ia} & $256^3$ & $4.10^{-3}$ & $0$ & $3.12$ &$-$ & $-$ & $1$  & $779$ & $3.12$& $-$ &$3.12$ & $-$ & $15$ \\
{\bf IIa} & $256^3$ & $4.10^{-3}$ & $1$ & $3.12$ &$3.85$ &$5.57$ & $1$  & $779$ & $3.12$ & $3.85$ & $3.12$&$5.57$ & $15$ \\
{\bf IIIa} & $256^3$ & $4.10^{-3}$ & $5$ & $3.12$ &$3.85$ &$5.57$ & $1$  & $779$ & $3.12$ & $3.85$ & $3.12$ &$1.11$& $15$ \\
{\bf IVa} & $256^3$  & $4.10^{-3}$ & $15$ & $3.12$ & $3.85$ & $5.57$ & $1$  &  $779$ &  $3.12$ & $3.85$ & $3.12$ & $0.37$ &$15$ \\
{\bf Va}  & $512^2\times64$ & $10^{-3}$ & $15$ & $3.12$ & $3.85$ & $5.57$ & $1$ &  $3120$ &  $3.12$ &$3.85$ & $3.12$ & $0.37$ &$15$ \\
{\bf VIa} & $512^2\times64$ & $10^{-3}$ & $30$ & $3.12$ & $3.85$ & $5.57$ & $1$  &  $3120$ &  $3.12$ & $3.85$ & $3.12$ & $0.18$ & $15$ \\
{\bf VIIa} & $512^2\times64$ & $10^{-6}$ & $15$ & $3.12$ & $3.85$ & $5.57$ & $1$  &  $3.12\times10^6$ &  $3.12$ & $3.85$ & $3.12$ & $0.37$ & $15$ \\
{\bf Ib} & $512^2\times64$ & $5.10^{-4}$ & $15$ &$1.27$&$1.90$&$2.04$ & $1$  & $2530$ & $1.26$ & $1.90$ &$1.26$& $0.13$ &$40$  \\
{\bf IIb} & $512^2\times64$ & $10^{-6}$ & $15$ &$1.27$&$1.90$&$2.04$&$1$& $3.16\times10^6$&$1.26$ &$1.90$ &$1.26$& $0.13$ &$40$  \\
\end{tabular}
\end{ruledtabular}
\end{table*}
%%%%%%%%%%%%%%%%%%%%%%%%%%%%%%%%%%%%%%%%%%%%%%

\subsection{Poloidal/toroidal decomposition}

In the presence of a large-scale magnetic field ${\bf B_0}$, Alfv\'en waves develop and propagate at 
Alfv\'en speed $B_0$ along the ${\bf B_0}$ direction. These waves may be decomposed into shear- and 
pseudo-Alfv\'en waves denoted, respectively, ${\bf z^\pm_1}$ and ${\bf z^\pm_2}$. The divergence-free 
condition implies that only two types of scalar field ($\psi^\pm$ and $\phi^\pm$) are needed to describe 
the incompressible MHD dynamics which are, respectively, the toroidal and poloidal fields. 
For the Fourier transforms of the involved fields, we have:
\begin{eqnarray}
{\bf \widehat{z}^{\pm}(k)} = {\bf \widehat{z}^\pm_1(k)} + {\bf \widehat{z}^\pm_2(k)} \, ,
\label{DECPT}
\end{eqnarray}
with
\begin{eqnarray}
{\bf \widehat{z}^\pm_1(k)} &=& i {\bf k}\times {\bf e_\|} \widehat{\psi}^\pm({\bf k}) \, , \\
{\bf \widehat{z}^\pm_2(k)} &=& - \frac{{\bf k}\times ({\bf k}\times {\bf e_\|})}{k} \widehat{\phi}^\pm({\bf k}) \, ,
\label{DECPT2}
\end{eqnarray}

where in our simulations $k_\perp = \sqrt{k_x^2+k_y^2}$ and $k=\sqrt{k_x^2+k_y^2+k_z^2}$. Here, 
$k_z \equiv k_\|$ and ${\bf e_\|}$ denotes the unit vector parallel to the ${\bf B_0}$ direction.
Hence, the shear-Alfv\'en waves correspond to a vector field perpendicular to the external magnetic 
field ${\bf B_0}$ whereas the pseudo-Alfv\'en waves is a vector field which may have a component along 
${\bf B_0}$; but both vector fields depend on the three coordinates of ${\bf k}$.
%Hence, the shear-Alfv\'en waves vary only in the perpendicular direction (with components only
%in the perpendicular planes) whereas the pseudo-Alfv\'en waves fluctuate in all directions.

\subsection{Initial conditions}
\label{IC}

We numerically integrate the three-dimensional incompressible MHD equations (\ref{mhd1})--(\ref{mhd2b}), 
in a $2\pi$-periodic box, using a pseudo-spectral code (including de-aliasing), and with spatial resolution 
from $256^3$ to $512^2 \times 64$ grid points according to the initial conditions (see Table \ref{table1}). 
The time marching uses an Adams-Bashforth / Cranck-Nicholson scheme, \ie a second-order 
finite-difference scheme in time (see \eg \cite{Galtier99}). 

\subsubsection{Runs Ia to IVa}

The initial kinetic and magnetic fluctuations are characterized by spectra at large-scales, \ie for 
$k=[1,8]$, proportional to $k^2 \exp(-k^2/4)$; for $k>8$, the spectra are exactly equal to zero. This 
condition means that for wavenumbers $k$ up to $2$, we have mainly a flat modal spectrum which 
prevents initially any favored wave vectors. No forcing is present during the simulations and the flows 
may evolve freely for time $t>0$. The associated kinetic, 
\be
E^v = {1 \over 2} <{\bf u}^2({\bf x})> \, ,
\ee
and magnetic, 
\be
E^b = {1 \over 2} <{\bf b}^2({\bf x})> \, ,
\ee 
energies are chosen initially equal, namely $E^v(t=0)=E^b(t=0)=0.5$. 
(Note that $< \cdot >$ means space  averaging.) 

The correlation between the velocity and magnetic field fluctuations, which is measured by the 
cross-correlation 
\be
\rho \equiv {2<{\bf u}({\bf x}) \cdot {\bf b}({\bf x)}> \over <{\bf u}^2({\bf x}) + {\bf b}^2({\bf x})>} \, ,
\label{rho}
\ee
is initially less than $1\%$. 

The initial (large-scale) kinetic and magnetic Reynolds numbers are about $800$ for the flows with 
$\nu= 4 \times 10^{-3}$ (see Table \ref{table1}), with $u_{rms}=b_{rms}=1$; the isotropic integral scale is
\be
L = 2\pi {\int{ (E^v(k)/k) dk} \over \int{E^v(k) dk}} \sim \pi \, .
\ee

A parametric study is performed according to the intensity of $B_0$. Four different values are used, 
namely $B_0 = 0, 1, 5$ and $15$. 
All these simulations are run up to a maximum computational time $t_{M}=15$, and correspond
to runs ${\bf Ia}$ to ${\bf IVa}$ described in Table~\ref{table1}.
%at which the loss of the total energy (kinetic plus magnetic) is about $95\%$ for the simulation 
%with $B=0$, $90\%$ for $B=1$, and $83\%$ for the $B=5$ and $B=15$ runs.

\subsubsection{Runs Va to VIIa}

Taking advantage of the strong reduction of the nonlinear transfers along ${\bf B_0}$ in highly magnetized  flows, a second set of direct numerical simulations is performed with a spatial resolution of $512^2$ grid 
points in the perpendicular plane to ${\bf B_0}$ and with only $64$ grid points in the parallel direction 
(runs ${\bf Va}$ and ${\bf VIa}$ in Table~\ref{table1}). For such runs, the initial conditions are the same as 
before with, however, a uniform magnetic field $B_0=15$ and $30$, and a smaller viscosity. 

Such simulations were analyzed in the past to explore the self-consistency of the reduced MHD model 
\cite{Oughton04} with the conclusion that small values of viscosities, adjusted according to the transverse 
dynamics, are not incompatible with the smaller spatial resolution in the parallel direction since the transfer 
toward small-scales is also reduced along the uniform magnetic field. 
We checked that the viscosity, $\nu=10^{-3}$, is indeed well adjusted. Note that such a small aspect ratio 
may reduce the number of resonant wave interactions which in turns may affect the dynamics (see \eg 
\cite{smith99}). However, in Alfv\'en wave turbulence, the resonant manifolds foliate wavevector space 
\cite{Galtier2000} which, in principle, prevent such a problem. 

In the same manner, another computation (run ${\bf VIIa}$) is made using an hyperviscous scheme, 
where the laplacian operator of the dissipative terms is replaced by a bi-laplacian, in order to enlarge 
the inertial range of the energy spectra.

\subsubsection{Runs Ib and IIb}
\label{bb}

Finally, to evaluate the influence of the initial conditions, a third set of runs is performed with a uniform 
magnetic field fixed to $B_0=15$, and with either a viscous (${\bf Ib}$) or an hyperviscous dissipation 
(${\bf IIb}$). In both simulations, we use a $512^2 \times 64$ grid points. The specific initial conditions 
of these runs correspond to a modal energy spectrum $E^\pm(\kpe,\kpa) = C(\kpa) \kpe^3$, for $\kpe$ 
and $\kpa$ $\in [0,4]$, the value of $C(\kpa)$ increasing with $\kpa$ to reach a maximum at $\kpa=4$. 
Note that this initial spectrum allows a transient period of cascade toward smaller scales during which 
energy is mainly conserved. Initially, the ratio between kinetic and magnetic energies is still fixed to $1$, 
whereas the cross-correlation coefficient is zero. A first set of results was given in \cite{Bigot07a}. 

For all the runs described in this Section, the computational parameters (initial Reynolds numbers, 
characteristic length scales and times...) are summarized in Table~\ref{table1}.

%%%%%%%%%%%%%%%%%%%%%%%%%%%%%%%%%%%%%%%%%%%%%%
\section{Temporal analysis}
\subsection{Energetic properties}
\subsubsection{Els\"asser ${\bf z}^{\pm}$ cartesian fields}

In this section, we study the temporal behavior of several global quantities to characterize the MHD 
flow dynamics and the influence of the  ${\bf B_0}$ strength on it. In all the following figures, time 
evolutions are shown from initial isotropic conditions up to time $t_M$ (the maximum computational 
time reached), for simulations {\bf Ia} to {\bf IVa} at moderate resolution ($256^3$ mesh points) and
$B_0=0,1,5$ and $15$, together with highly magnetized flows {\bf Va} and {\bf VIa} at $B_0=15$ and 
$30$, using $512^2 \times 64$ spatial resolution (see Table~\ref{table1}). 

\begin{figure}[ht]
\resizebox{88mm}{!}{\includegraphics{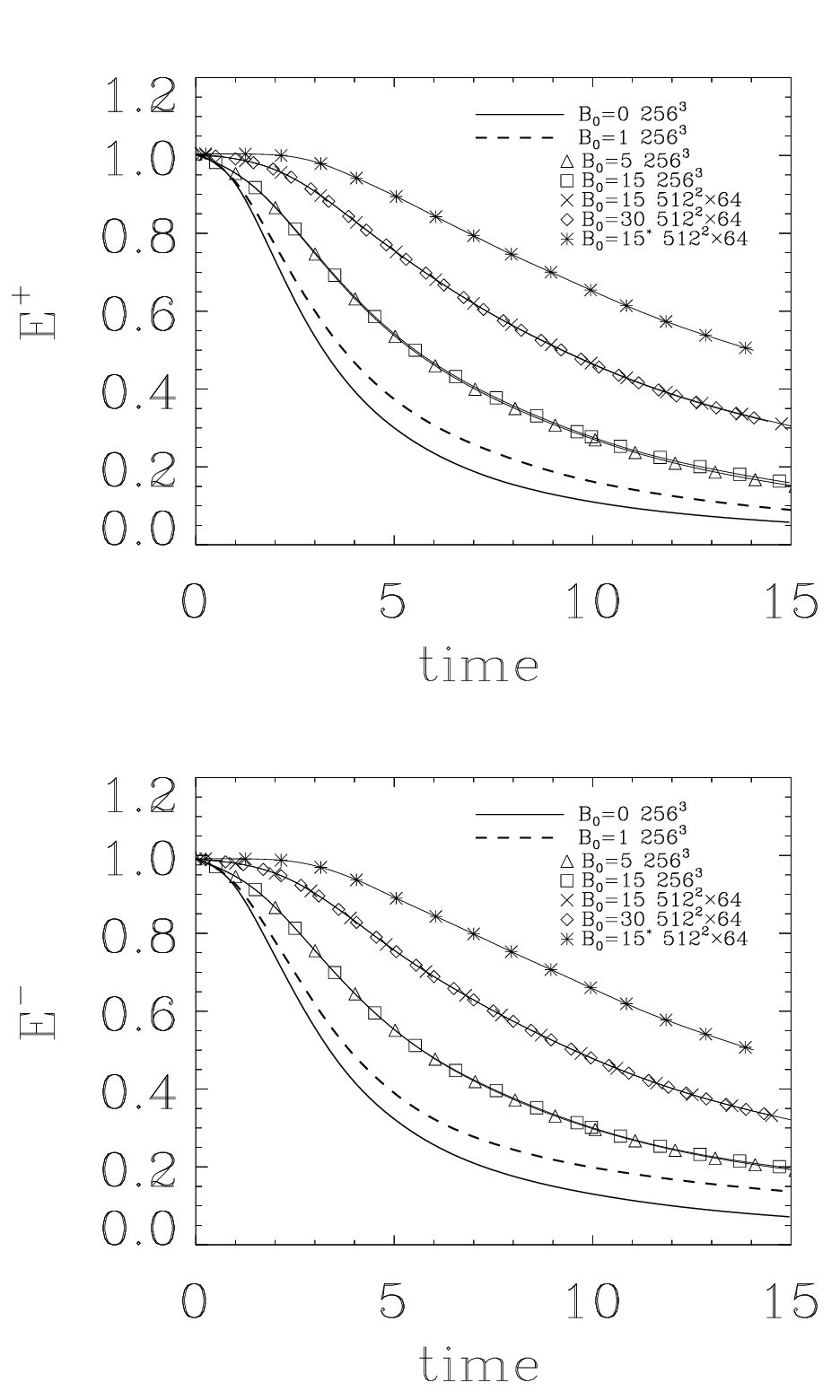}}
\caption{Temporal evolution of energies $E^+$ (top) and $E^-$ (bottom) for $B_0=0,1,5,15$  (runs $\bf Ia$ 
to $\bf IVa$; $256^3$) and $B_0=15,30$ (runs $\bf Va$ and $\bf VIa$; $512^2\times64$). 
The hyperviscous run $\bf VIIa$ with $B_0=15$ ($512^2\times64$) is also given up to $t=14$. 
\label{FigtmpE}}
\end{figure}
\begin{figure}[ht]
%\resizebox{88mm}{!}{\includegraphics{FigtmpOm+hv.eps}}
\resizebox{88mm}{!}{\includegraphics{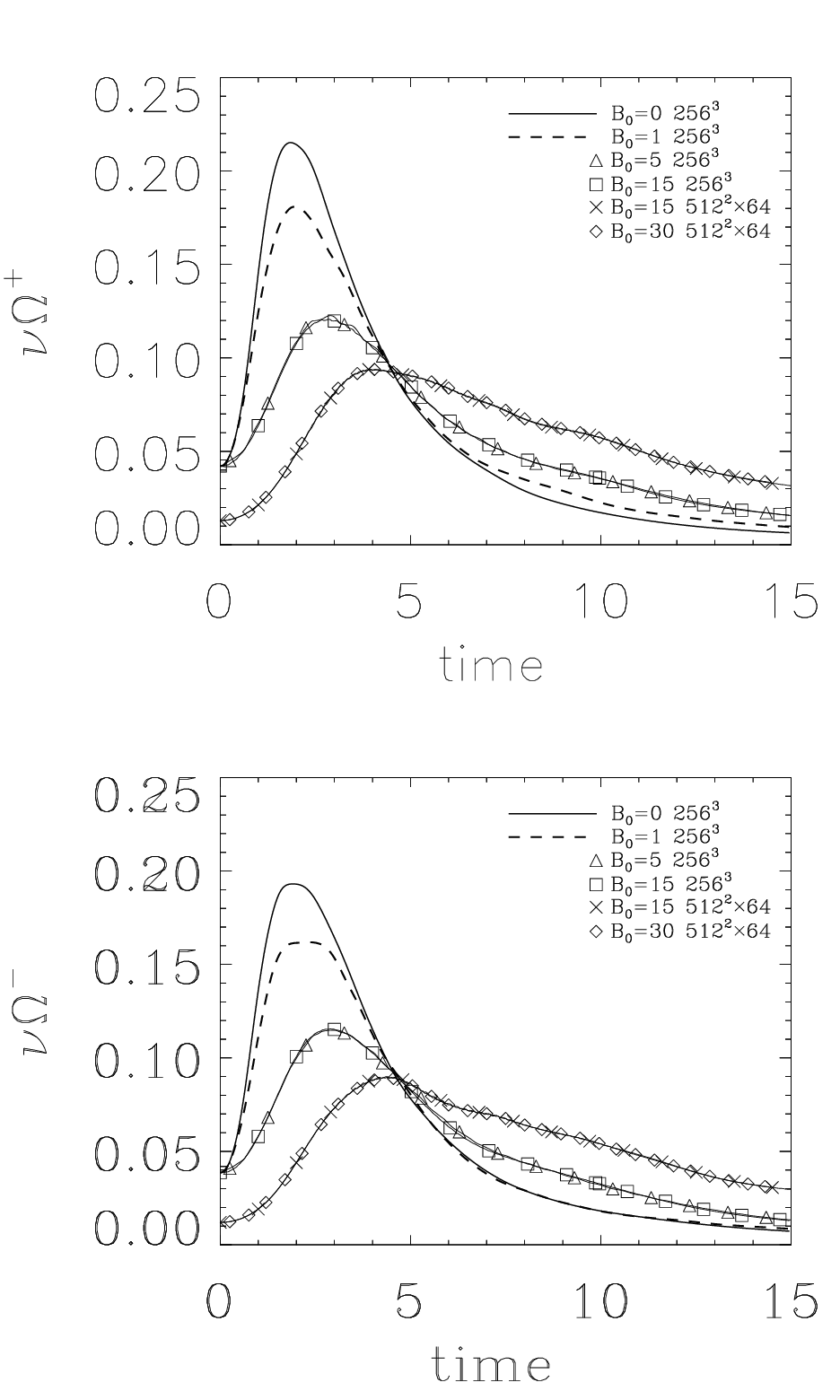}}
\caption{Temporal evolution of the global dissipation $\nu \Omega^+$ (top) and $\nu \Omega^-$ (bottom); 
same viscous runs as in Figure~\ref{FigtmpE}. 
\label{Figtmpdiss}}
\end{figure}
We first consider the evolution of the  Els\"asser energies, 
\be E^\pm(t) =\frac{1}{2}<{\bf z}^{\pm^2}({\bf x})>(t) \, ,
\ee
displayed in Figure \ref{FigtmpE}. Note that, for 
periodic boundary conditions, these energies are two independent invariants of the inviscid MHD 
equations (\ref{mhd3}), with or without the presence of a uniform magnetic field. Energies $E^+(t)$ 
and $E^-(t)$ present a similar behavior for a given $B_0$. For runs {\bf Ia} to {\bf IIIa}, where ${\bf B_0}$ 
intensity is increased, we clearly see a slowdown of the energy decay.  
On one hand, this slowing down reflects the energy transfer inhibition along the ${\bf B_0}$ direction, and 
thus, the flow inability to create, in the parallel direction, smaller and smaller scales up to the dissipative 
ones. Hence, the energy dissipation mainly takes place in transverse planes which are led to play a more efficient role as the flow magnetization is increased. On the other hand, energy transfers themselves could 
also be weakened (in the transverse planes) since the MHD cascade of energy to smaller scales is 
produced by successive interactions of oppositely directed waves. Indeed, for higher ${\bf B_0}$ 
intensities, the waves become faster and thus the time duration of individual collision of $z^\pm$ waves 
decreases. Therefore it takes much more collisions between (fast) Alfv\'en wave packets (as 
measured by the ratio between the nonlinear turnover time on the linear wave period; $\tau_{NL}/\tau_A$) 
to have an efficient energy cascade process.
One could also note that, for a given flow, a saturation effect occurs according to $B_0$ intensities.
Indeed, the $E^\pm(t)$ evolutions are quite similar for flows at $\nu=4.10^{-3}$ with $B_0=5$ and $15$ 
(runs $\bf IIa$ and $\bf IVa$, respectively), as well as for flows at $\nu=10^{-3}$ with $B_0=15$ and 
$30$ (runs $\bf Va$ and $\bf VIa$, respectively).  
The hyperviscous run $\bf VIIa$ is also shown but only up to $t=14$. We see that the initial plateau 
is wider and almost flat because of the larger inertial range and the higher Reynolds number. Then, 
we see a decay of energy which is slower than for the other viscous runs.

The $B_0$ saturation effect is also visible on the time evolution of the global dissipation of the flow,
\be
\nu \Omega^\pm(t)=\nu <[\nabla \times {\bf z}^\pm]^2({\bf x})>(t) \, , 
\ee
displayed in Figure~\ref{Figtmpdiss}. 
The early time dynamics, near the first inflection point, is almost inviscid; it corresponds to the small-scale 
generation (\eg at times $t \leq 1$ in the $B_0=0$ simulation). As the $\bf B_0$ intensity is increased, this 
small-scale development is slightly retarded which means that the duration of the essentially inviscid phase 
increases. Moreover, the maximum of the dissipation is substantially reduced, and occurs at later times, namely $t \sim 2$ for flows with $B_0=0$ and $1$, $t \sim 3$ with $B_0=5$, $15$ and $t \sim 4$ for the 
less viscous flows with $B_0=15$, $30$. 
Altogether, in physical space, this corresponds to the creation of more elongated structures along 
$\bf B_0$ as the flow is more magnetized, with a smaller dissipation on the whole, and, in spectral space,
to higher inhibition of parallel energy transfers, as already explained. 
One can also note that the dissipation peak is smoothed in the less viscous flows ($\nu=10^{-3}$), 
meaning an almost constant dissipation between $t \sim 3$ and $t \sim 5$ in  runs $\bf Va$ and $\bf VIa$ 
with a more extended range of small scales. 
Finally, note a different evolution between case $\bf IVa$ (with $\nu=4.10^{-3}$) and $\bf Va$ (with 
$\nu=10^{-3}$) whereas the uniform field $B_0$ is the same. A factor $4$ of difference is visible initially 
which may be attributed mainly to a decrease of factor $4$ of the viscosity. In this case, the time delay to 
reach the maximum may be explained by a wider inertial range in $\kpe$ and therefore a longer time 
needed to reach the dissipative scales (an effect also seen in Figure \ref{FigtmpE} with a wider initial 
plateau where energy is roughly conserved).
%We also display the hyperviscous run for which the dissipation is multiplied by a factor 1000. In this 
%case, the shape is quite different with the formation of a plateau before the decay phase. 
\begin{figure}
\resizebox{88mm}{!}{\includegraphics{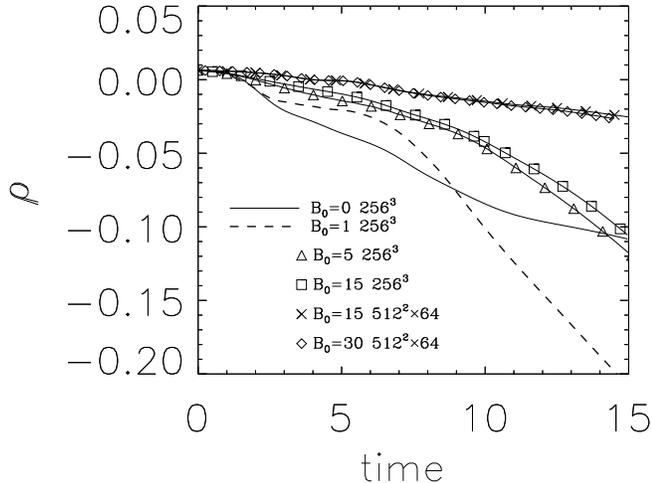}}
\caption{Temporal evolution of the cross-correlation coefficient $\rho$; same runs as in 
Figure~\ref{FigtmpE}.}
\label{Figtmprho}
\end{figure}

Figure \ref{Figtmprho} shows the cross-correlation coefficient (\ref{rho}) between velocity and 
magnetic fields which also reads, in terms of the  Els\"asser energies, as
\be
\rho (t)= {E^+(t) - E^-(t) \over E^+(t) + E^-(t)} \, . 
\ee
It measures the relative amount of the two $z^\pm$ species. Indeed, $\rho(t) \rightarrow \pm 1$ means 
that $E^\mp=0$, and hence only one type of waves is excited, whereas when  $\rho(t) \rightarrow 0$,
they are as many $z^+$ as $z^-$ counterpropagating waves, with the same amount of energy. Initially, 
$\rho(t=0) \sim 0$ (\ie less than $1\%$), and stays so during the flow inviscid phases. Close to the times 
at which the maximum of dissipation occurs in the different flows, $\rho(t)$ deviates from zero with a lesser 
departure as the flow is more magnetized, from $B_0=1$ up to $B_0=30$, because the field lines are 
rigidified by the ambiant magnetic field, and the dissipation is delayed.
In the case of $ B_0=0$, the temporal evolution of the cross-correlation coefficient is globally different, 
due to the absence of a guiding magnetic field and different dissipative processes. Note however that 
all flows evolve toward an excess of $E^-$ energy.

\begin{figure}[ht]
\includegraphics[width=88mm]{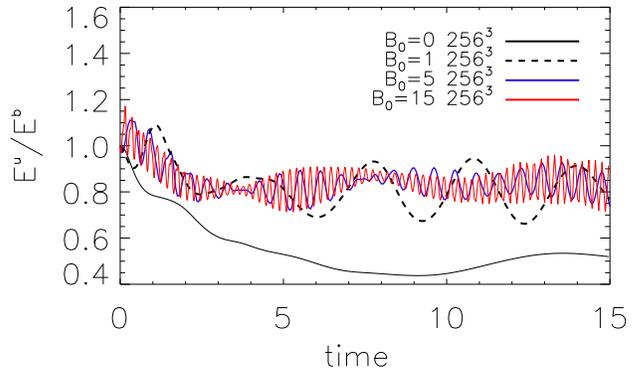}
\caption{Time evolution of the Alfv\'en ratio $r_A$ for runs ${\bf Ia}$ to ${\bf IVa}$.
\label{figtmp_rA}}
\end{figure}
Apart from this, the prevalence of the Alfv\'en wave fluctuations can be measured by the so-called 
Alfv\'en ratio 
\be
r_A(t)= {E^v(t) \over E^b(t)}={<{\bf v}^2>(t) \over <{\bf b}^2>(t)} \, ,
\ee 
between kinetic and magnetic energies. For example in the wave turbulence regime we have an 
equipartition (at the level of the kinematics \cite{Galtier2000}) between kinetic and magnetic energies. 
Its departure from unity suggests the presence of non Alfv\'enic fluctuations. Indeed, the energy of an 
individual Alfv\'en wave is equipartitioned between its kinetic and magnetic components, averaged 
over a wave period, with thus a ratio $r_A=1$. 
In presence of an external magnetic field, exchanges between magnetic and velocity fluctuations, 
due to Alfv\'en waves, produce oscillations as shown on $r_A(t)$ in Figure~\ref{figtmp_rA}.
The period of these oscillations is given by the Alfv\'en time $\tau_A \sim 5$, ${1}$ and ${0.4}$ 
(see Table \ref{table1}) which are 
found by a simple analysis based on the values $B_0=1$, $5$ and $15$, respectively, and the values of 
the characteristic parallel length scale $L^\parallel \sim 5.57$ for runs $\bf IIa$ to $\bf IVa$.
Although, initially the magnetic and kinetic energies are chosen equal, $E^v(t=0)=E^b(t=0)=0.5$, the 
magnetic energy stabilizes around twice the kinetic energy, after $t \sim 5$, for the non-magnetized flow 
($B_0=0$). While for the magnetized flows, whatever the $\bf B_0$ intensity is, the magnetic energy 
saturates to about $1.25$ lower than the kinetic energy level after time $t \sim 2$. This result may be 
compared with solar wind data where the same tendency is found with a domination of the magnetic 
energy. (This comparison is however not direct since outward propagating Alfv\'en waves are initially 
dominant.) This Alfv\'en ratio seems to find a limit of about $1/2$ at several astronomical units which 
might be explained by the decreasing importance of the large-scale magnetic field at larger heliocentric 
distances (see \eg \cite{bruno05}). 

\begin{figure*}[ht]
\begin{tabular}{cccc}
\resizebox{80mm}{!}{\includegraphics{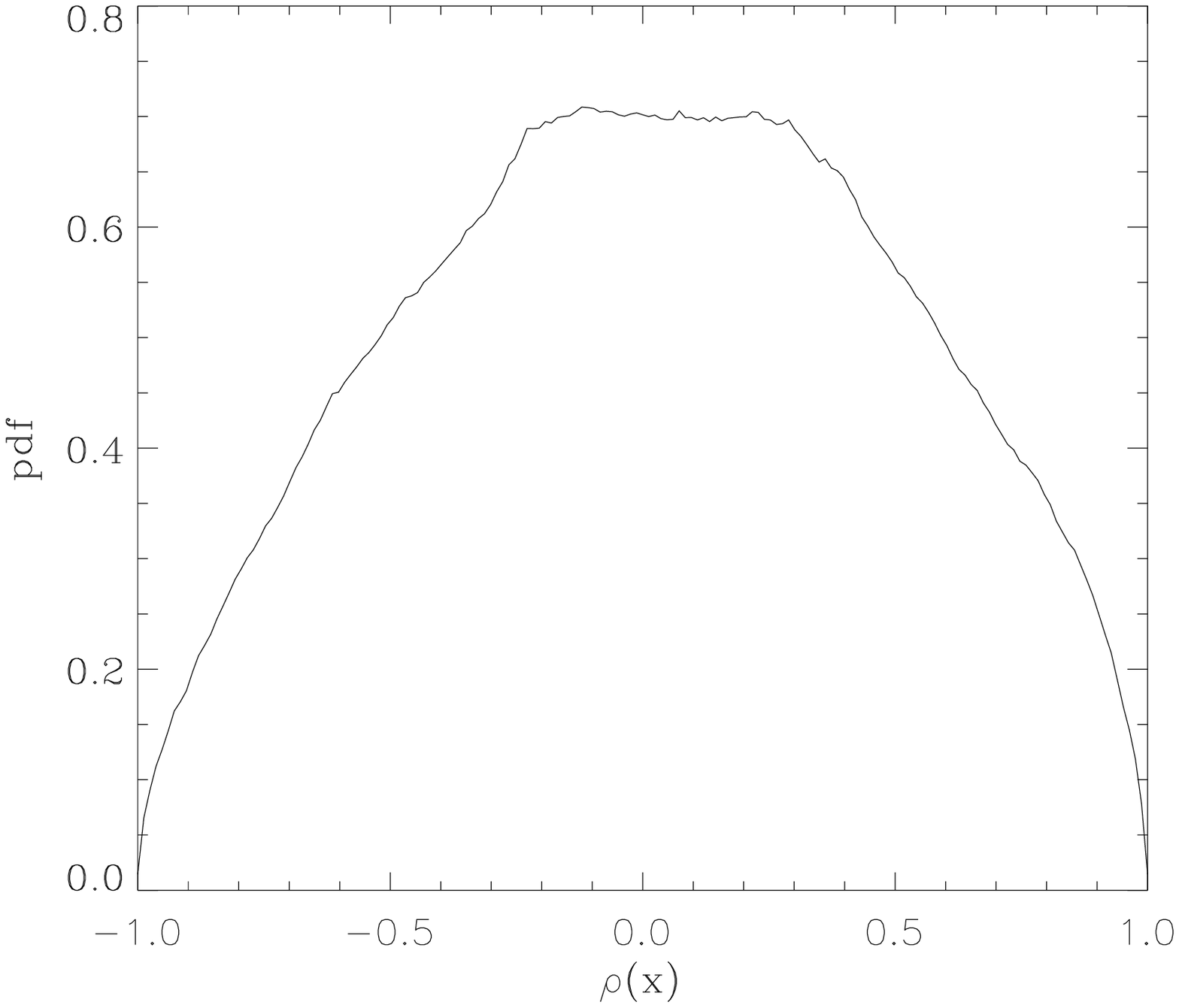}}
\resizebox{80mm}{!}{\includegraphics{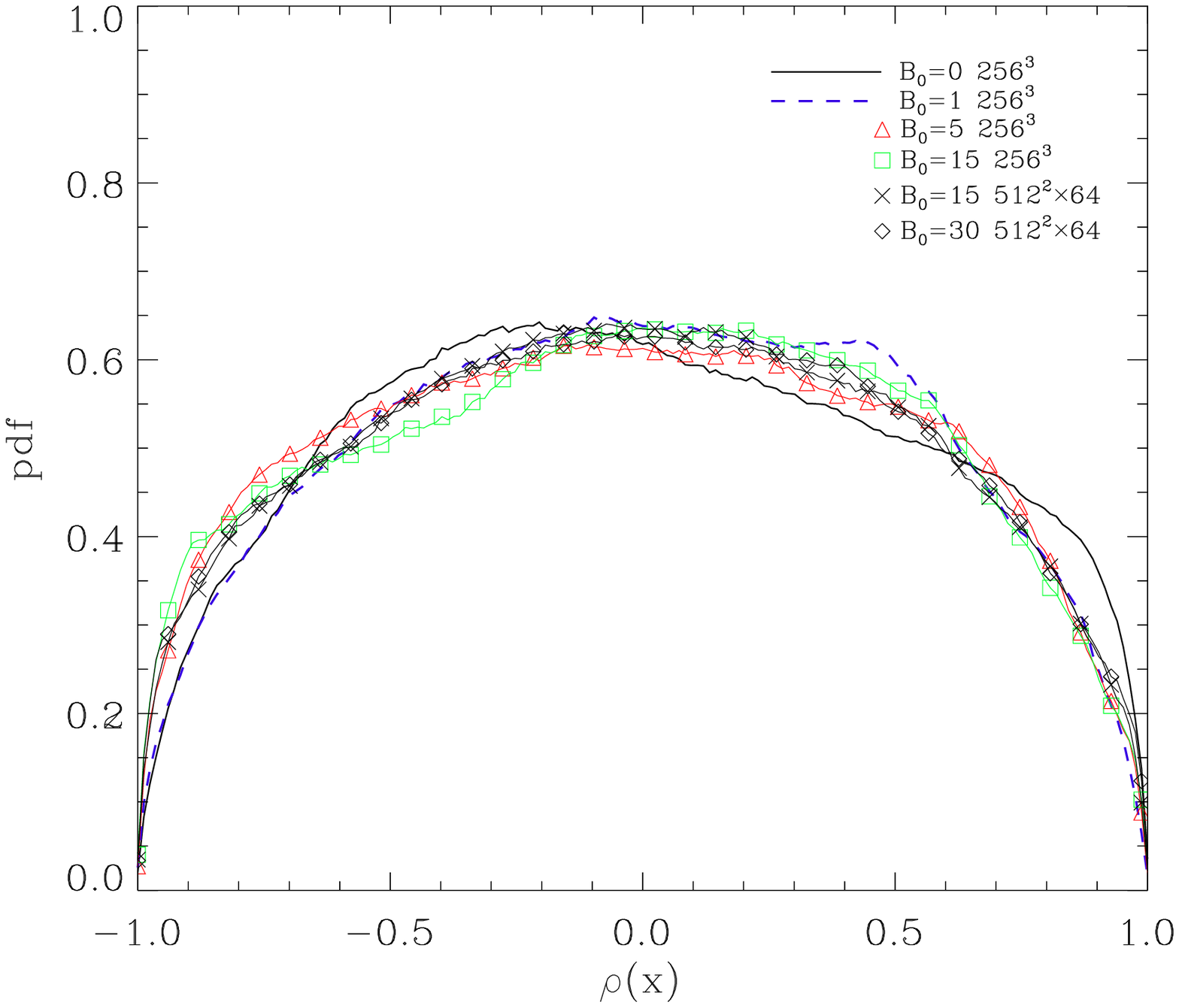}}\\
\resizebox{80mm}{!}{\includegraphics{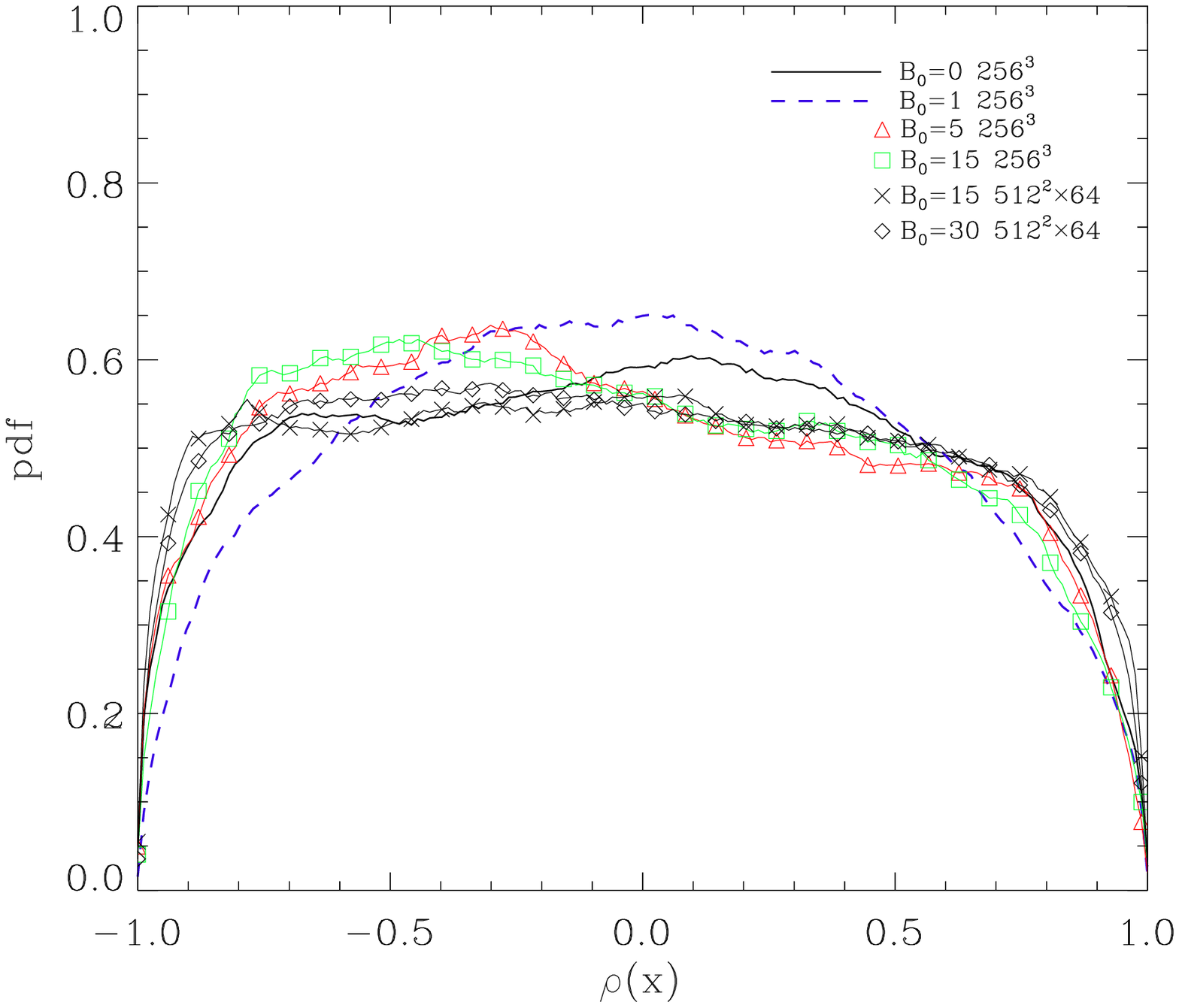}}
\resizebox{80mm}{!}{\includegraphics{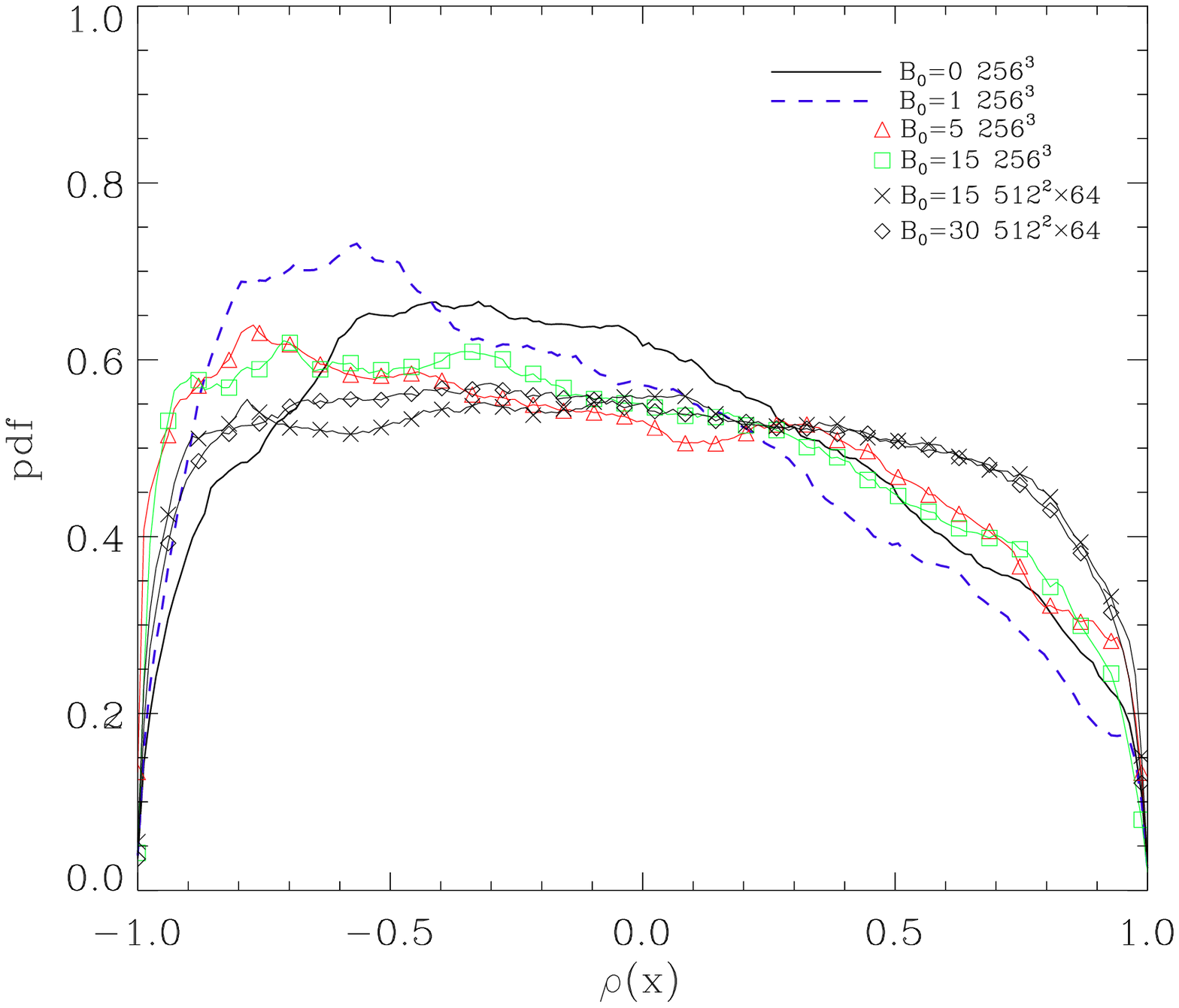}}
\end{tabular}
\caption{Probability distribution functions of the cross-correlation (runs ${\bf Ia}$ to ${\bf VIa}$) 
initially (top left), at the maximum of the global dissipation (top right), when $77$\% of the total energy
is dissipated (bottom left) and at the final time (bottom right). 
\label{Fig_pdf_cor}}
\end{figure*}
Figure \ref{Fig_pdf_cor} displays the pdf of the cross-correlation for different times (same runs as in 
Figure \ref{Figtmprho}). As expected, we start initially with a distribution clearly centered around zero. 
As the time increases, we see a distribution shifted towards negative values to finally be centered 
around $-0.4$ for the non-magnetic case. The case $B_0=1$ is even more shifted with a maximum 
of the distribution around $-0.6$. The strongly magnetic cases are mainly characterized by the formation 
of extended plateaux centered around the negative values. This result means that although the 
cross-correlation coefficient (\ref{rho}) is close to zero for strongly magnetized flows (see Figure 
\ref{Figtmprho}), a wide range a values is often reached locally.

\subsubsection{Shear- and pseudo-Alfv\'en wave decomposition}

In the presence of an external magnetic field, it is convenient to describe the flow dynamics in terms 
of shear- and pseudo-Alfv\'en waves, or in other words to use, respectively, the toroidal and poloidal 
components of the ${\bf z}^\pm$ fields (see equation (\ref{DECPT})). Indeed, the Alfv\'en waves 
dynamics for the stronger magnetized flows have crucial consequences on the turbulent properties. 
We will use here the shear- and pseudo-Alfv\'en wave decomposition to analyze our numerical 
simulations and, therefore, we will not considered the $B_0=0$ case anymore. 

\begin{figure}[ht]
\resizebox{88mm}{!}{\includegraphics{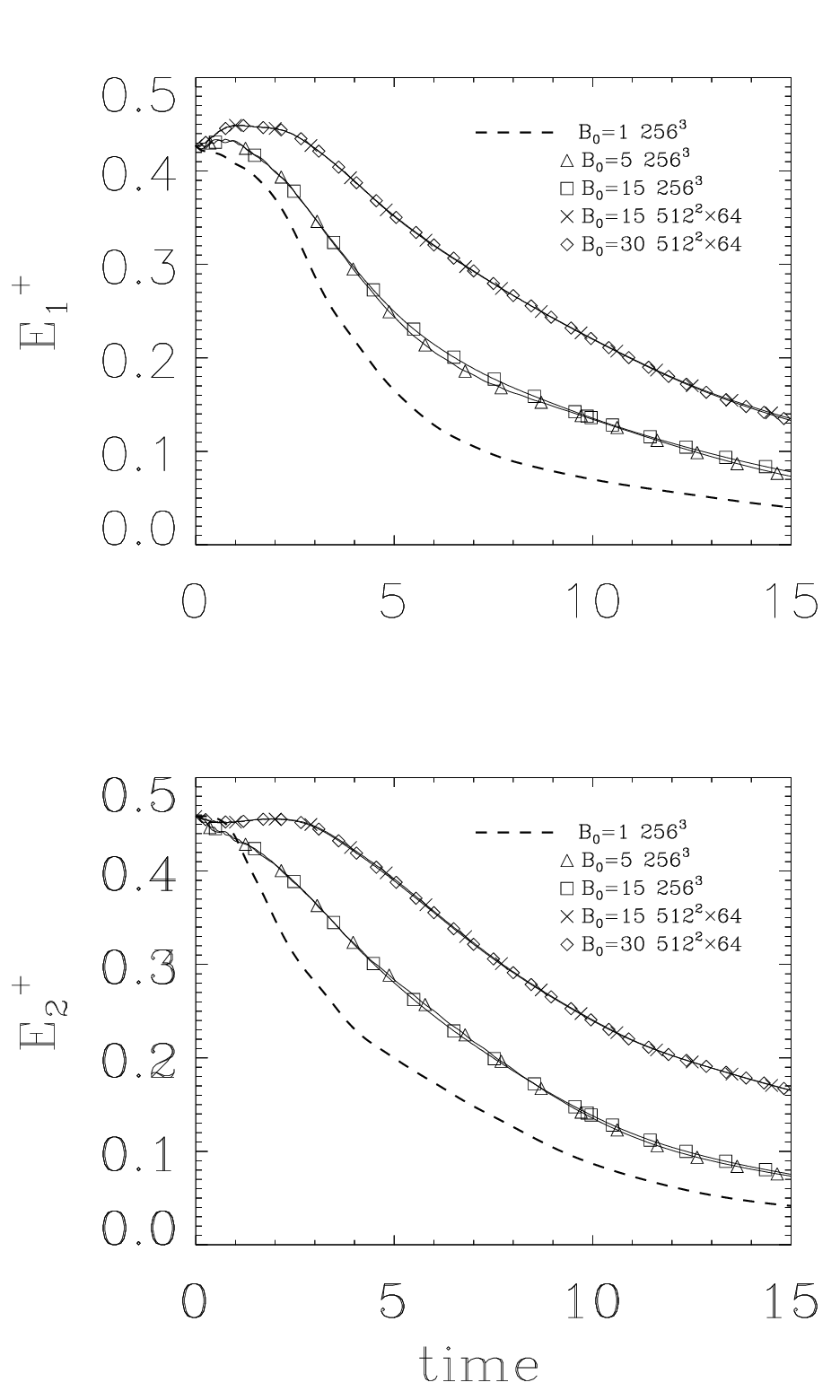}}
\caption{Temporal evolution of energies $E^+_1$ (top) and $E^+_2$ (bottom) of the shear-
and pseudo-Alfv\'en waves for runs ${\bf IIa}$ to ${\bf VIa}$.
\label{FigtmpEnAlz}}
\end{figure}
In Figure \ref{FigtmpEnAlz}, we show the temporal evolutions of energies $E^{+}_1$ and $E^{+}_2$ 
associated, respectively, to the shear-Alfv\'en and pseudo-Alfv\'en waves; they are defined as 
\be
E_{1,2}^+ (t)=< {\bf z}_{1,2}^{{+}^2}>(t) \, ,
\ee
and are not inviscid invariants. 
Note that $E^\pm \neq E_1^\pm + E_2^\pm$ because the energy contained into the $k_\perp=0$ modes 
are not included in the toroidal/poloidal decomposition (although it is, of course, in the original cartesian 
fields). 
%; thus, we cannot classify this type of waves as a shear- or a pseudo-Alfv\'en waves. 
%We only show the energies $E_{1,2}^+$ (the evolution of energies $E_{1,2}^-$ being similar). 
First, we observe a slowdown of the energy decay when the intensity of $B_0$ increases. It is a 
behavior similar to the one found in Figure \ref{FigtmpE} for the energies $E^{\pm}$. With such a 
decomposition, a similar behavior is also found for runs ${\bf IIIa}$ and ${\bf IVa}$, and runs ${\bf Va}$ 
and ${\bf VIa}$. The important new information is about the initial increase of energies which is more 
pronounced for runs ${\bf Va}$ and ${\bf VIa}$, and for the shear-Alfv\'en waves. These energies are 
in fact pumped from the $\kpe=0$ mode (the total energy is not an increasing function). 
%This major difference with Figure \ref{FigtmpE} is mainly linked to the nature of these quantities which 
%are not conserved in the inviscid case. 
Note that the same behavior is found for the $-$ polarity. 

\begin{figure}[ht]
\resizebox{88mm}{!}{\includegraphics{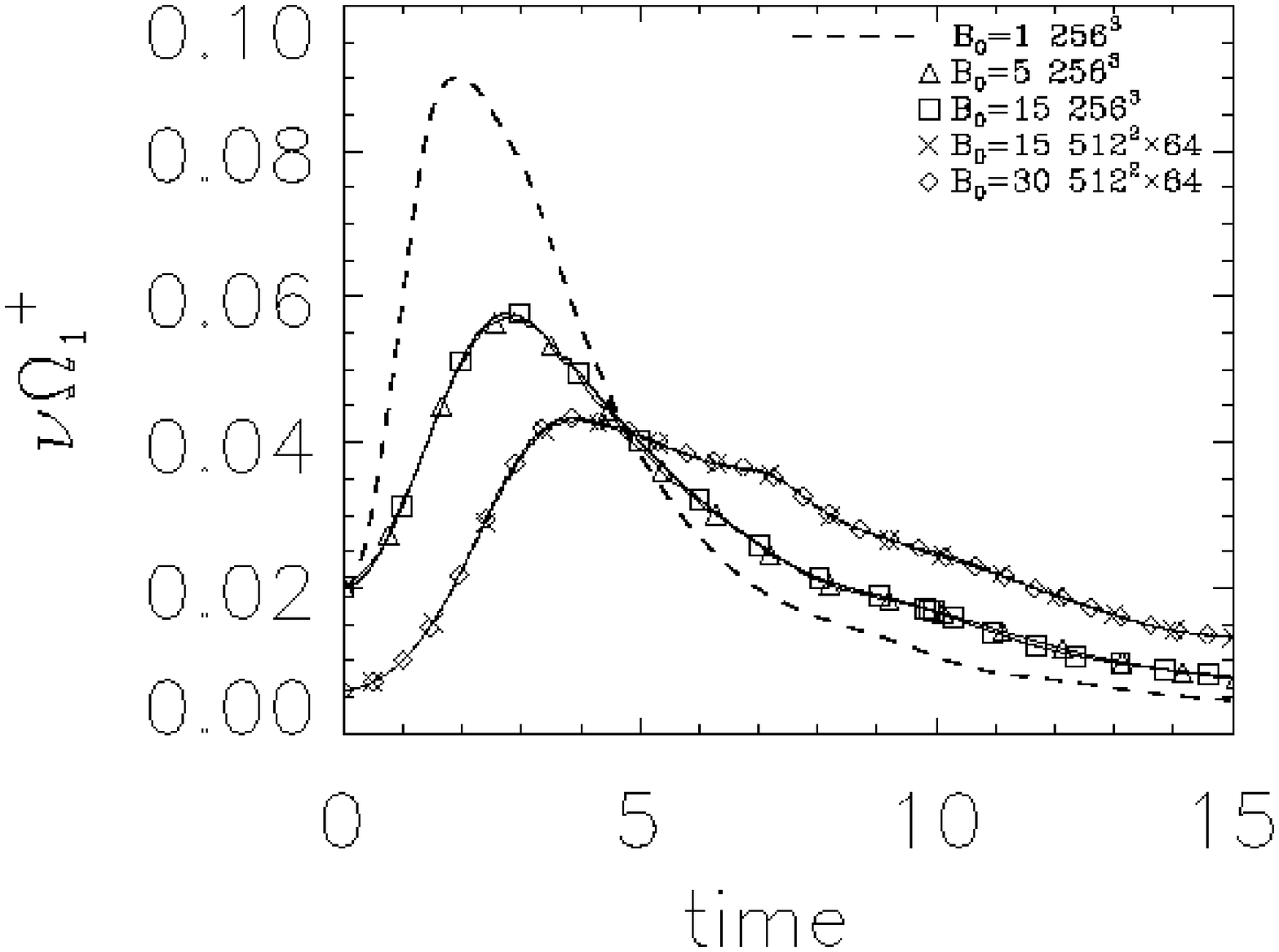}}\\
\resizebox{88mm}{!}{\includegraphics{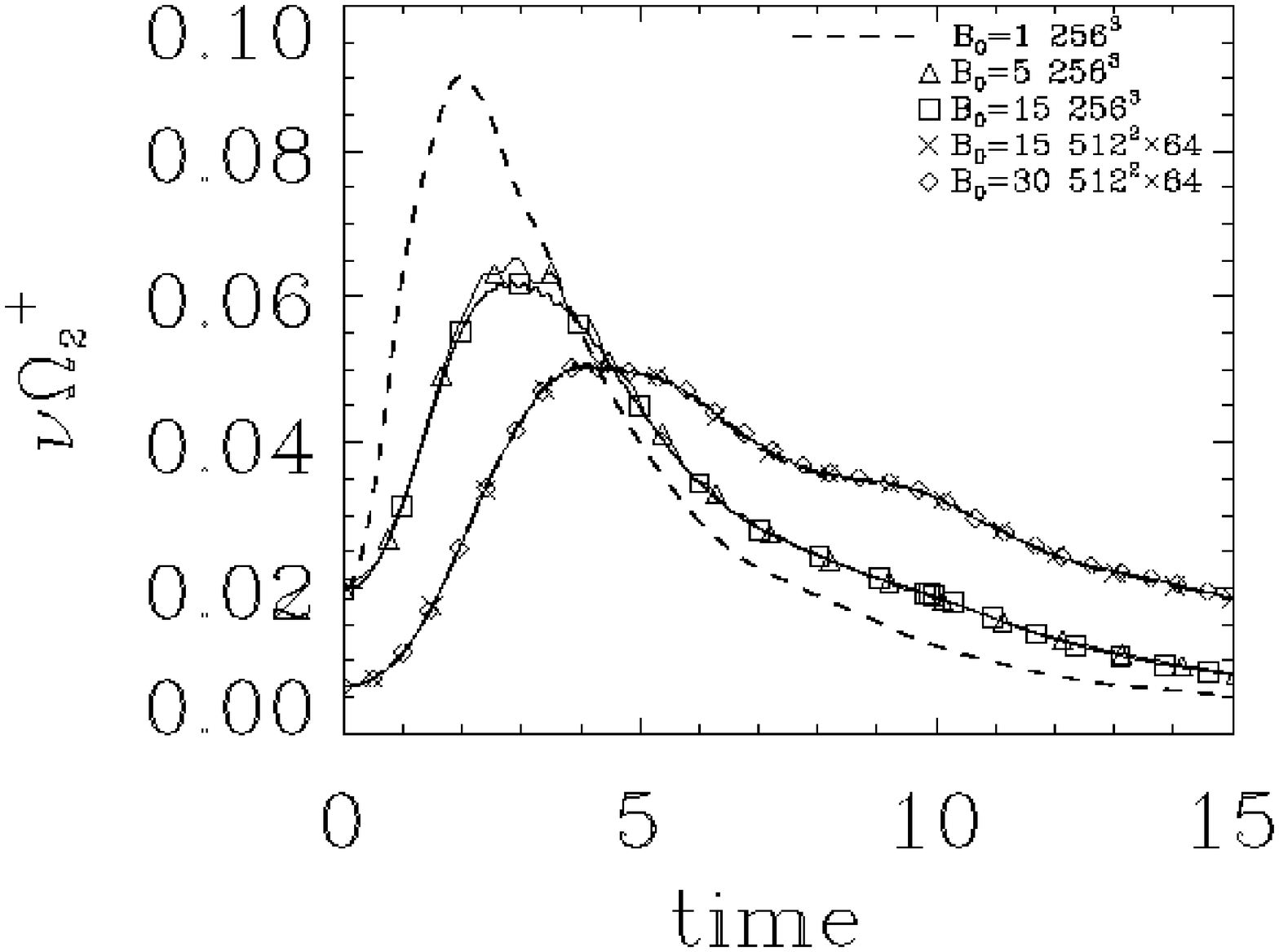}}
\caption{Temporal evolution of the dissipations $\nu \Omega^+_1$ (top) and $\nu \Omega^+_2$ (bottom) 
of shear- and pseudo-Alfv\'en waves for runs ${\bf IIa}$ to ${\bf VIa}$. 
\label{FigtmpdissAls}}
\end{figure}
Figures \ref{FigtmpdissAls} presents the temporal evolution of the dissipations 
\be
\nu \Omega_{1,2}^+(t) = \nu <[\nabla \times ({\bf z}_{1,2}^+)]^2>(t) \, ,
\ee
for, respectively, the shear and pseudo-Alfv\'en waves (only the $+$ polarity is shown since the same 
behavior is found for the $-$ polarity). No clear difference is found between the type of dissipation. We 
also note no significant difference with Figure \ref{Figtmpdiss} except a factor two in magnitude because 
here we do not see the total dissipation for a given polarity but either the shear- or the pseudo-Alfv\'en 
waves contribution. 

\begin{figure}[ht]
\begin{tabular}{ll}
\resizebox{88mm}{!}{\includegraphics{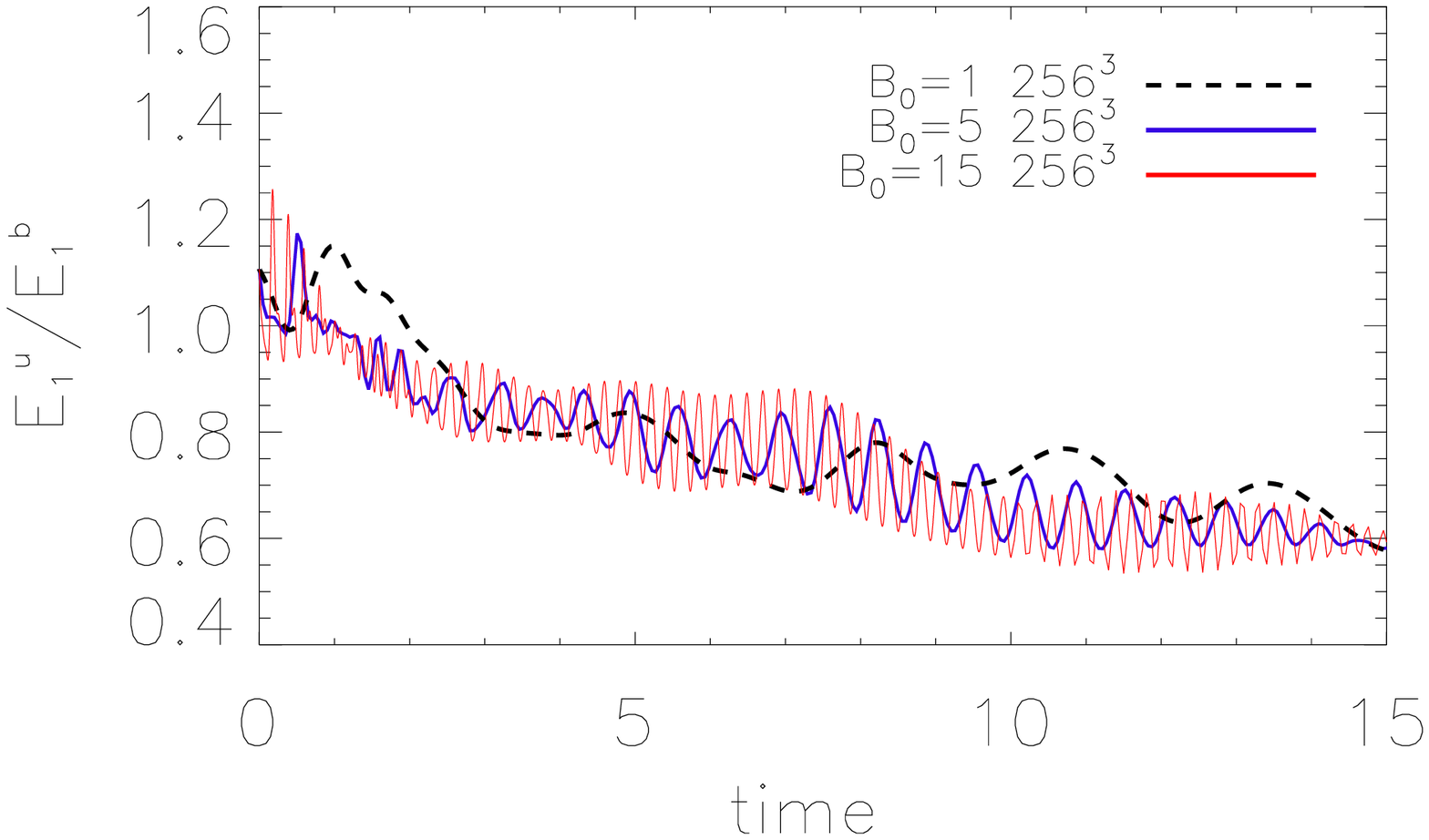}}\\
\resizebox{88mm}{!}{\includegraphics{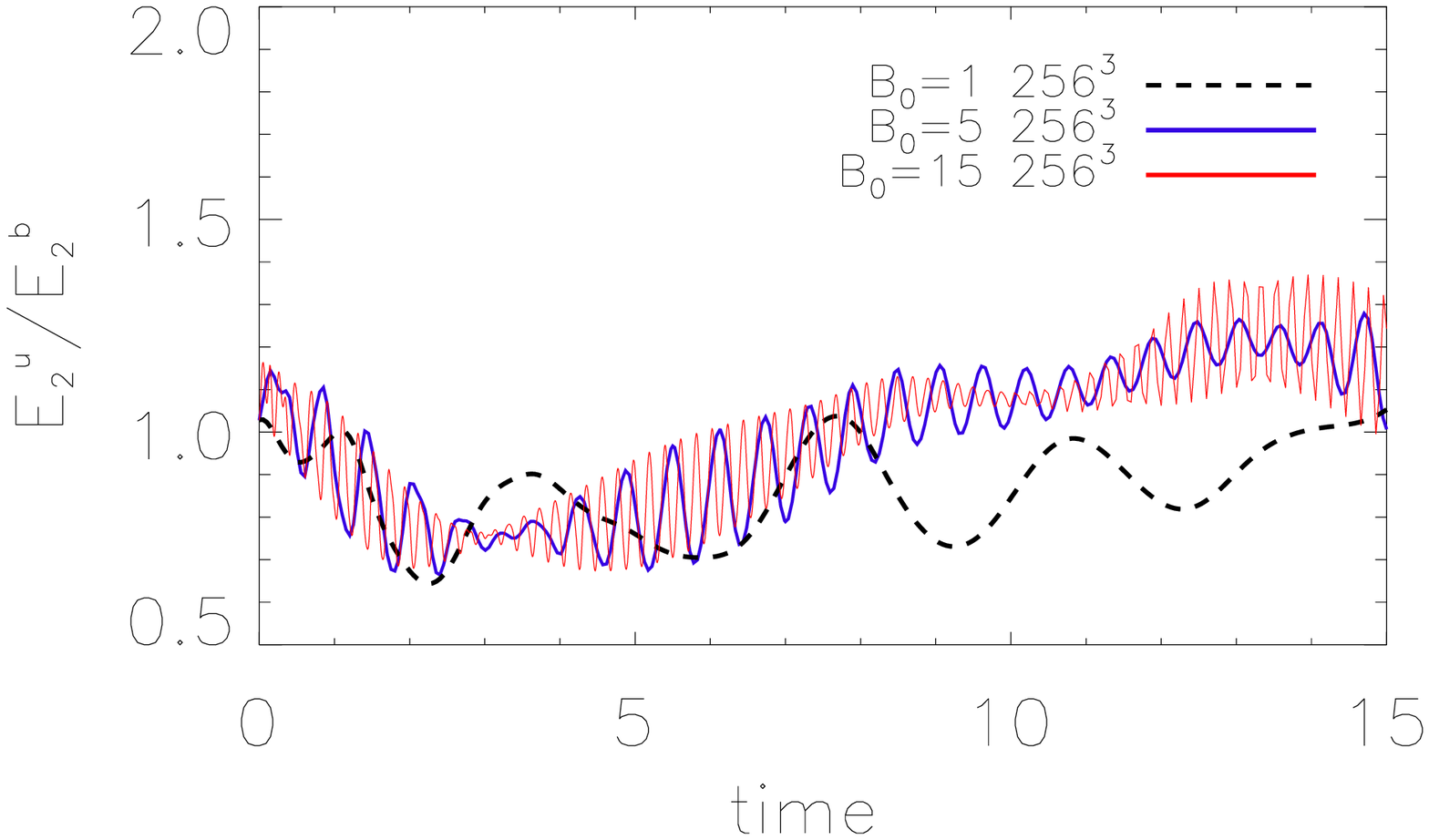}}
\end{tabular}
\caption{Temporal evolution of Alfv\'en ratio for shear- and pseudo-Alfv\'en waves 
(runs ${\bf IIa}$ to ${\bf IVa}$).}
\label{Figtmp_rA_Al}
\end{figure}
\begin{figure}[ht]
\begin{tabular}{ll}
\resizebox{85mm}{!}{\includegraphics{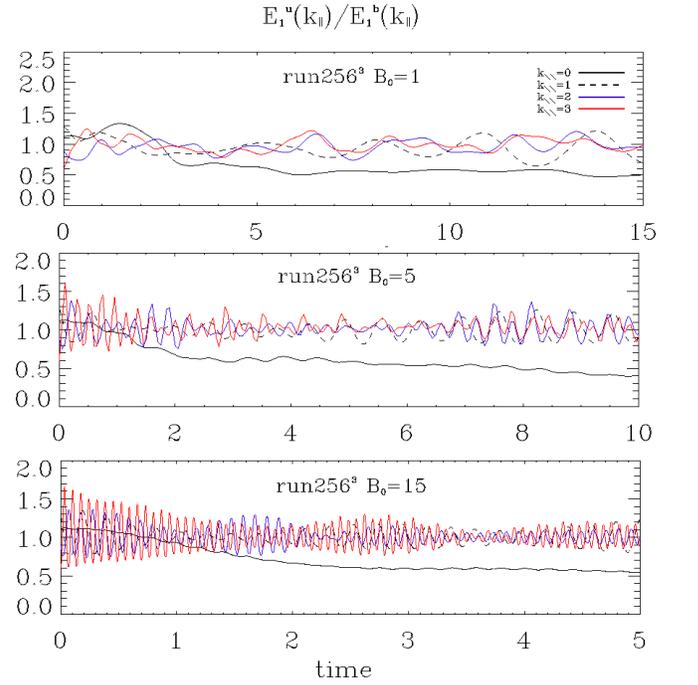}}
\end{tabular}
\caption{Temporal evolution of the spectral Alfv\'en ratio for shear-Alfv\'en waves 
at a given parallel wavenumber (runs ${\bf IIa}$ to ${\bf IVa}$).}
\label{Figtmp_rA_Al2}
\end{figure}
Figure \ref{Figtmp_rA_Al} presents the temporal evolution of the alfv\'enicity (or Alfv\'en ratio) 
\be
r_{A{_{1,2}}}(t)={E_{1,2}^v(t) \over E_{1,2}^b(t)} \, , 
\ee
for the shear- and pseudo-Afv\'en waves respectively, with 
\be
E_{1,2}^{v} = {1 \over 2} < ({\bf z}^+_{1,2}+{\bf z}^-_{1,2})^2> \, ,
\ee
and 
\be
E_{1,2}^{b} = {1 \over 2} < ({\bf z}^+_{1,2}-{\bf z}^-_{1,2})^2> \, .
\ee
This plot is particularly interesting since it shows that shear-Alfv\'en waves and pseudo-Alfv\'en waves 
behave differently with an Alfv\'en ratio of about one for the latter and significantly smaller than one for 
the former. Since for strongly magnetized flows the perpendicular fluctuations are mainly made of 
shear-Alfv\'en waves and the parallel one made of pseudo-Alfv\'en waves we have here a prediction 
that can be compared with measurements made in natural plasmas like in the solar wind. 
Additionally, we observe the same oscillations as in Figure \ref{figtmp_rA} where the same type 
of analysis on the time-scales may be made. 

Figure \ref{Figtmp_rA_Al2} displays the temporal evolution of the spectral alfv\'enicity for
shear-Alfv\'en waves 
\be
r_{A_{\Large 1}}(k_\parallel,t)=\frac{E_1^u(k_\parallel,t)}{E_1^b(k_\parallel,t)}, 
\ee
with $k_\parallel=0,1,2$ and $3$ (run {\bf IIa} to {\bf IVa}). 
The initial Afv\'en ratio is close to unity for every parallel wavenumbers, then a different 
behavior is found for the 2D state ($k_\parallel=0$) which deviates strongly from the equipartition 
and tends approximately to $1/2$ independently of the $B_0$ intensity. For the 3D modes, the spectral 
Alfv\'en ratio oscillates around unity meaning a tendency towards equipartition between the kinetic 
and magnetic energies. This tendency is stronger for stronger magnetized flows. Thus the 3D modes 
follow the dynamics expected in wave turbulence in which an exact equipartition happens 
\cite{Galtier2000}. It is actually the 2D state which explains the behavior found previously in Figure 
\ref{Figtmp_rA_Al} where a discrepancy from the equipartition was observed. Therefore, Figure
\ref{Figtmp_rA_Al} is not in contradiction with the wave turbulence regime and offer a new possible 
interpretation of observations in natural plasmas like the solar wind. Note that the same type of results 
is found for pseudo-Alfv\'en waves (not shown) with a deviation from the equipartition for the 2D state.

%%%%%%%%%%%%%%%%%%%%%%%%%%%%%%%%%%%%%%%%%%%%%%
\subsection{Characteristic length- and time-scales}

\begin{figure}[ht]
\begin{tabular}{cc}
\resizebox{88mm}{!}{\includegraphics{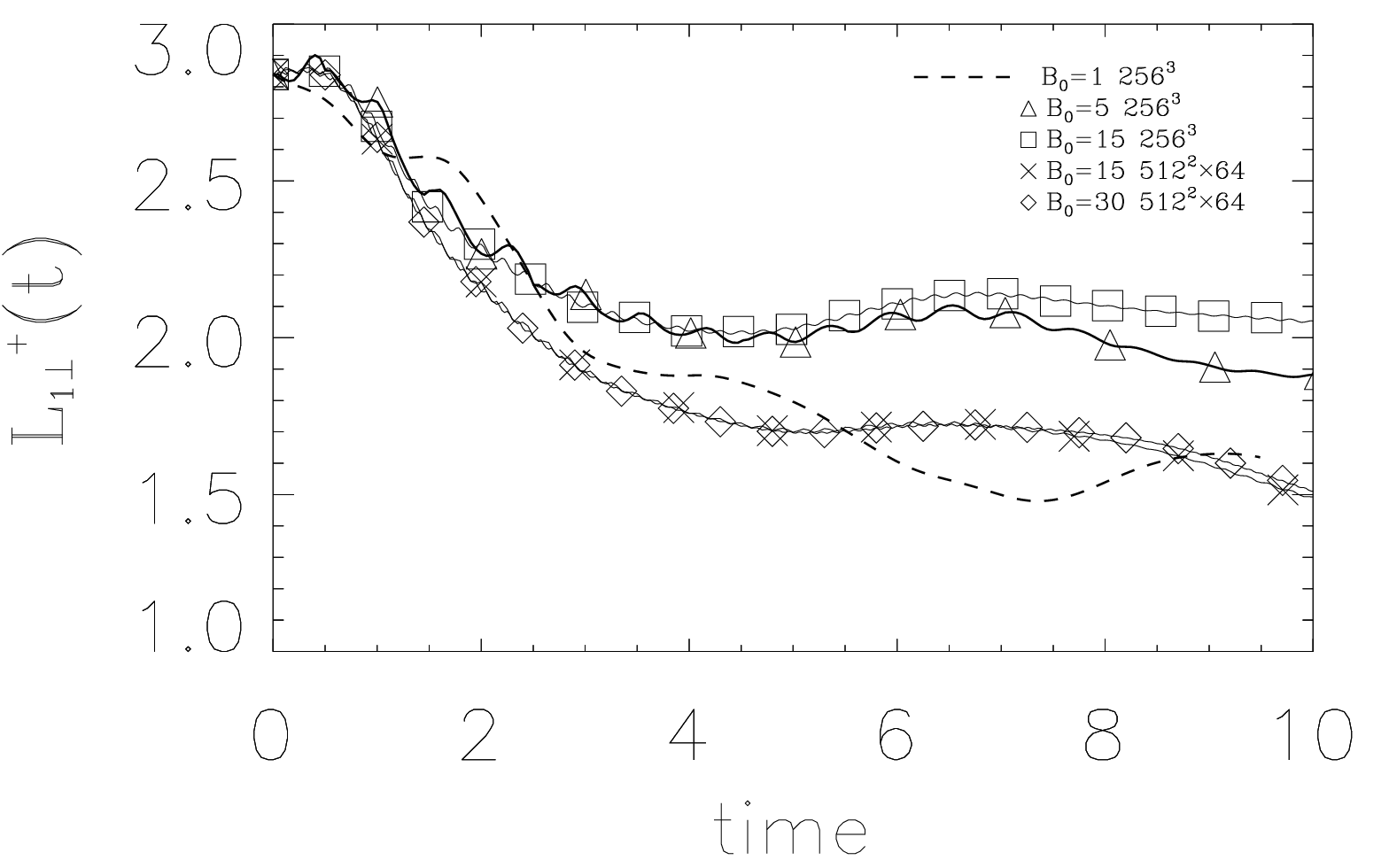}}\\
\resizebox{88mm}{!}{\includegraphics{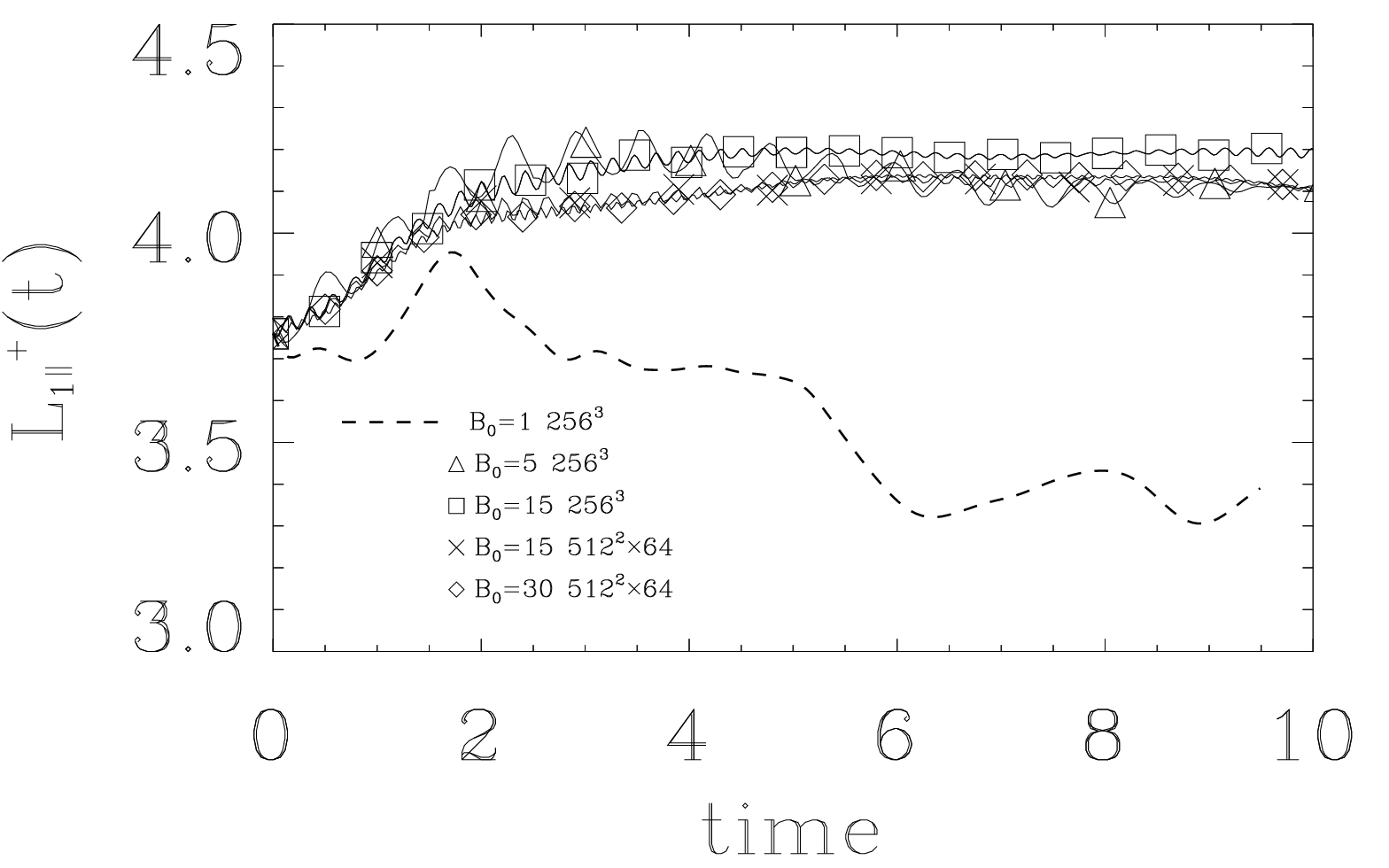}}
\end{tabular}
\caption{Temporal evolution, up to $t=10$, of perpendicular (top) and parallel (bottom) integral length-scales 
for shear-Alfv\'en ($+$) waves; same runs as in Figure \ref{FigtmpdissAls}. 
\label{figLint}}
\end{figure}
\begin{figure}[ht]
\begin{tabular}{cc}
\resizebox{88mm}{!}{\includegraphics{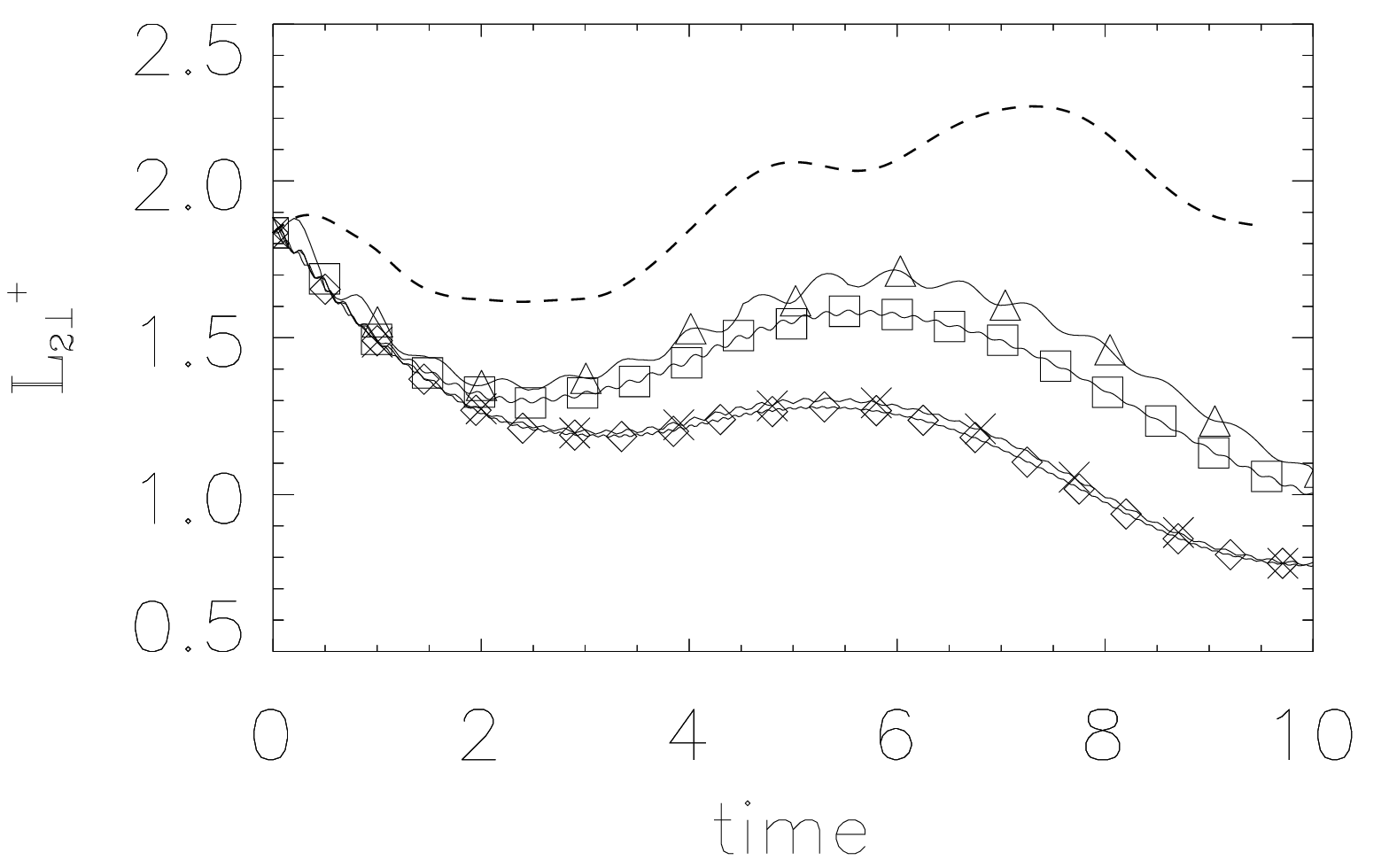}}\\
\resizebox{88mm}{!}{\includegraphics{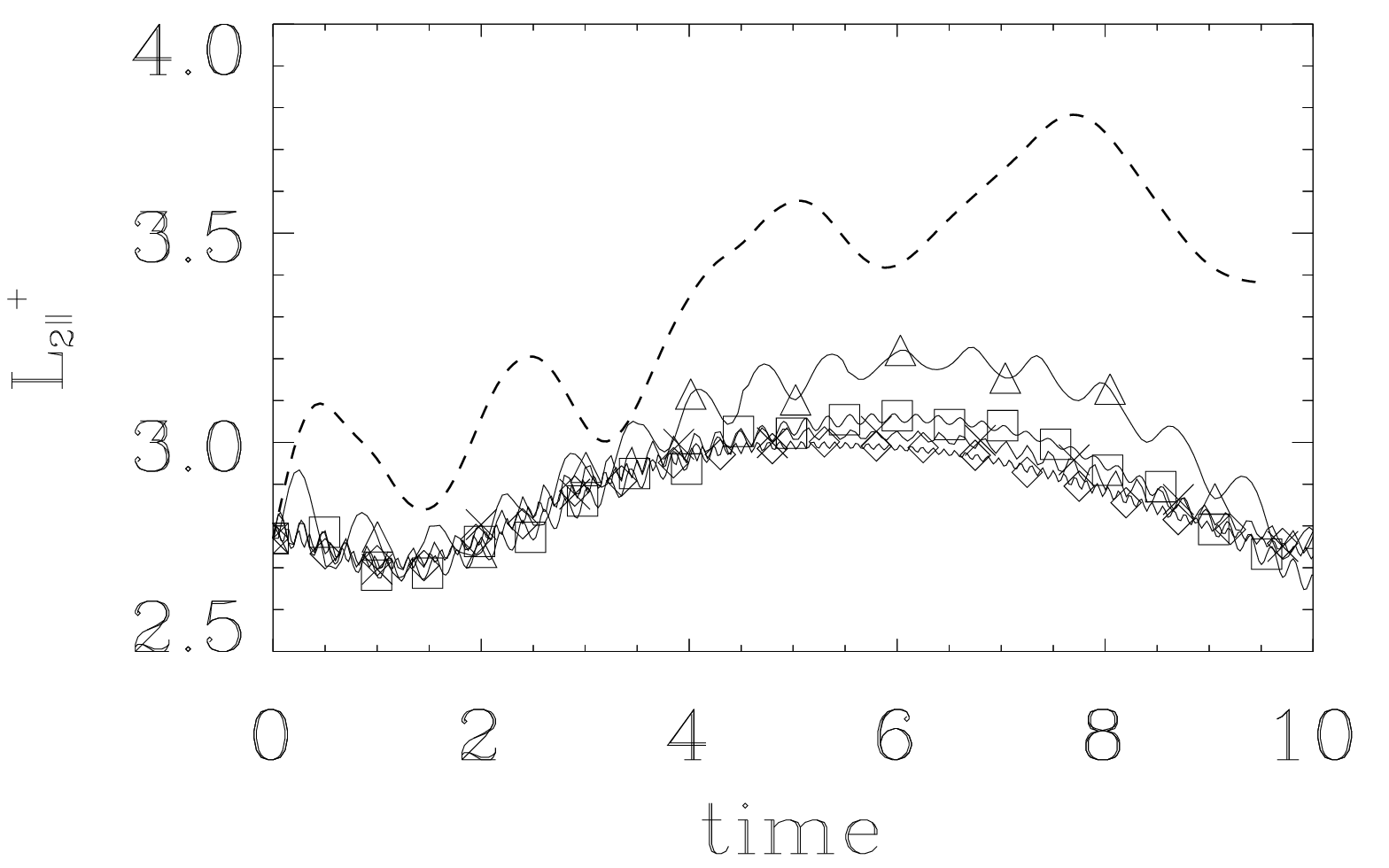}}
\end{tabular}
\caption{Temporal evolution, up to $t=10$, of perpendicular (top) and parallel (bottom) integral length-scales 
for pseudo-Alfv\'en ($+$) waves; same runs and legend as in Figure \ref{figLint}. 
\label{figLint2}}
\end{figure}
Figures \ref{figLint} and \ref{figLint2} present the time evolutions of the perpendicular integral 
length-scales, defined as
\be
L_{\perp_{1,2}}^+={\int E_{1,2}^+(\kpe,\kpa) / \kpe \, d\kpe d\kpa \over E_{1,2}^+}  \, ,
\ee
and the parallel integral length-scales 
\be
L_{\parallel_{1,2}}^+={\int E_{1,2}^+(\kpe,\kpa) / \kpa \, d\kpe d\kpa \over E_{1,2}^+} \, ,
\ee
for, respectively, the shear- and pseudo-Afv\'en waves. We first note, for shear-Alfv\'en waves, a 
decrease of the perpendicular scales and an increase of the parallel one afterward we observe a 
saturation. These behaviors may be interpreted as a direct cascade in the perpendicular direction 
and a possible inverse cascade in the parallel one. The saturation phase with length-scales 
approximately frozen means that the spectra are well developed. The case $B_0=1$ deviates from
this analysis because the mean field is not strong enough to impose a full anisotropic dynamics; 
it can be compared with a previous study made for pure isotropic turbulence \cite{Galtier99}.
For pseudo-Alfv\'en waves, the situation is less clear even if we still observe globally the same 
behavior as before in the initial phase. It is the saturation phase which is the most different with an 
apparent oscillation that can be related to the period found from the previous analysis made for 
Figure \ref{figtmp_rA}.

\begin{figure*}[ht]
\begin{tabular}{lclc}
\resizebox{180mm}{!}{\includegraphics{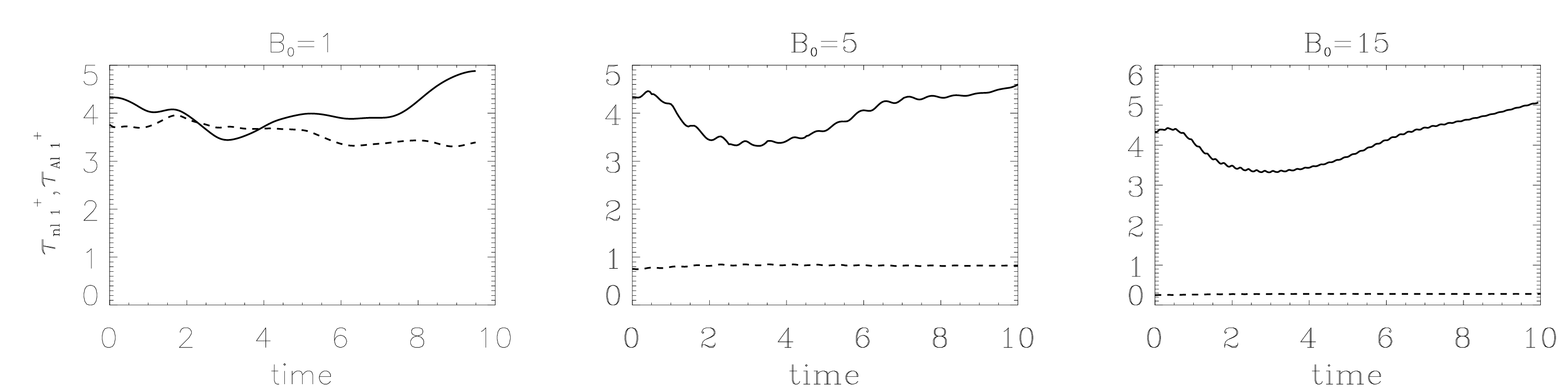}}\\
\resizebox{180mm}{!}{\includegraphics{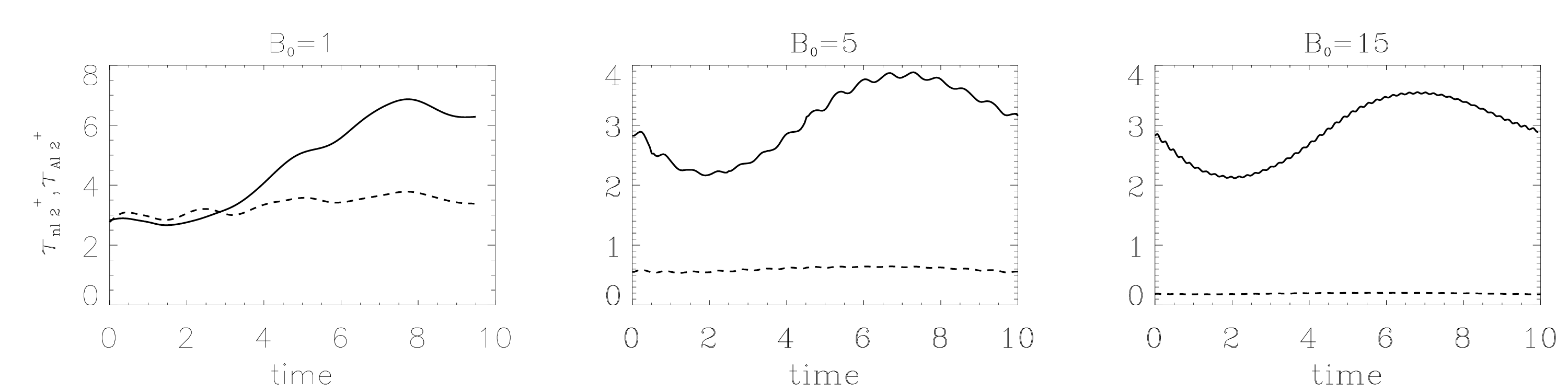}}
\end{tabular}
\caption{Temporal evolution, up to $t=10$, of nonlinear (solid line) and Alfv\'en (dashed line) time-scales 
for shear- (top) and pseudo-Alfv\'en waves (bottom) for runs ${\bf IIa}$ to ${\bf IVa}$ (from left to right). 
\label{figTimes}}
\end{figure*}
Figure \ref{figTimes} presents the temporal evolution of the nonlinear time 
\be
\tau^+_{NL_{1,2}}= {L_{\perp_{1,2}}^+ \over z^-_{rms_{1,2}}} \, ,
\label{t1}
\ee
and the Alfv\'en time 
\be
\tau^+_{A_{1,2}}={L_{\parallel_{1,2}}^+ \over B_0} \, ,
\label{t2}
\ee
based on the shear- and pseudo-Afv\'en waves dynamics. We mainly observe a decrease of the 
Alfv\'en time when the strength of the uniform field $B_0$ increases whereas the nonlinear time 
is not strongly affected. Note that the profiles of the nonlinear times for the case $B_0=5$ and 
$B_0=10$ look similar with oscillations whose periods are approximately the same as before. 
This is simply due to the definition used to build the nonlinear time which includes the previous 
length-scales. It is also this definition which explains the initial decrease ($t \le 2$) of the nonlinear 
time since the perpendicular integral length-scales follow the same behavior. 

\begin{figure*}[ht]
\begin{tabular}{cccc}
\resizebox{59mm}{!}{\includegraphics{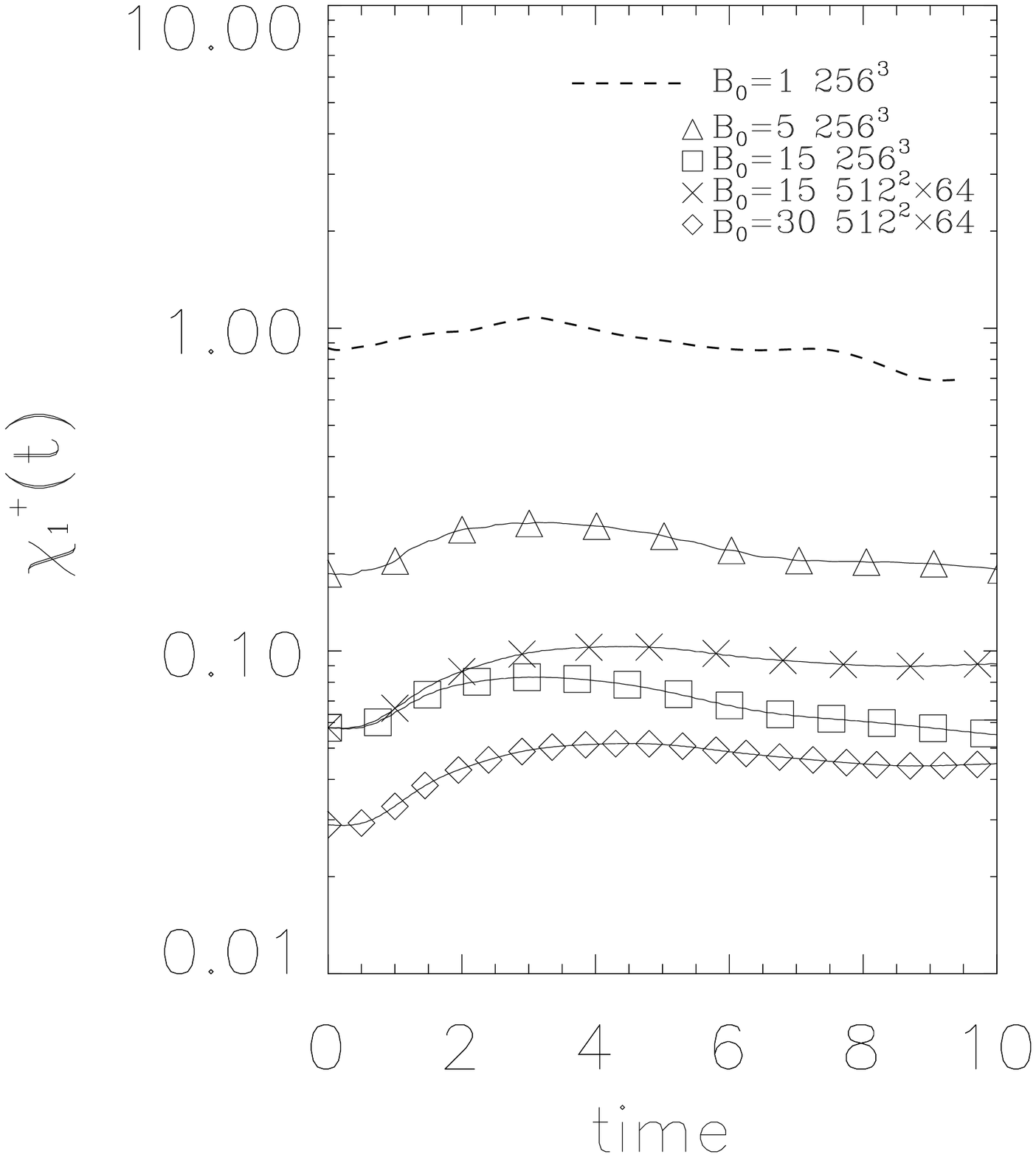}}
\resizebox{59mm}{!}{\includegraphics{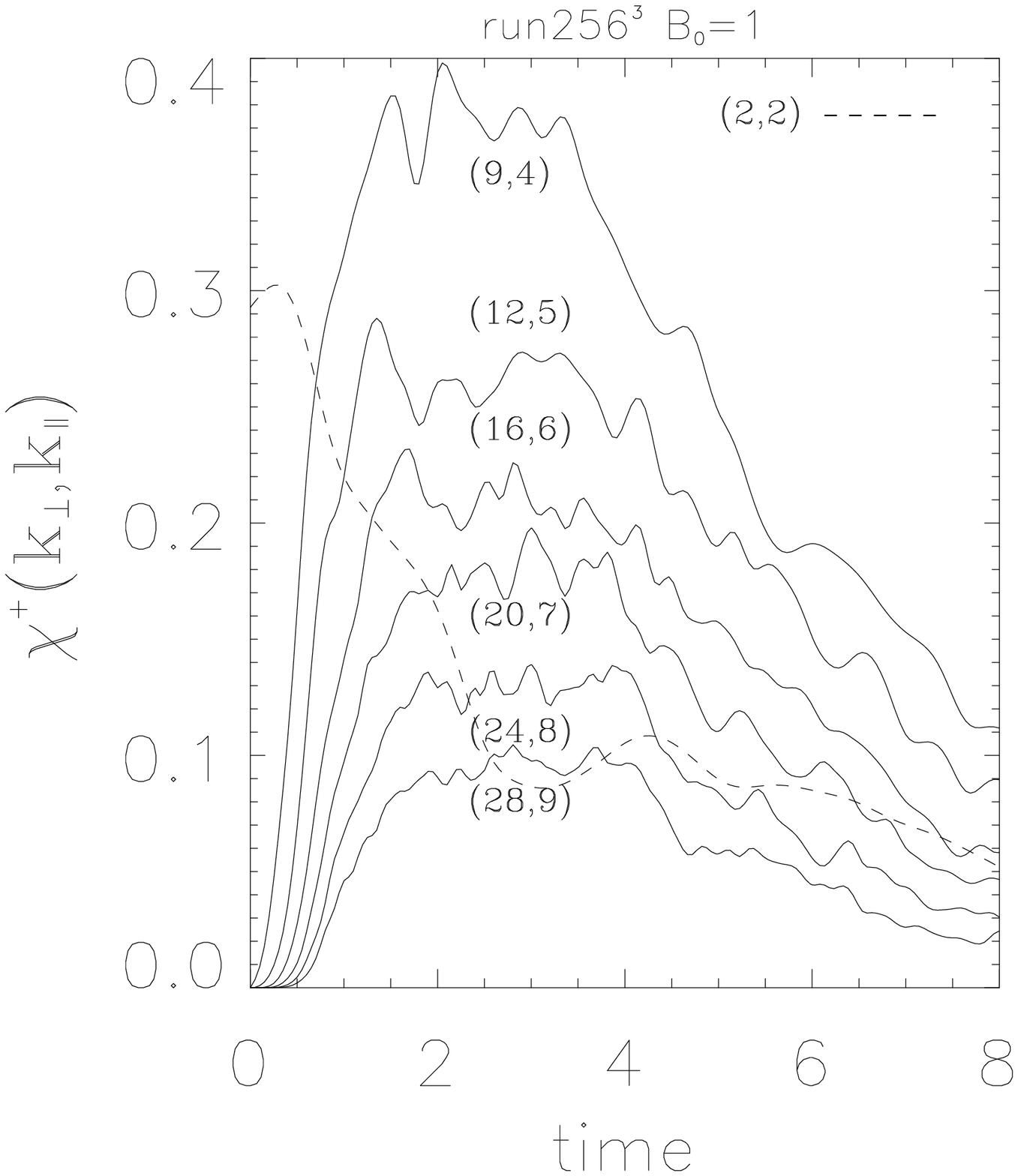}}
\resizebox{59mm}{!}{\includegraphics{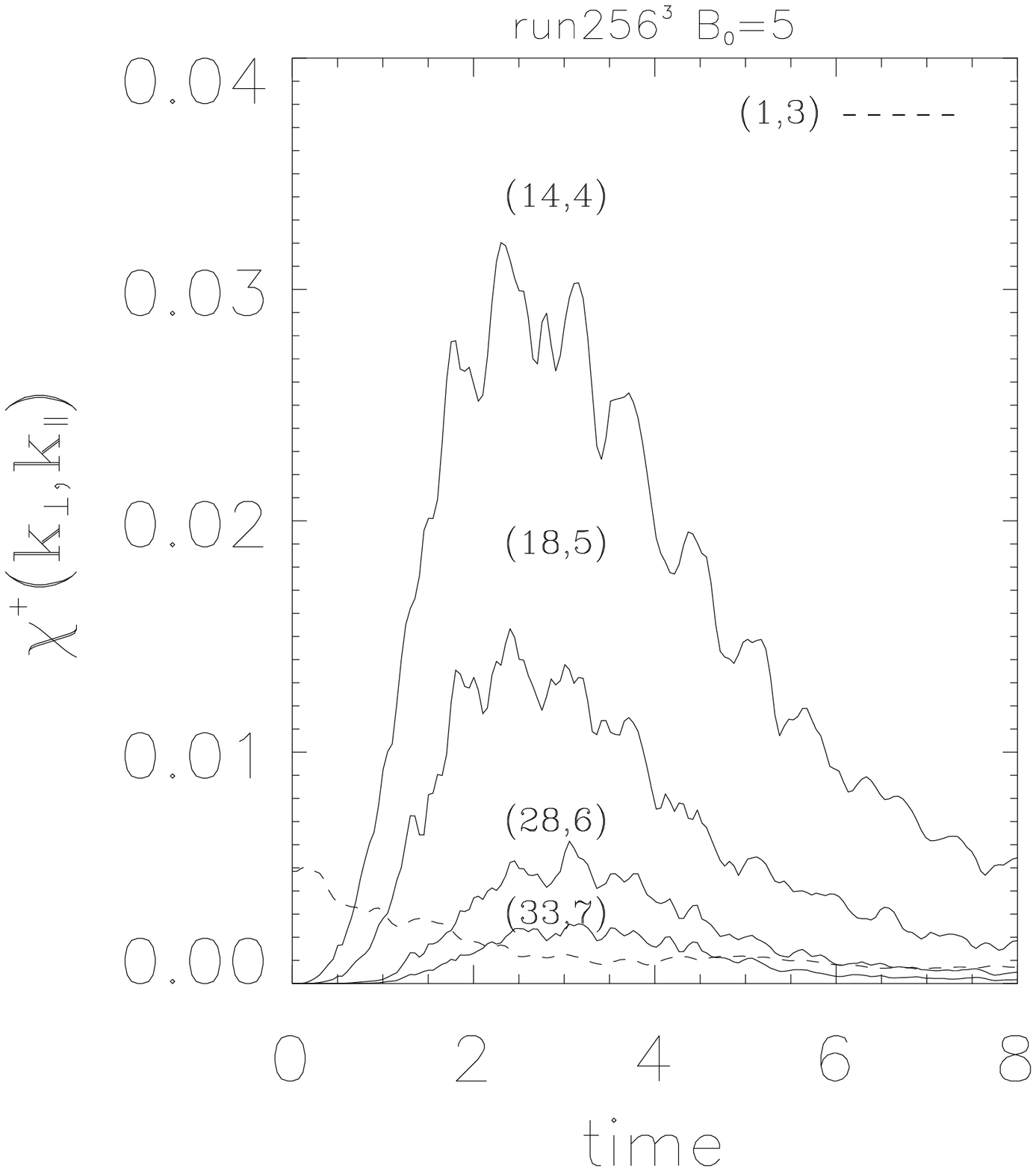}}\\
\resizebox{59mm}{!}{\includegraphics{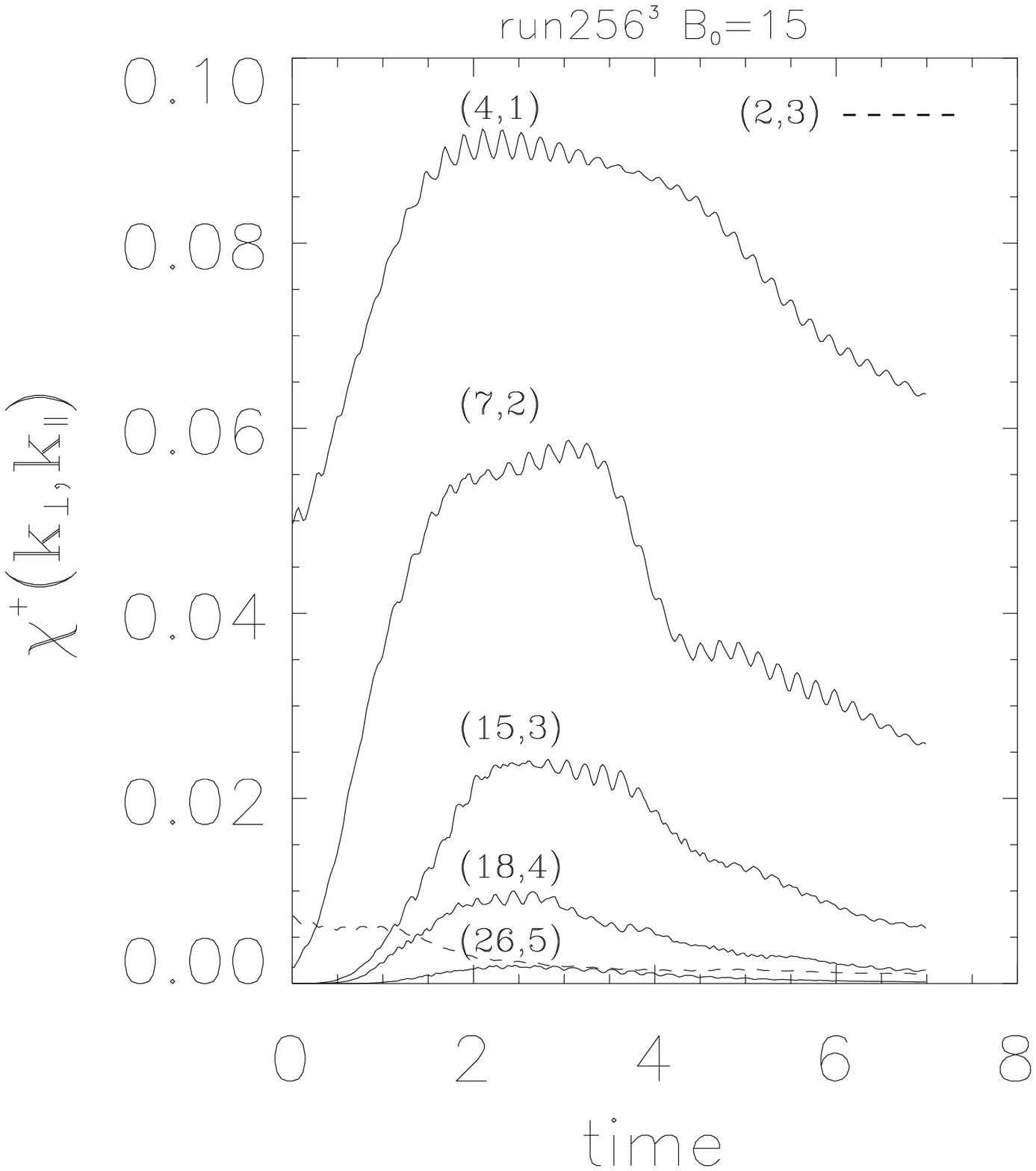}}
\resizebox{59mm}{!}{\includegraphics{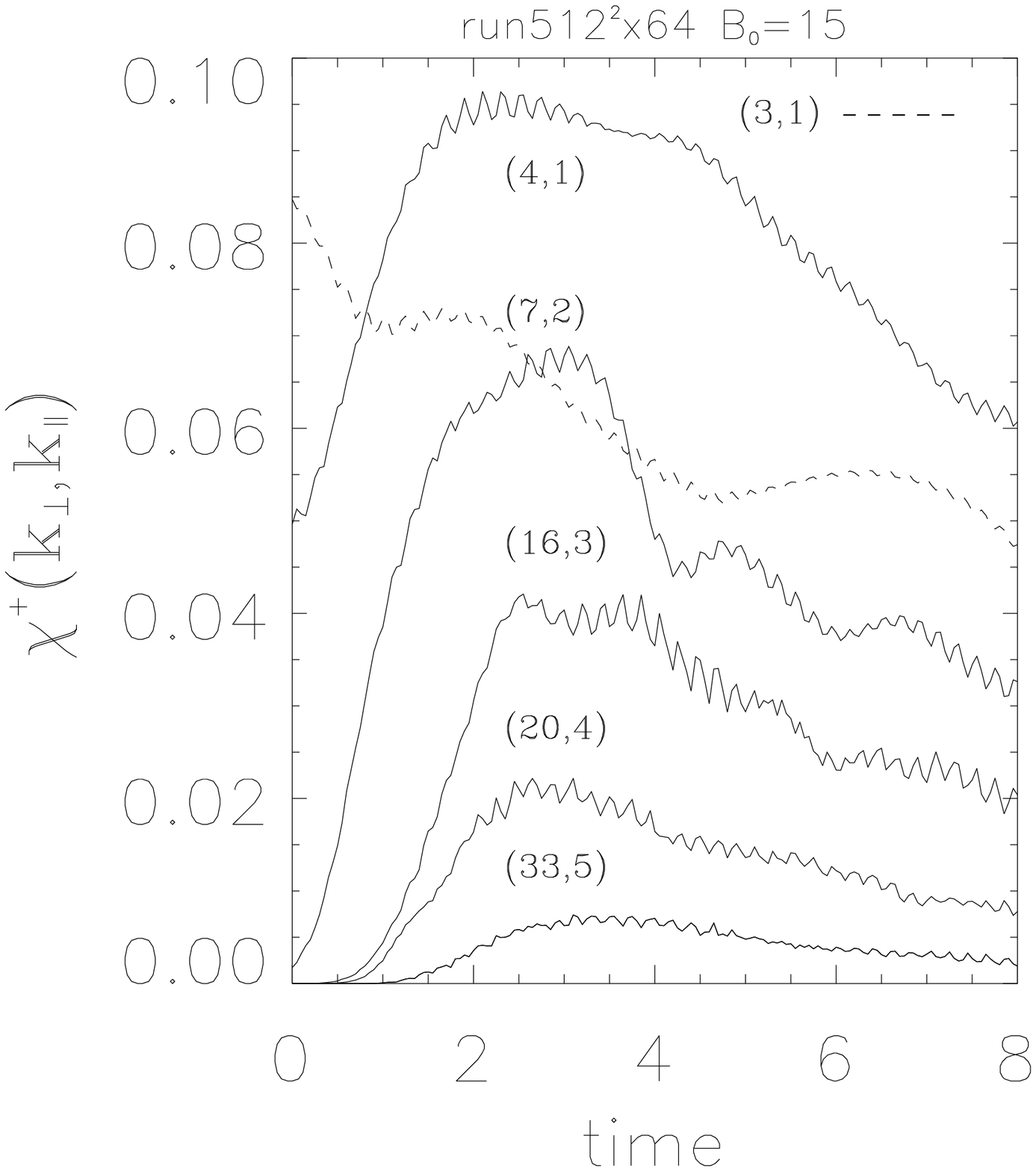}}
\resizebox{59mm}{!}{\includegraphics{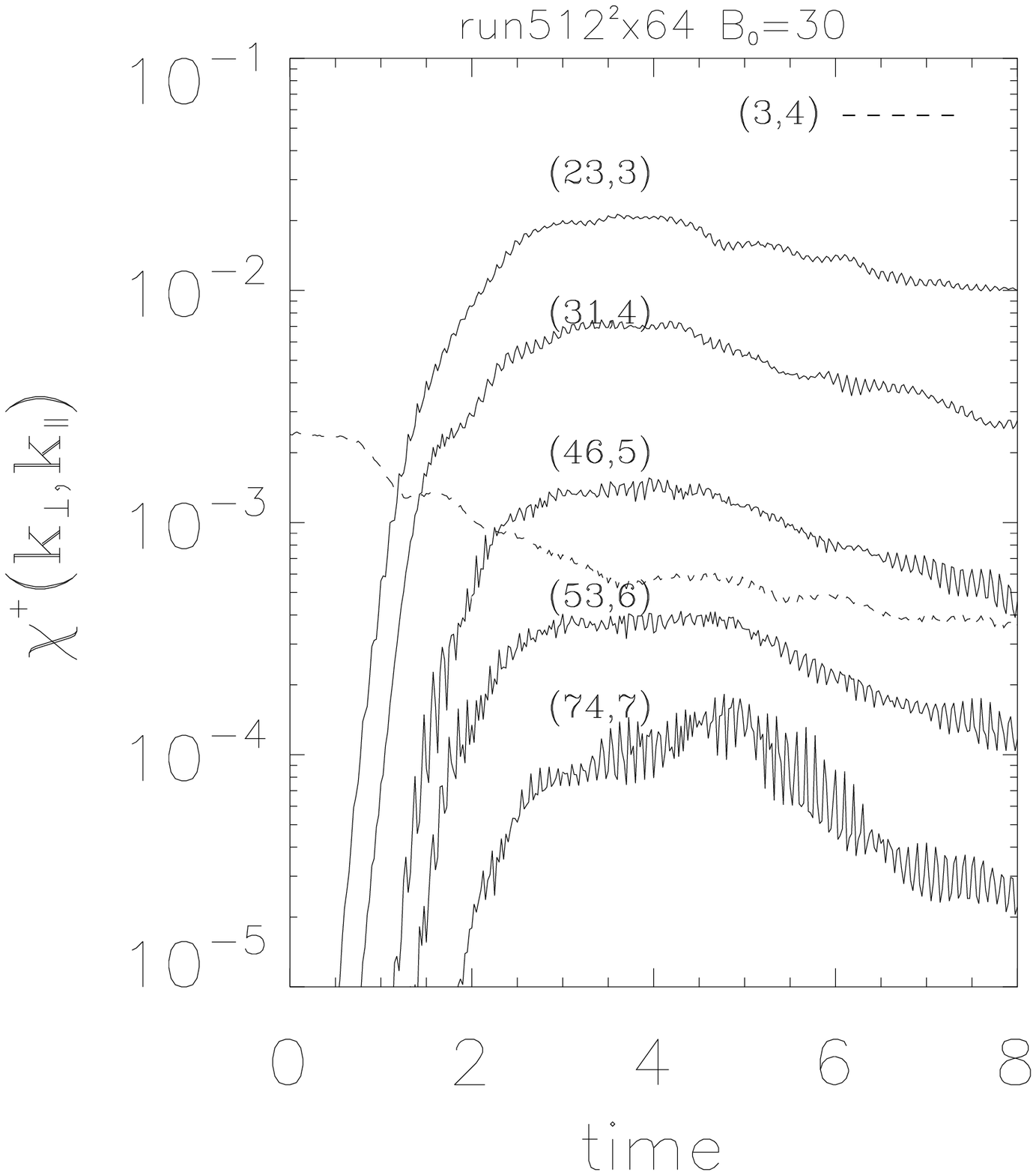}}
\end{tabular}
\caption{Temporal evolution for shear-Alfv\'en waves of the ratio between the Alfv\'en and nonlinear 
time-scales based on the anisotropic IK phenomenology (top-left; for runs ${\bf IIa}$ to ${\bf VIa}$) and 
the spectral ratio between the same times (runs ${\bf IIa}$ to ${\bf VIa}$). Note the use of a logarithmic 
coordinate in the top-left and bottom-rigth panels. 
\label{figChil}}
\end{figure*}

Figure \ref{figChil} shows the temporal evolution (top-left) of the time-scales ratio 
\be 
\chi^+_1(t) = {\tau^+_{A_1}(t) \over \tau^+_{NL_1}(t)} \, ,
\label{def1}
\ee
between the Alfv\'en (\ref{t2}) and eddy turnover (\ref{t1}) times. Only the case of shear-Alfv\'en waves 
is shown since the same behavior is found for pseudo-Alfv\'en waves. This new plots give a quantitative 
estimate of the balance between the time-scales that we discussed in the introduction. We clearly see 
that the balance is sub-critical ($\chi^+_1(t)$ stays well below unity) as the strength of $B_0$ increases 
with a value which remains about constant during the time of the simulation. 
\begin{figure}[h]
\begin{tabular}{cccc}
\resizebox{69mm}{!}{\includegraphics{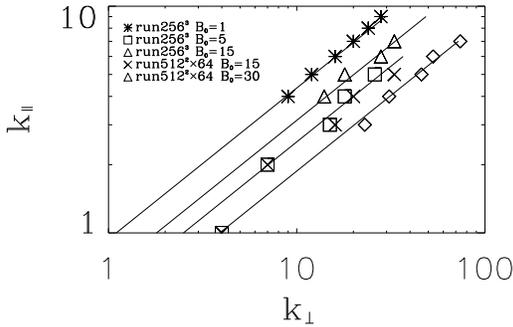}}
\end{tabular}
\caption{Couples of points extracted from Figure \ref{figChil} well fitted by the anisotropic scaling law 
$\kpa \sim \kpe^{2/3}$. 
\label{fig_ani}}
\end{figure}
Figure \ref{figChil} displays also the spectral ratio between the Alfv\'en and nonlinear time-scales for 
shear-Alfv\'en waves. It is defined as
\be
\chi_1^+(\kpe,\kpa) (t) = {\kpe z^-_{\ell_1}(t) \over \kpa B_0} \, ,
\ee
with
\be
z^-_{\ell_1}(t) = \sqrt{E_1^-(\kpe,\kpa) \kpe \kpa} \, .
\ee
The previous definition (\ref{def1}) is based on a global estimate of the time-scales. This new definition 
is more precise since it allows to take into account the scale at which the times are defined. Then, each 
time evolution is associated with a couple of (spectral) scales ($\kpe$,$\kpa$). 
Different couples have been tried and only those for which the ratio $\chi_1^+(\kpe,\kpa)$ displays an 
extended plateau have been reported. It is basically for times between $t=2$ and $t=4$, a range of time 
during which the small-scales have been produced and the nonlinear interactions are still important. 
Note that we still observe oscillations that can be explained in terms of Alfv\'en times scales. 

In Figure \ref{fig_ani}, we report each couple ($\kpe$,$\kpa$) and show the anisotropic scaling 
law $\kpa \sim \kpe^{2/3}$ as a reference. We see that such law fits well the points which 
means that the sub-critical balance (observe again here with $\chi_1^+(\kpe,\kpa) < 1$) is still well 
described by the anisotropic scaling law (\ref{aniso1}). This property may be understood by a heuristic 
model \cite{Galtier2005} where the time-scale ratio $\chi$ is supposed to be constant at all scales but 
not necessarily equal to unity which allows to use the IK phenomenology instead of the Kolmogorov one 
\cite{GS95}. (Note that the same behavior is found when $\chi_1^-(\kpe,\kpa)$ is considered.) 

The question of the validity of the anisotropic scaling law $\kpa \sim \kpe^{2/3}/B_0$ (we use here 
the formulation given in \cite{Galtier2005} which includes the uniform magnetic field) beyond the 
inertial range, and in particular at larger scales, may be addressed from these numerical simulations. 
A first answer is given in Figure \ref{fig_ani} with the couple ($\kpe=4$,$\kpa=1$) which is at the 
largest scales of the system but does not follow the anisotropic law.

%%%%%%%%%%%%%%%%%%%%%%%%%%%%%%%%%%%
%\subsection{Global isotropic coefficients}

%The isotropy level of the velocity and magnetic field fluctuations is 
%estimated by means of coefficients computed as in \cite{curry}.
%For each wavevector ${\bf k}$, an orthonormal reference frame is defined as
%(${\bf k}/|{\bf k}|$, ${\bf e}_1({\bf k})/|{\bf e}_1({\bf k})|$, ${\bf e}_2({\bf k})/|{\bf e}_2({\bf k})|$),
%with ${\bf e}_1({\bf k})= {\bf k} \times {\bf z}$ and ${\bf e}_2({\bf k})={\bf k} \times {\bf e}_1({\bf k})$,
%where ${\bf z}$ is the vertical unit wavevector. In that frame, since the incompressibility condition,
%say for the velocity field, yields
%${\bf k} \cdot {\bf u}({\bf k}) = 0$, ${\bf u}({\bf k})$ is only determined by its two components
%${\bf u}_1({\bf k})$ and ${\bf u}_2({\bf k})$. The isotropy coefficient is then defined
%as $C^{\bf u}_{iso}=\sqrt{<|{\bf u}_1|^2>/<|{\bf u}_2|^2>}$, with thus a unit value for fully isotropic flows.
%A similar coefficient can be based on the magnetic field, $C^{\bf b}_{iso}$, characterizing the
%small scale isotropy.

%\AD{Figure} for the 4 values of $B_0$ : isotropy coefficient ${\bf u}, {\bf b}, {\bf z}^\pm$ or  
%${\bf w}, {\bf j}, {\bf w}^\pm$ (where $ {\bf w}= \nabla \times {\bf u}$)  

%%%	QUE FAIT-ON DE CETTE SECTION ?????

%%%%%%%%%%%%%%%%%%%%%%%%%%%%%%%%%%%
\subsection{Generalized anisotropy angles}

\begin{figure}[ht]
\includegraphics[width=85mm]{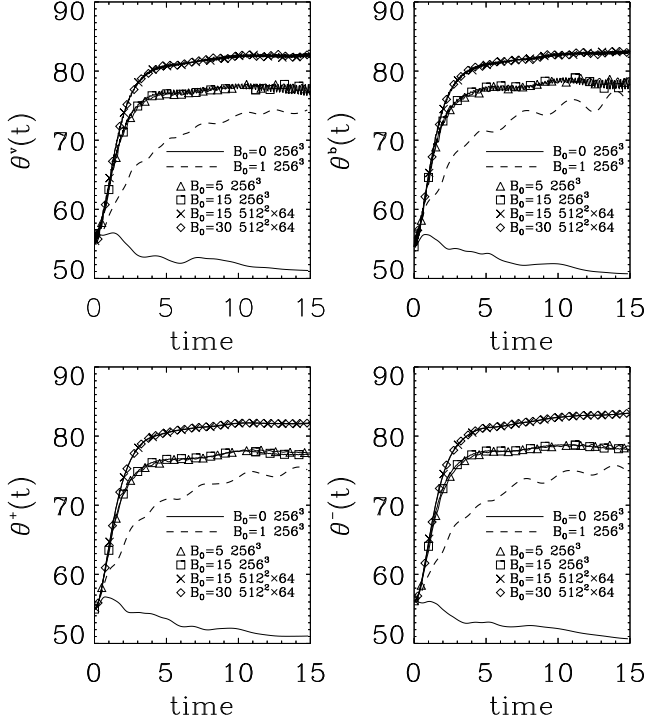}
\caption{Temporal evolution of generalized Shebalin angles for the velocity, magnetic, and $z^\pm$ 
fields (runs ${\bf Ia}$ to ${\bf VIa}$). 
\label{figSheba}}
\end{figure}
\begin{figure}[ht]
\includegraphics[width=85mm]{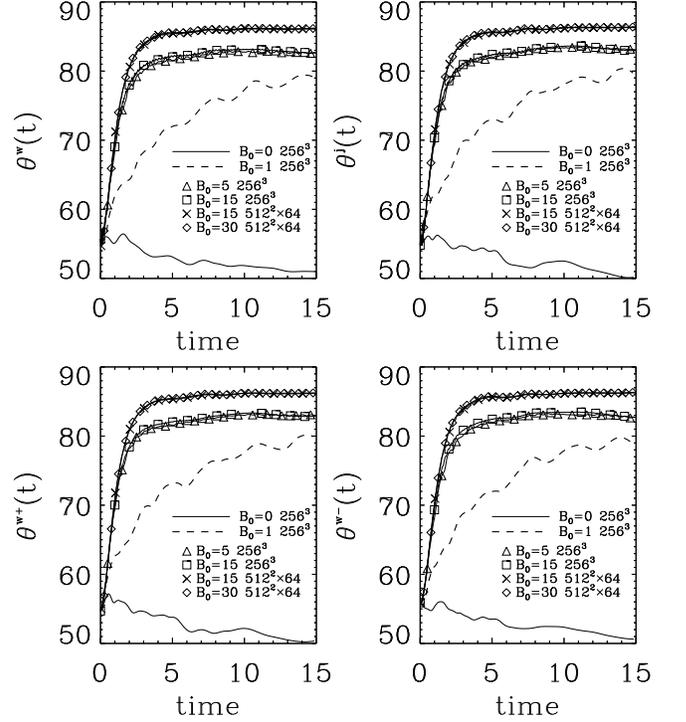}
\caption{Temporal evolution of generalized Shebalin angles for the vorticity, current density and 
$\omega^\pm$ vorticity fields (runs ${\bf Ia}$ to ${\bf VIa}$). 
\label{figShebaw}}
\end{figure}
To quantify the degree of anisotropy associated with the flow, we use the generalized Shebalin 
angles (see \cite{Shebalin,Oughton94}, and reference therein), defined as 
\be
\tan^2\theta_{\bf q} = {{\sum \kpe^2 |{\bf q}({\bf k},t)|^2} \over {\sum \kpa^2 |{\bf q}({\bf k},t)|^2}} \, ,
\label{angleS}
\ee
where ${\bf q}$ is a vector field, like ${\bf v}$, ${\bf b}$ or ${\bf z}^{\pm}$ in Figure \ref{figSheba}. 
We start initially with a 3D isotropic flow for which $\theta_{\bf{q}} \sim 54,74^{\circ}$. 
Figure \ref{figSheba} shows that the temporal evolution of the different angles (for the different fields) 
is the same with a behavior depending mainly on the intensity of ${\bf B_0}$. 
For $B_0=0$ the energy transfer is similar in all directions and the temporal evolution of Shebalin 
angles remains almost constant, close to its initial value. For $B_0=5$ and $B_0=15$, the Shebalin 
angles quickly increase and stabilize around $78^{\circ}$. Thus, and as expected, the anisotropy 
develops with ${\bf B_0}$. However, the flow is not totally confined in planes perpendicular 
to ${\bf B_0}$ like for a purely bi-dimensional fluid for which the Shebalin angle is $90^{\circ}$. 
Note that a stronger anisotropy is produced for runs ${\bf Va}$ and ${\bf VIa}$ (for which the Reynolds 
number is higher) with angles up to $83^{\circ}$. This is explained by the wider range of $\kpe$ 
available for such runs. 

In Figure \ref{figShebaw}, we report the generalized Shebalin angles for the vector fields 
${\bf j}$, ${\bf w}$ and ${\bf w}^{\pm}$, where ${\bf w}^\pm=\nabla\times{\bf z}^\pm$. 
The same behavior as before is found with apparently a slightly stronger anisotropy (with angles 
closer to $90^{\circ}$) for the highest values of $B_0$. This is explained by the fields used which 
are built on the rotational of the previous fields shown in Figure \ref{figSheba} and thus to a higher 
dependence of relation (\ref{angleS}) in perpendicular wavenumbers (in $\kpe^4$ instead of 
$\kpe^2$).

%%%%%%%%%%%%%%%%%%%%%%%%%%%%%%%%%%%%%%%%%%%%%%%%
\section{Spectral analysis}
\subsection{Reduced spectra}

\begin{figure*}[ht]
\begin{tabular}{lllllll}
\resizebox{70mm}{!}{\includegraphics{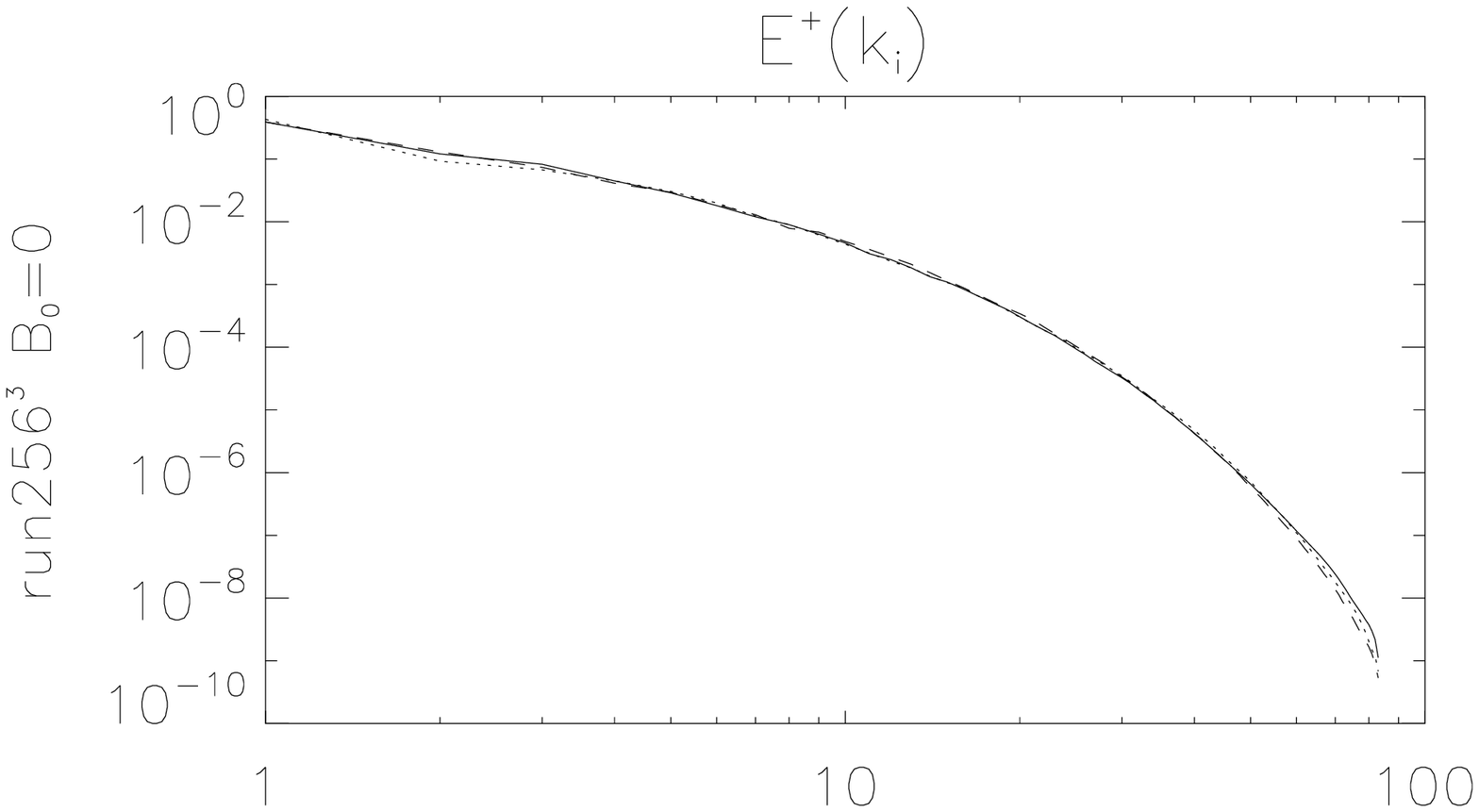}}
\resizebox{62mm}{!}{\includegraphics{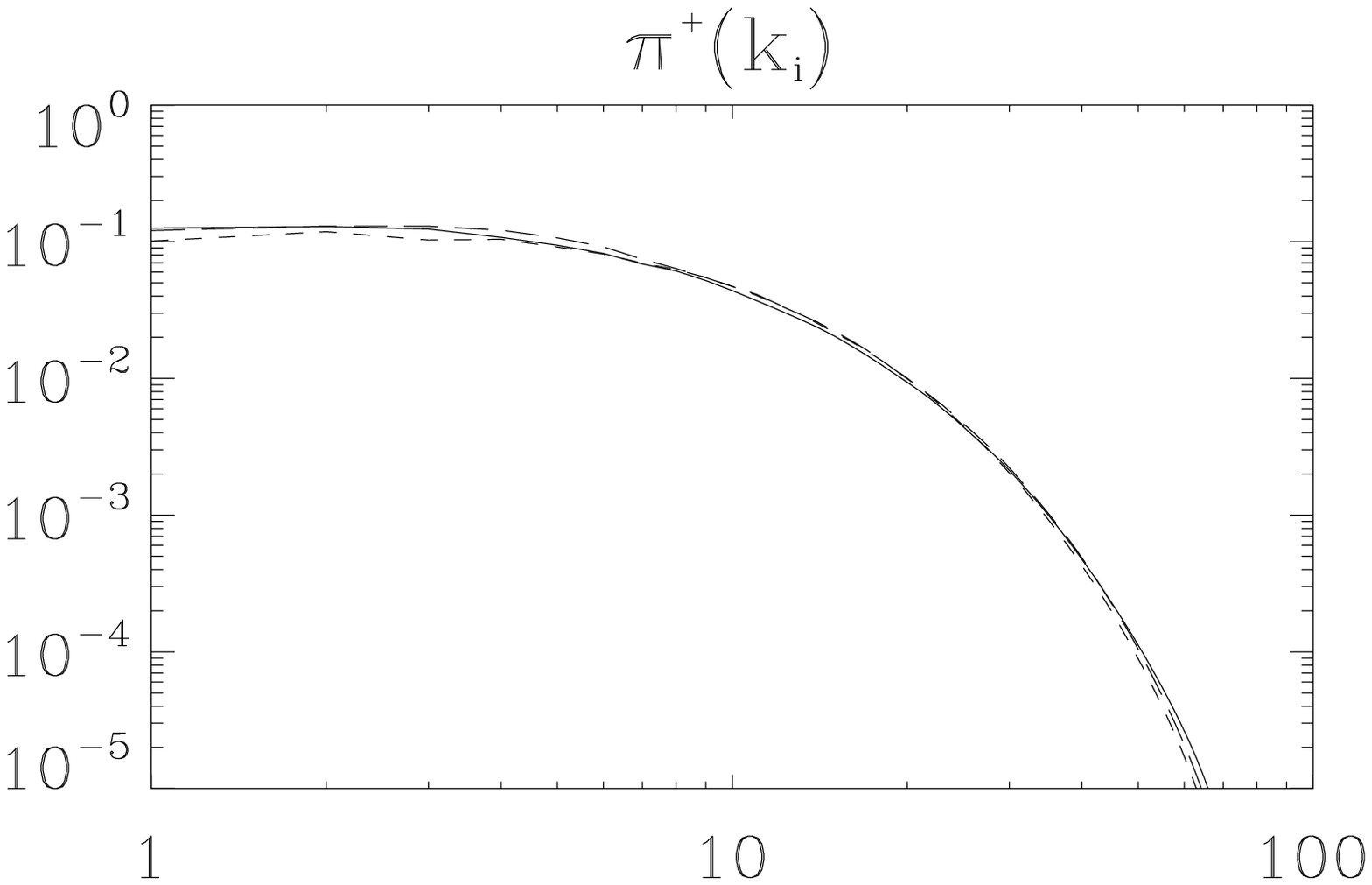}}\\
\resizebox{70mm}{!}{\includegraphics{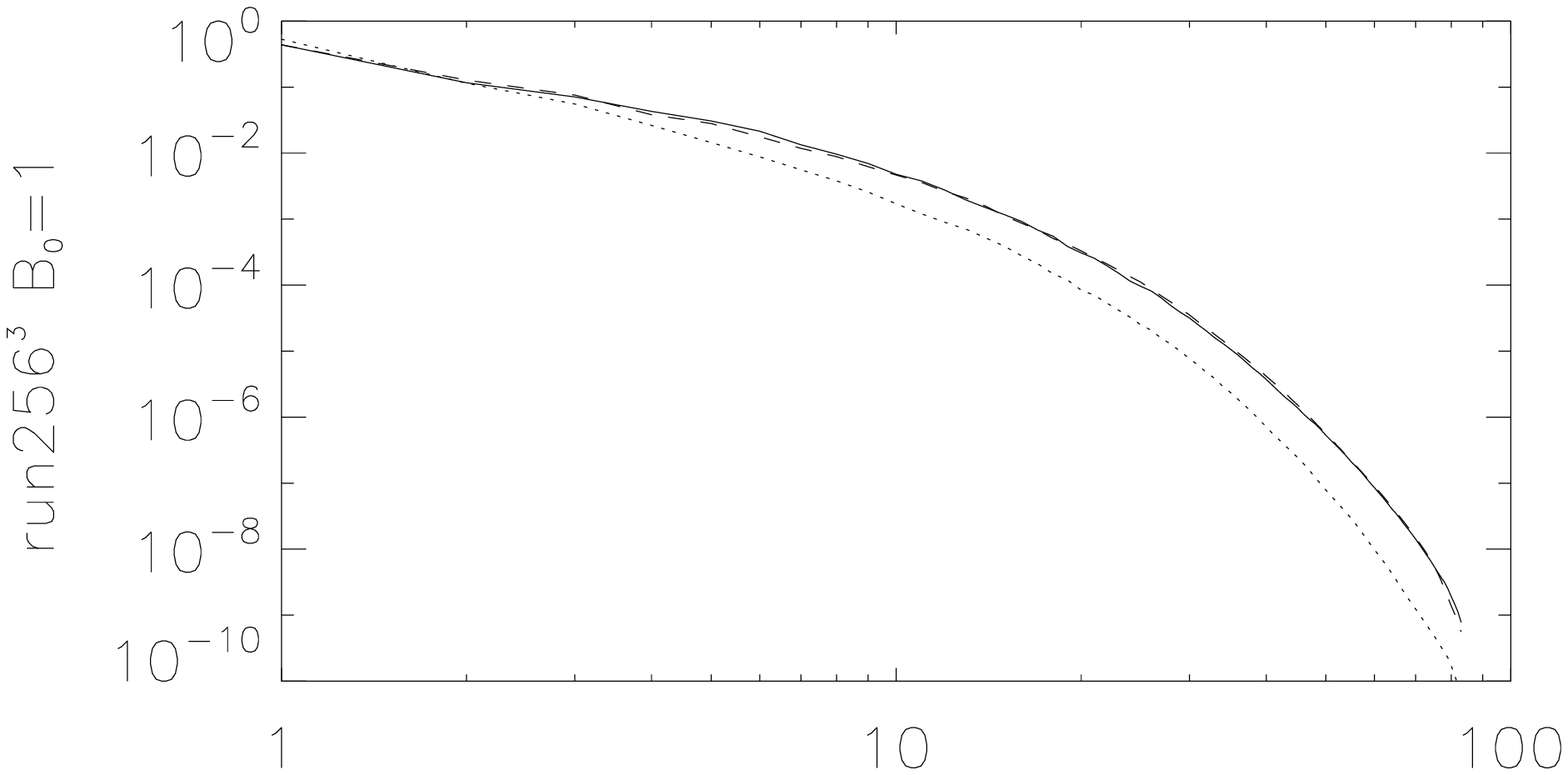}}
\resizebox{62mm}{!}{\includegraphics{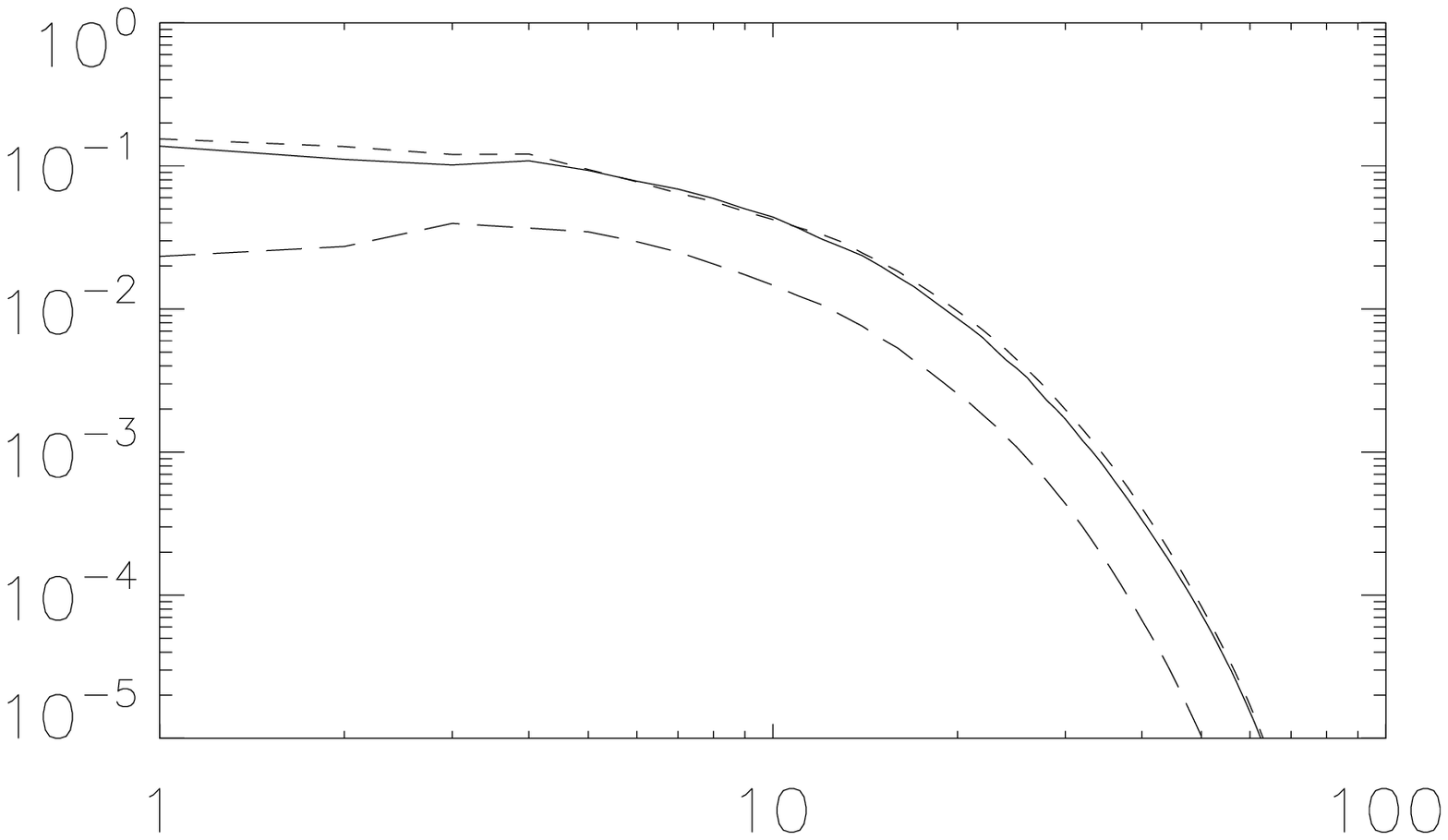}}\\
\resizebox{70mm}{!}{\includegraphics{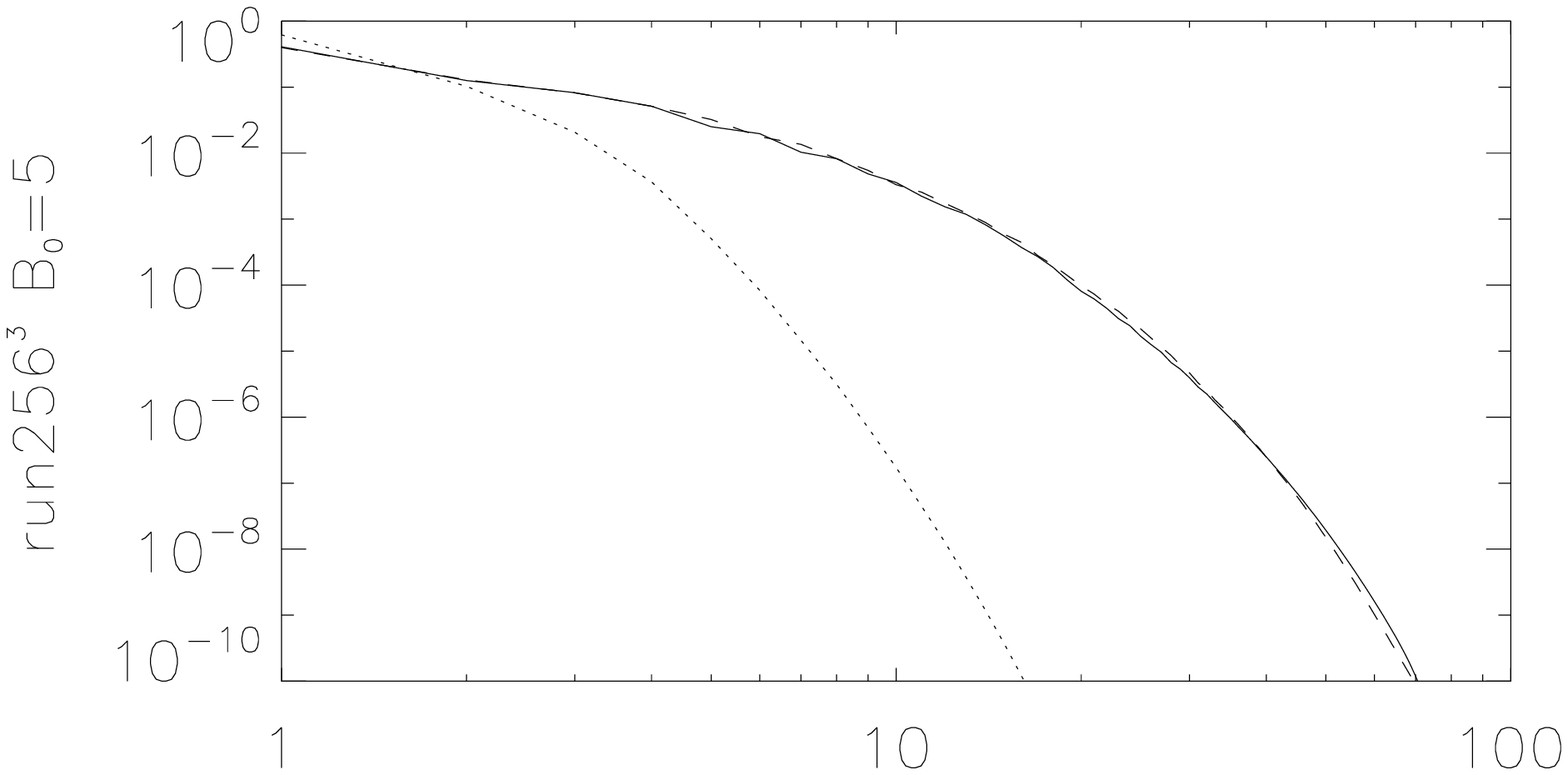}}
\resizebox{62mm}{!}{\includegraphics{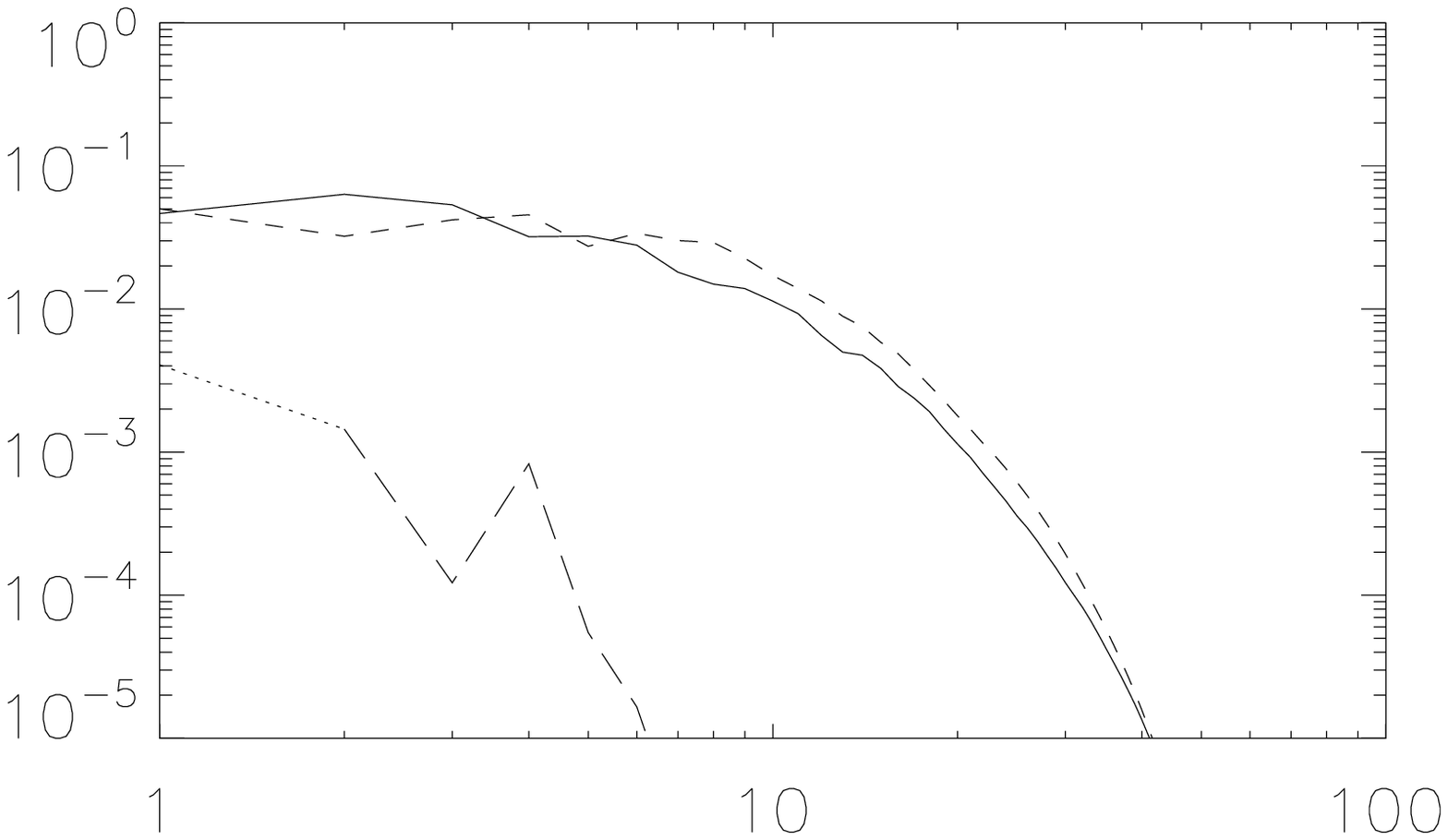}}\\
\resizebox{70mm}{!}{\includegraphics{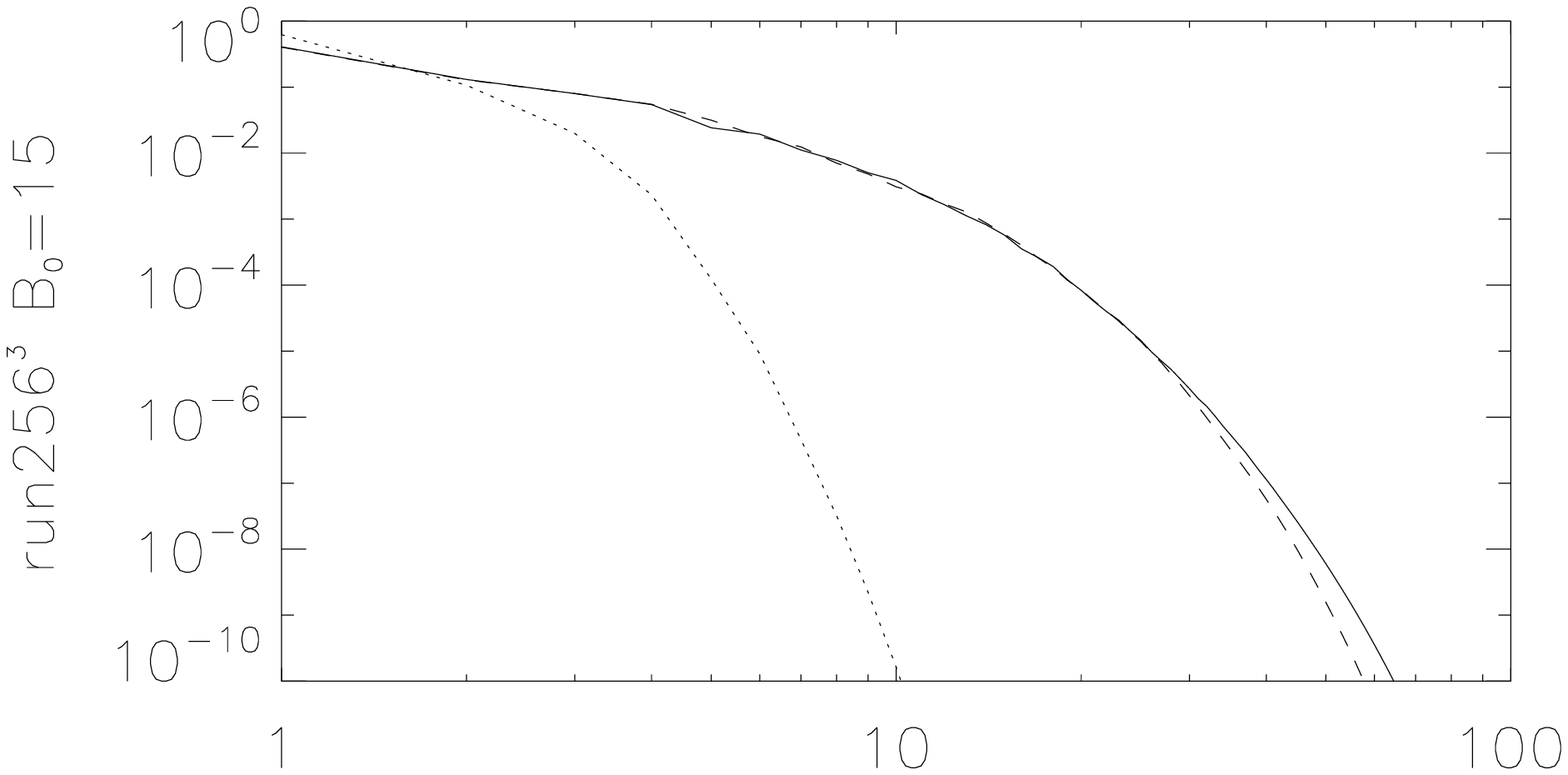}}
\resizebox{62mm}{!}{\includegraphics{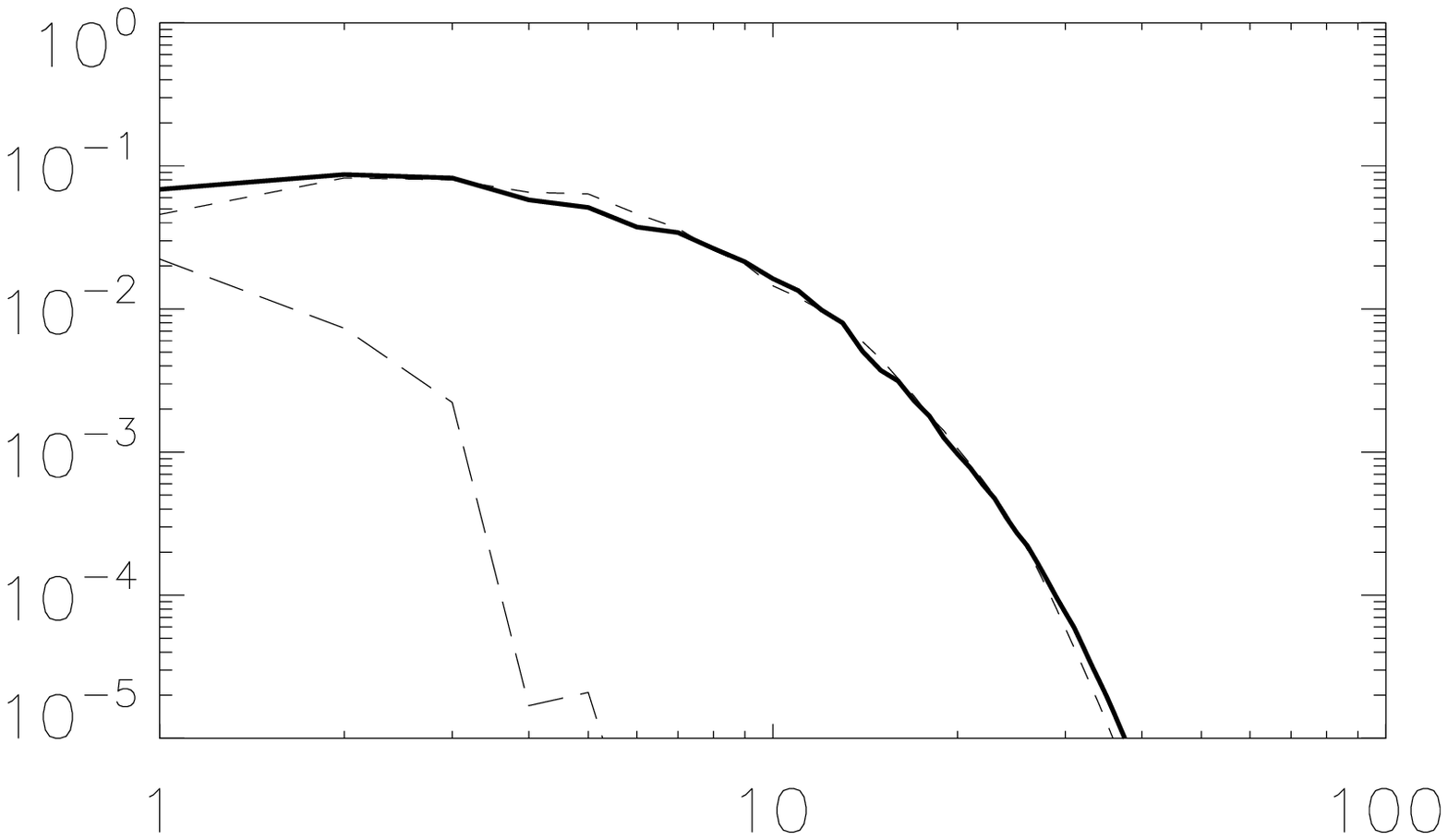}}\\
\resizebox{70mm}{!}{\includegraphics{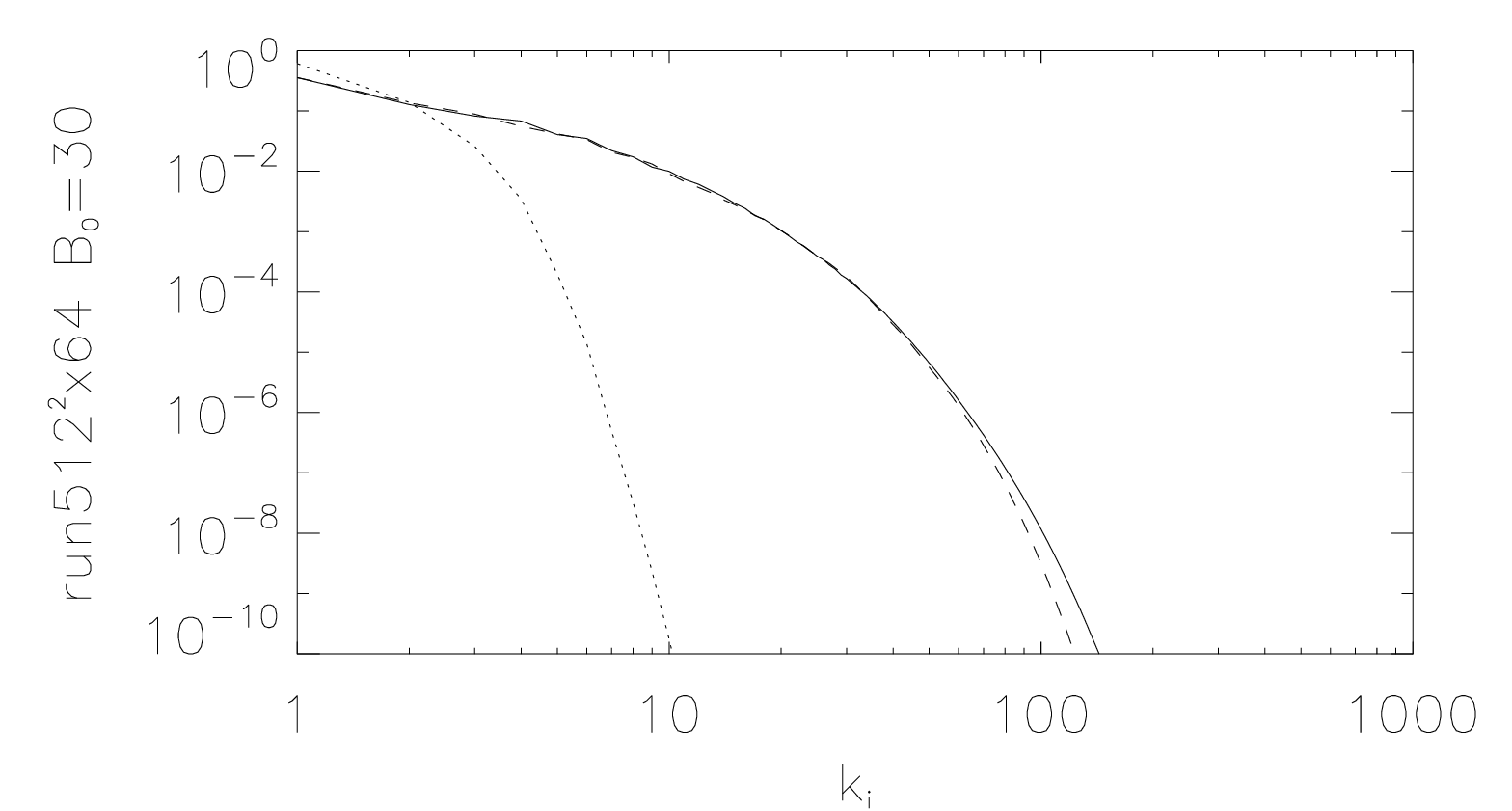}}
\resizebox{62mm}{!}{\includegraphics{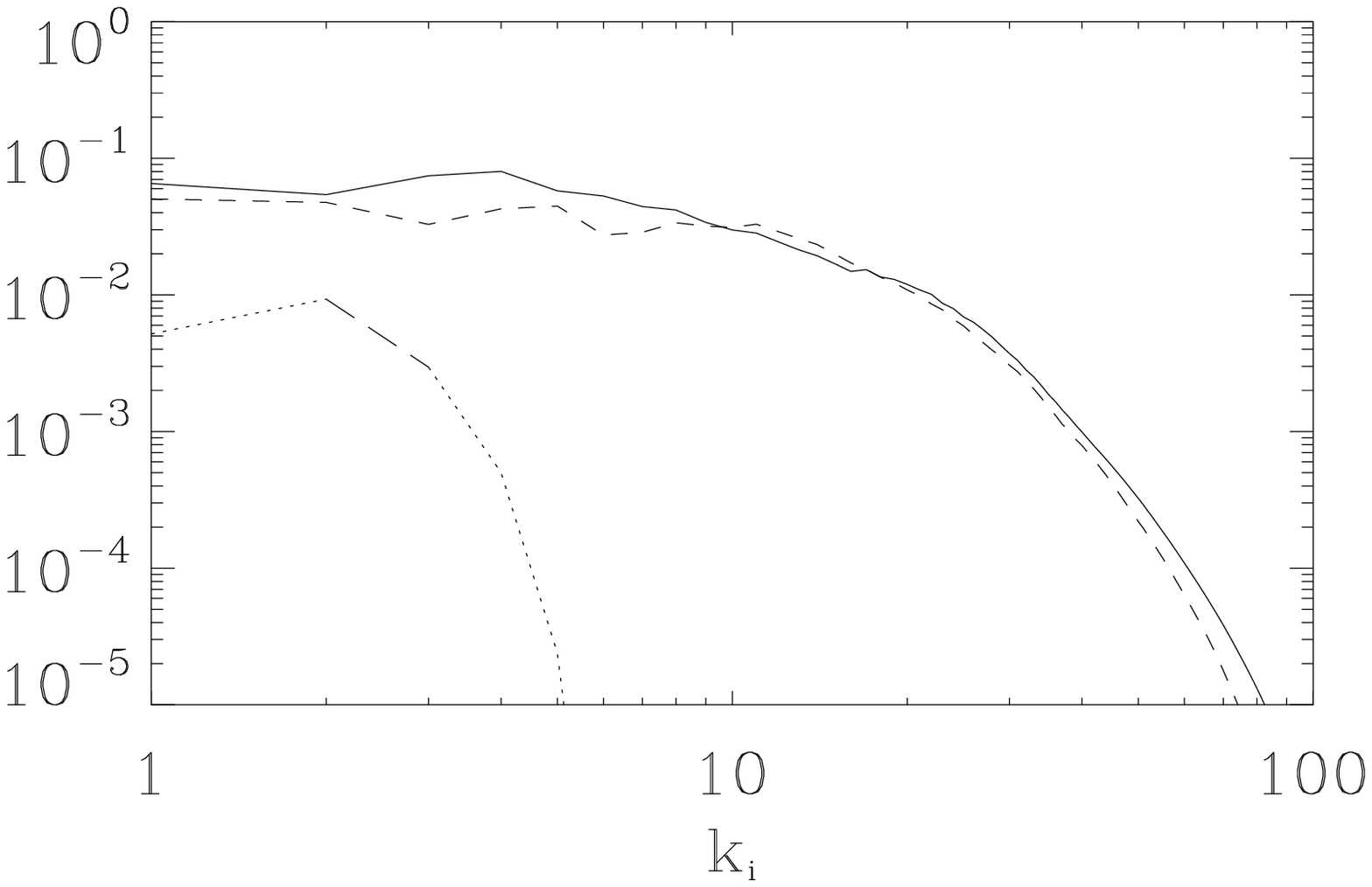}}
\end{tabular}
\caption{Reduced spectra (left) $E^+$ built from the cartesian fields ${\bf z}^+$ and the associated 
reduced energy fluxes (right) $\Pi^+$ for the variables $k_x$ (solid line), $k_y$ (dashed line) and $k_z$ 
(dotted line for $E^+$ and long-dashed line for $\Pi^+$). In the latter case when a negative flux is found, 
the absolute value is taken (dotted line). (Runs ${\bf Ia}$ to ${\bf IVa}$ and run ${\bf VIa}$, from top to 
bottom.)
\label{figSpecEP1D}}
\end{figure*}

Figure \ref{figSpecEP1D} (left) displays the one dimension (reduced) spectra 
\begin{eqnarray}
E^+(k_x)&=&\int E^+({\bf k})dk_y \, dk_z \, , \\
E^+(k_y)&=&\int E^+({\bf k})dk_x \, dk_z \, , \\
E^+(k_z)&=&\int E^+({\bf k})dk_x \, dk_y \, ,
\end{eqnarray} 
for different $B_0$ intensity, with the same initial condition, and at times where the spectra are the 
most extended (\ie $t\sim2$ for runs ${\bf Ia}$ and ${\bf IIa}$, $t\sim3$ for ${\bf IIa}$ and ${\bf IVa}$,
$t\sim4$ for ${\bf Va}$ and ${\bf VIa}$). It basically illustrates the 
different spectral transfers in the perpendicular and parallel directions when the strength of the 
uniform magnetic field increases, whereas the $x$ and $y$ dependence is roughly the same. 
The equivalent spectra with polarity $-$ is not shown since it gives the same picture. Note that 
the scaling at large-scales is not in contradiction with the initial condition discussed in Section 
\ref{IC} which concerns the modal spectrum. 

\subsection{Energy fluxes}

In Figure \ref{figSpecEP1D} (right) the associate reduced energy fluxes are given in $k_x$, $k_y$ 
and $k_z$. They are built from the cartesian ${\bf z}^+$ fields. A constant flux is only found at 
the largest scales of the system. The presence of a negative flux is sometimes observed for a 
uniform field $B_0 \ge 5$. This property may be linked to the increase of the parallel length-scale 
seen in Figure \ref{figLint}. In this case, the flux is clearly not constant which means that it is likely 
the result of a non-local interaction rather than an inverse cascade. Note that the same behavior is 
also found four the $-$ polarity. 

The locality or nonlocality of the energy flux and transfer of runs ${\bf Ia}$ to ${\bf IVa}$ has been 
investigated recently \cite{Alexakis07b} by means of different geometrical wave number shells. It is 
shown that the interactions between the two counterpropagating Els\"asser waves may become nonlocal 
for strong magnetized flows. In particular, the energy flux in the $\kpe$ direction is mainly due to modes 
which interact with the plane $\kpa=0$ (with local interactions), while the weaker cascade in the parallel 
direction is due to modes which interact with $\kpa=1$ (with possible nonlocal interactions) 
\cite{Alexakis07b,Alexakis07a}. 
This property has been interpreted as a signature of a transition towards the weak  turbulence regime 
during which the number of effective modes in the energy cascade is reduced.

\subsection{Anisotropic spectra}

\begin{figure*}[ht]
\begin{tabular}{ccc}
\resizebox{58mm}{!}{\includegraphics{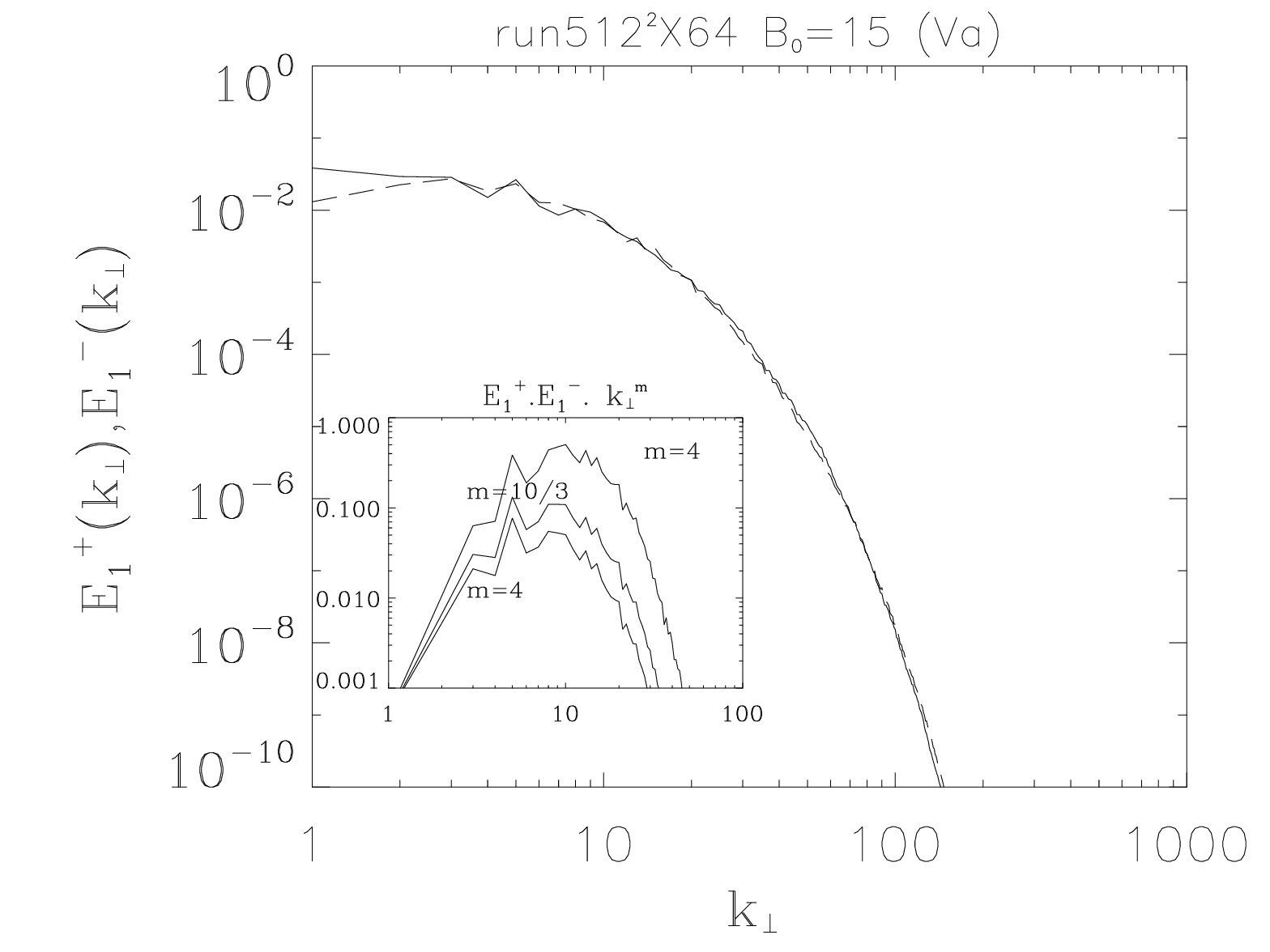}}
\resizebox{58mm}{!}{\includegraphics{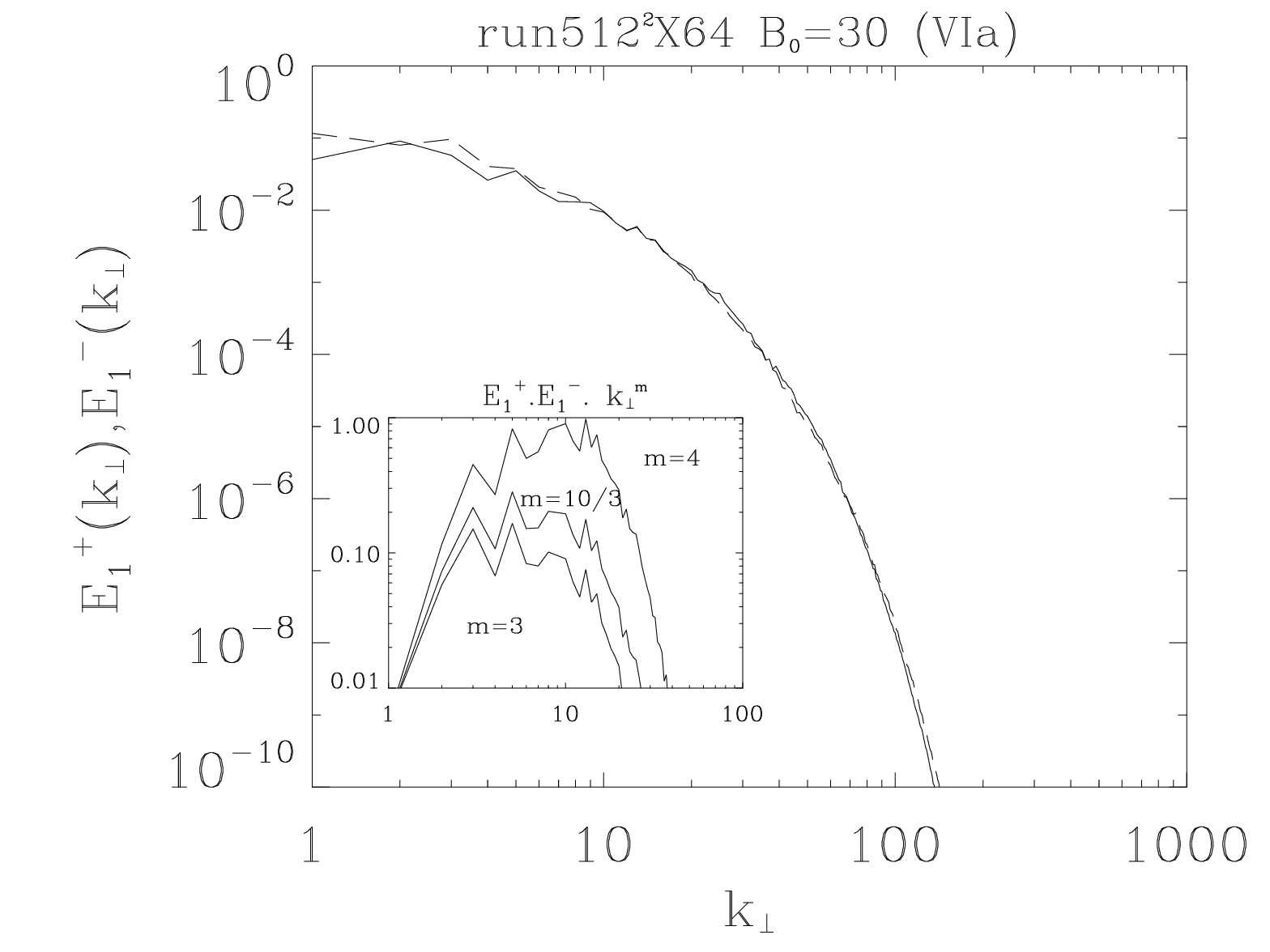}}
\resizebox{58mm}{!}{\includegraphics{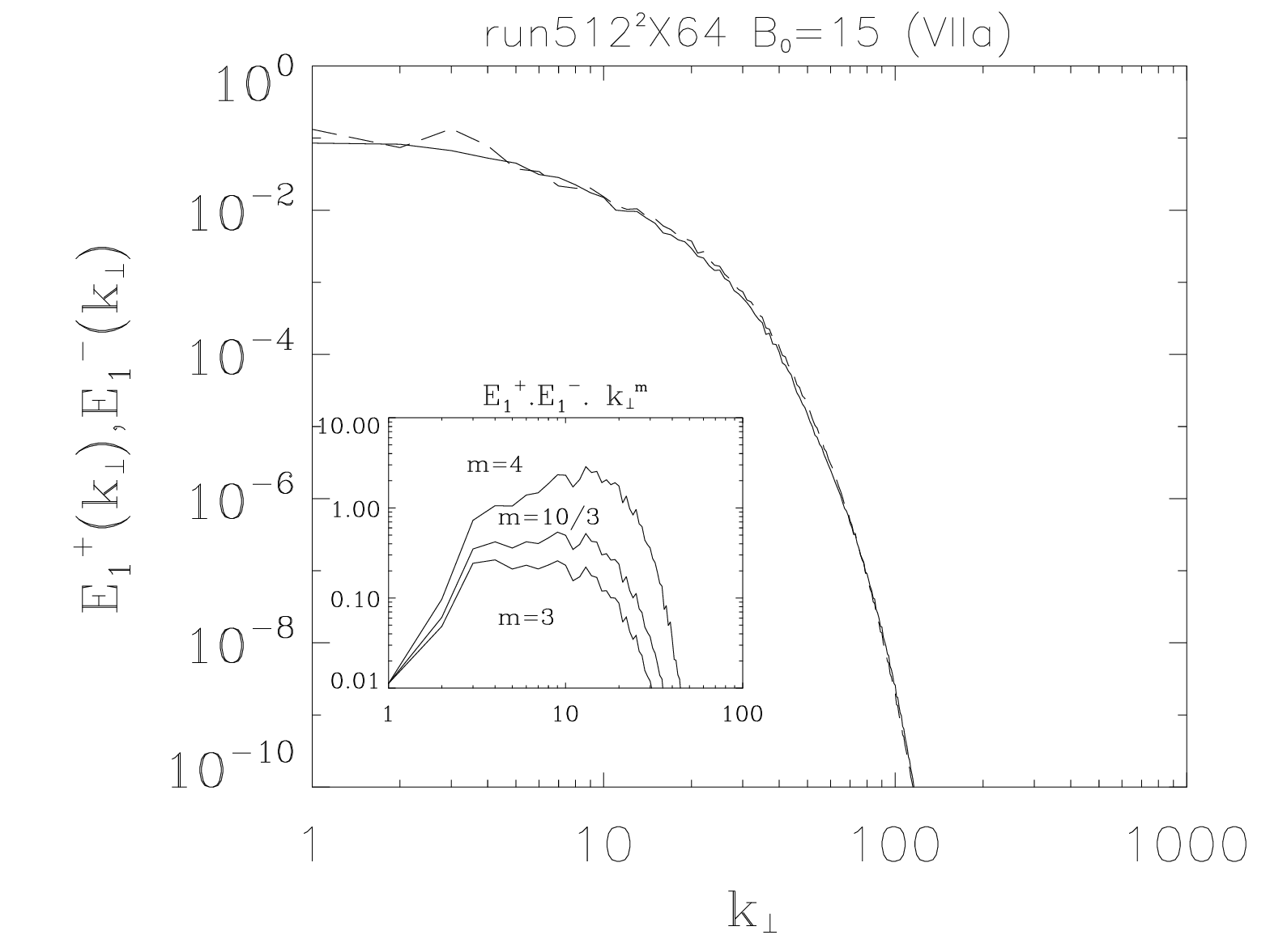}}\\
\resizebox{58mm}{!}{\includegraphics{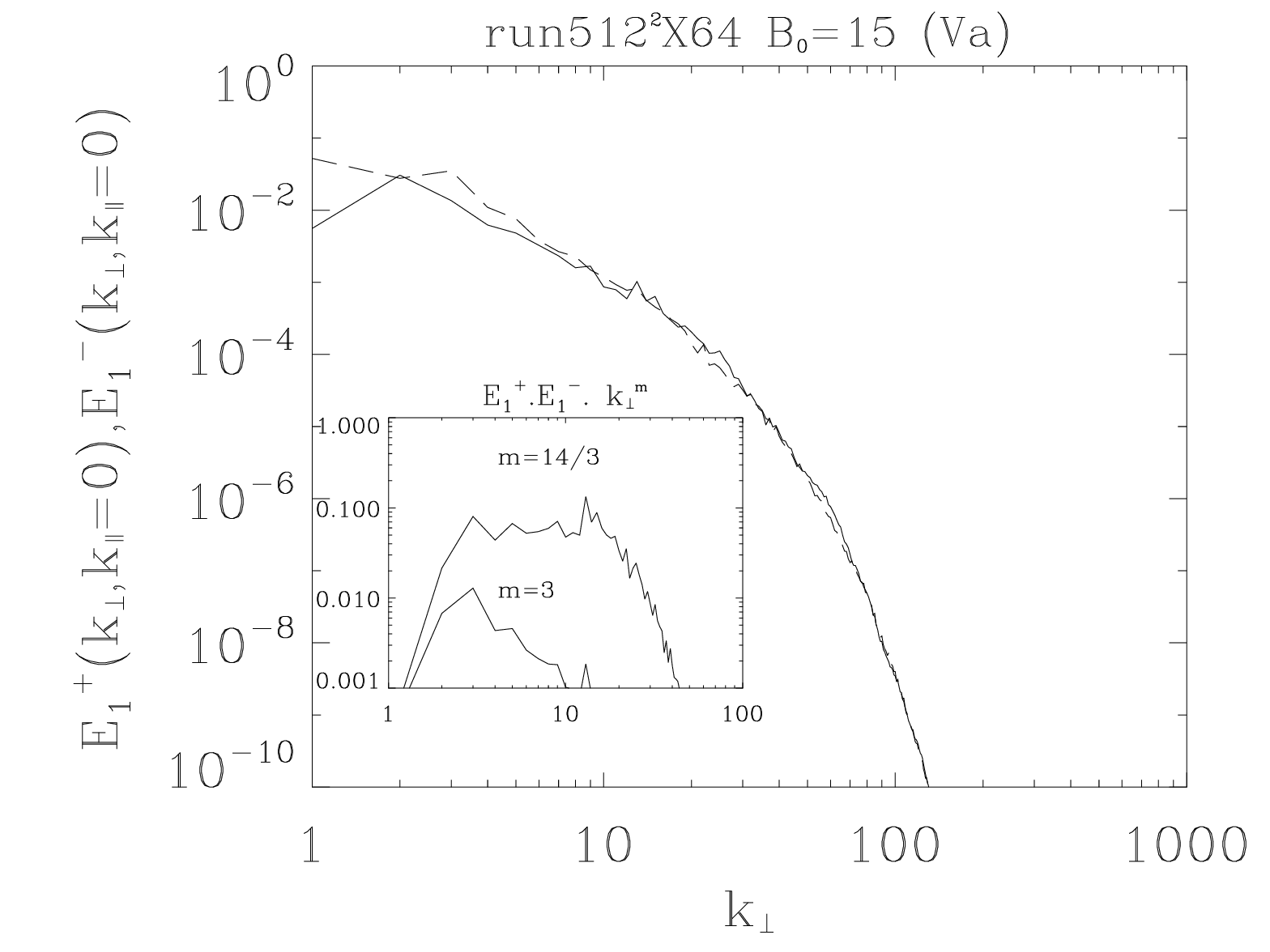}}
\resizebox{58mm}{!}{\includegraphics{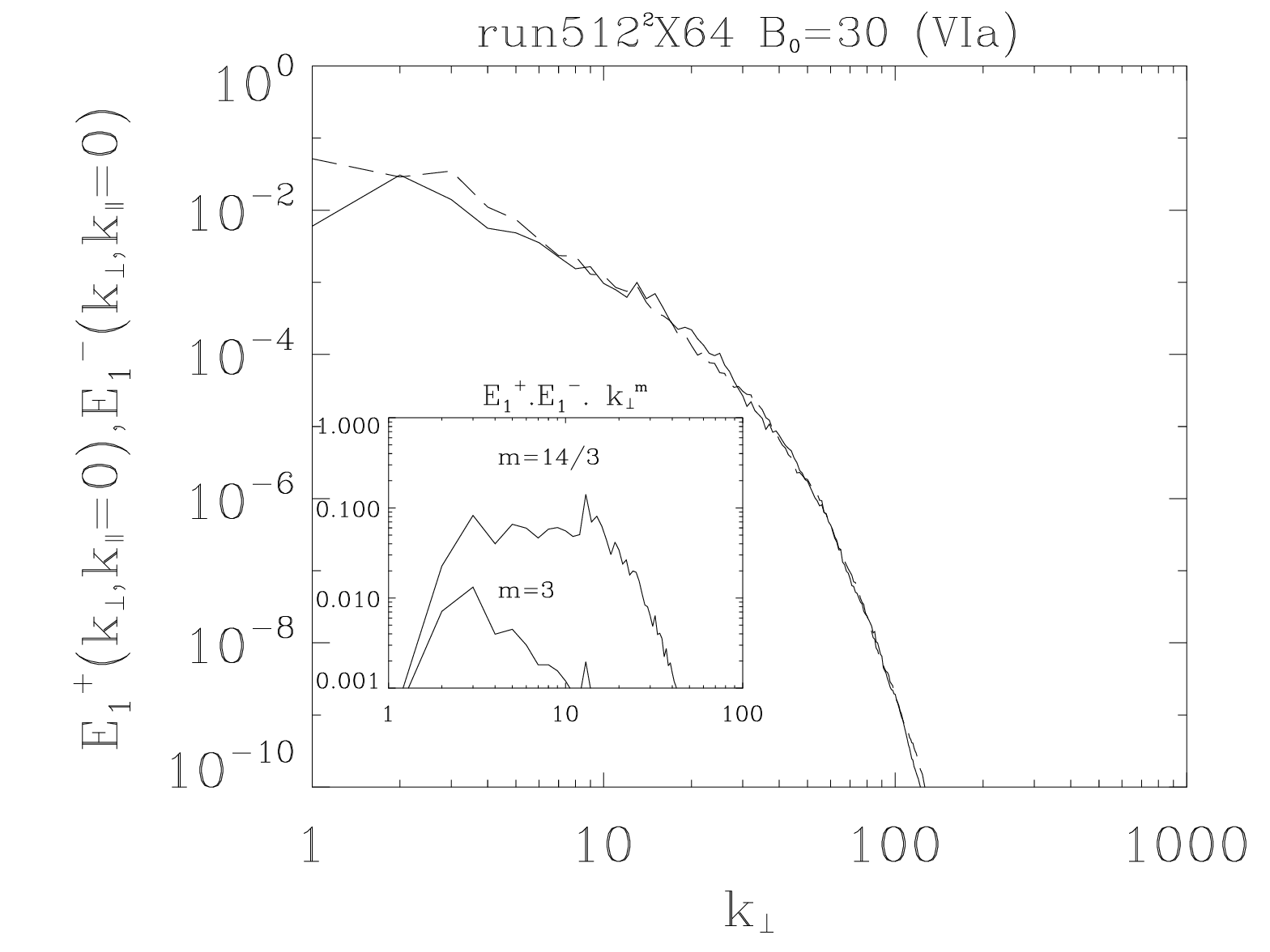}}
\resizebox{58mm}{!}{\includegraphics{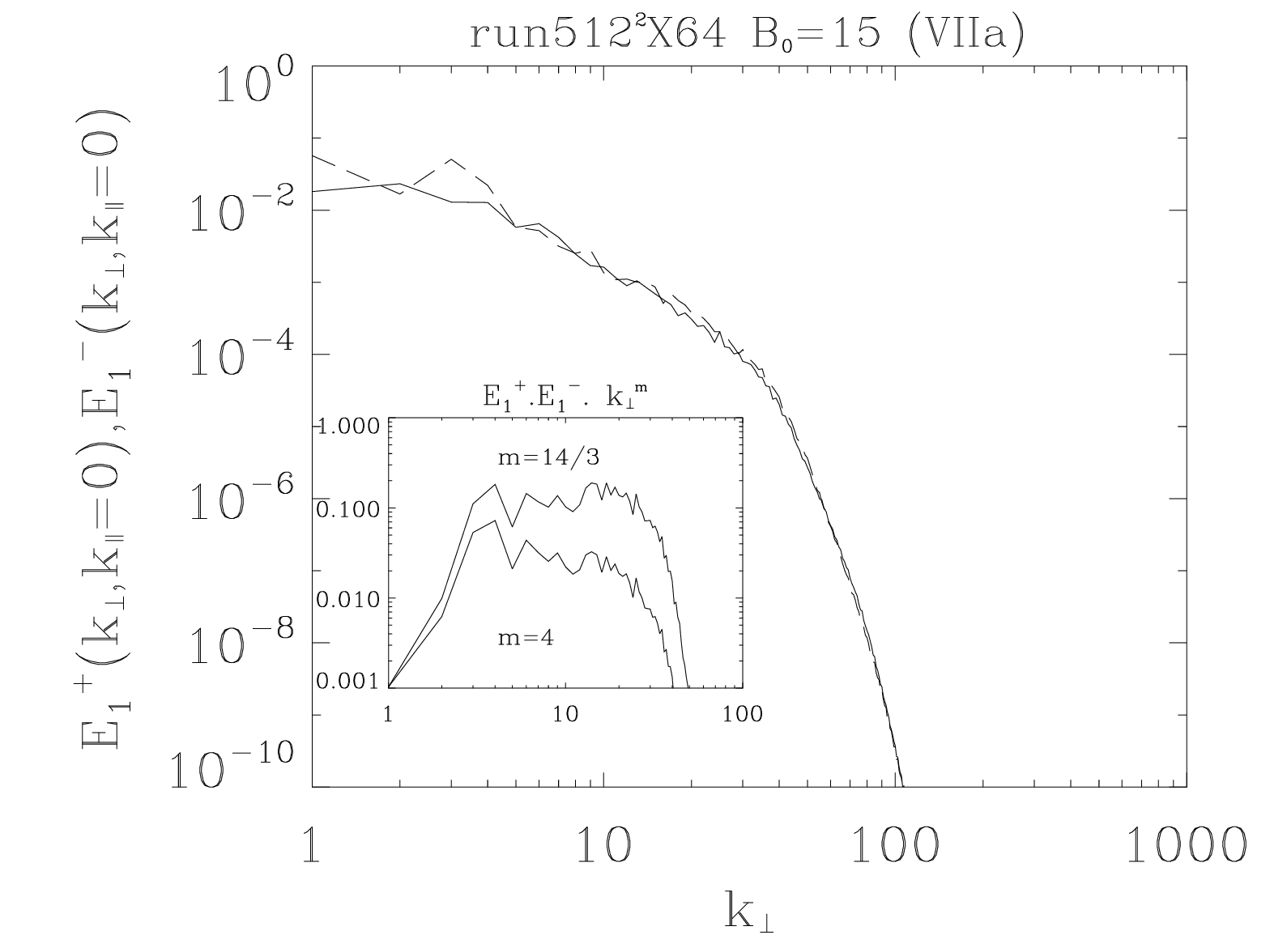}}\\
\resizebox{58mm}{!}{\includegraphics{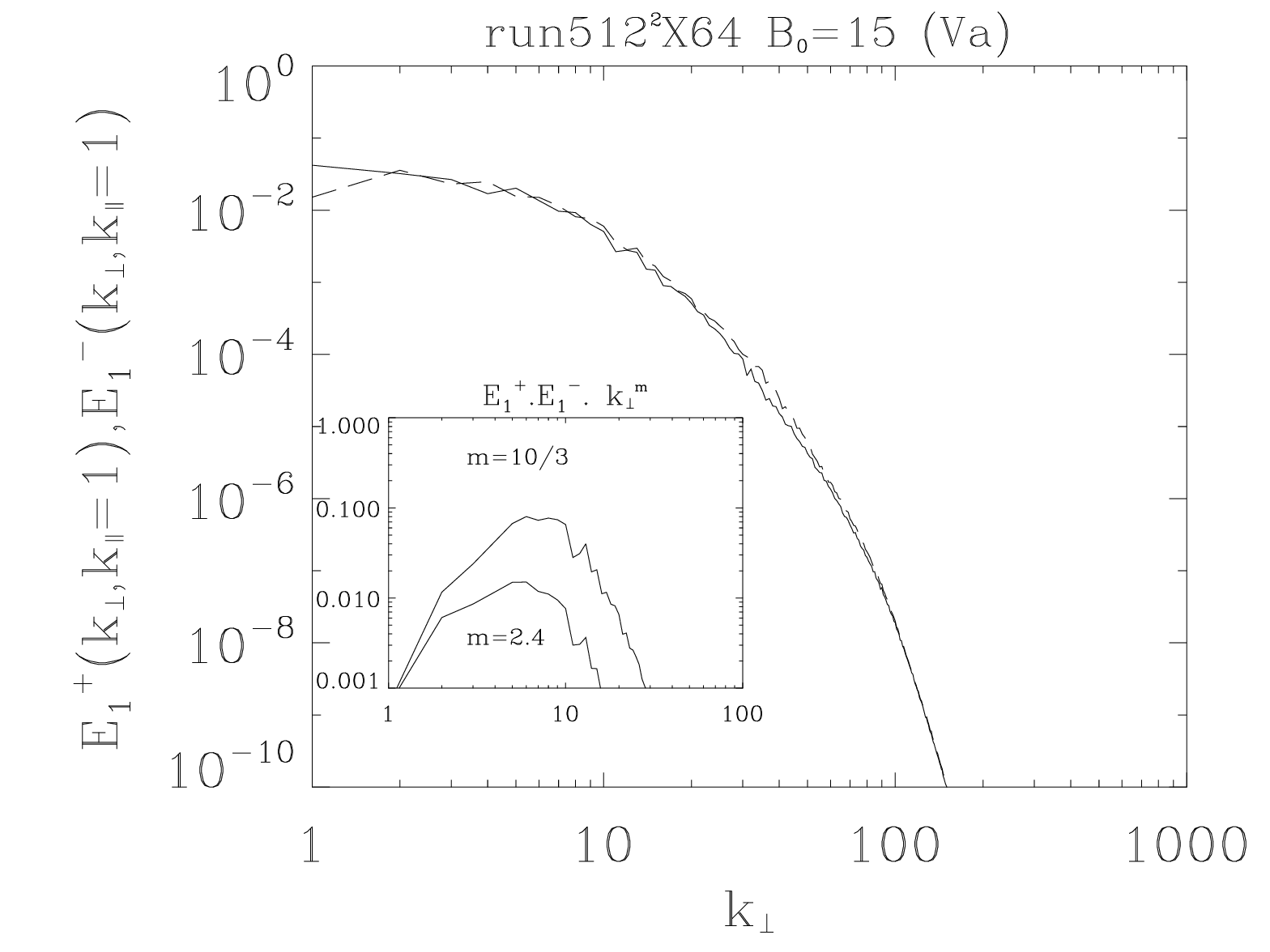}}
\resizebox{58mm}{!}{\includegraphics{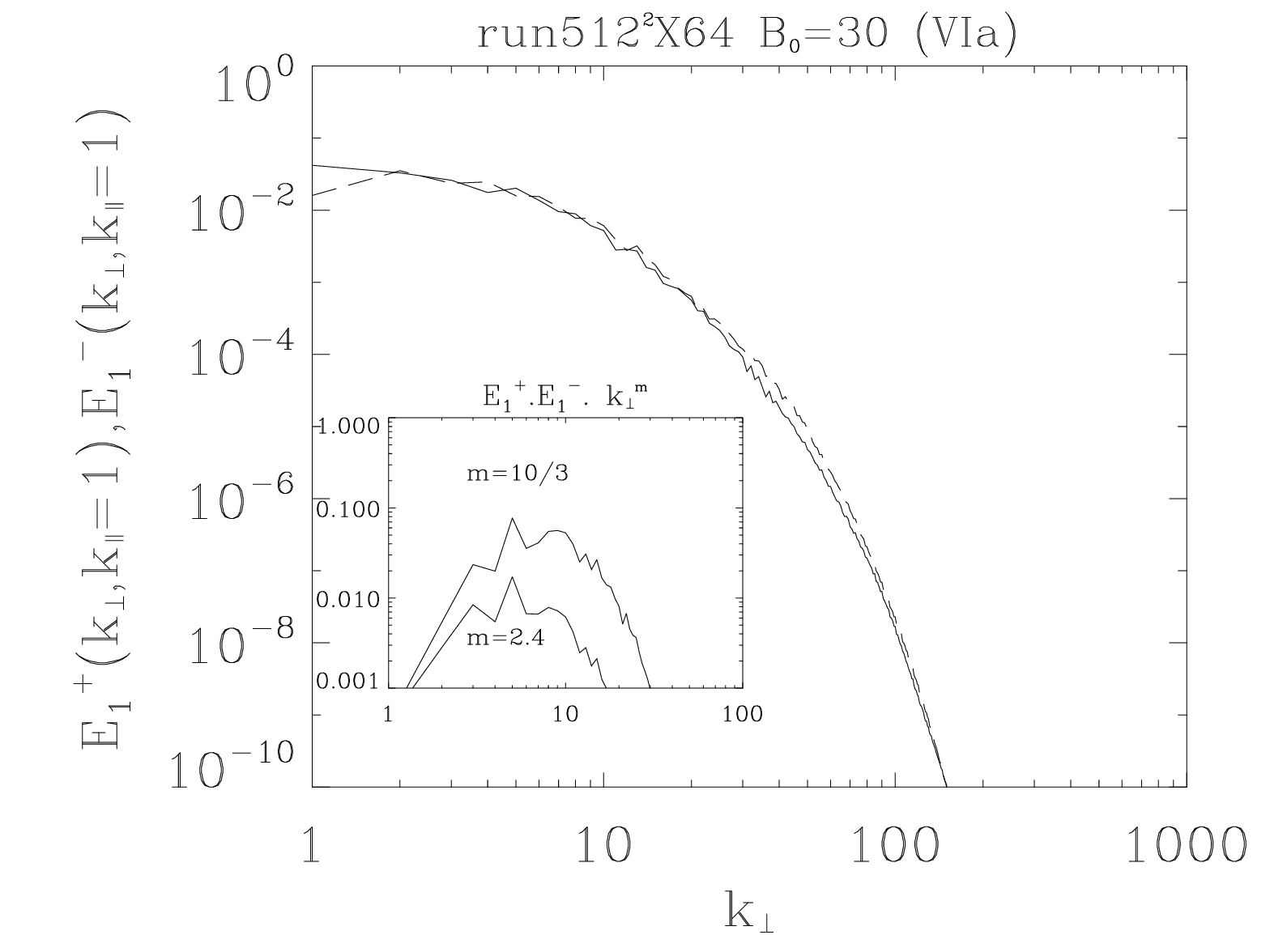}}
\resizebox{58mm}{!}{\includegraphics{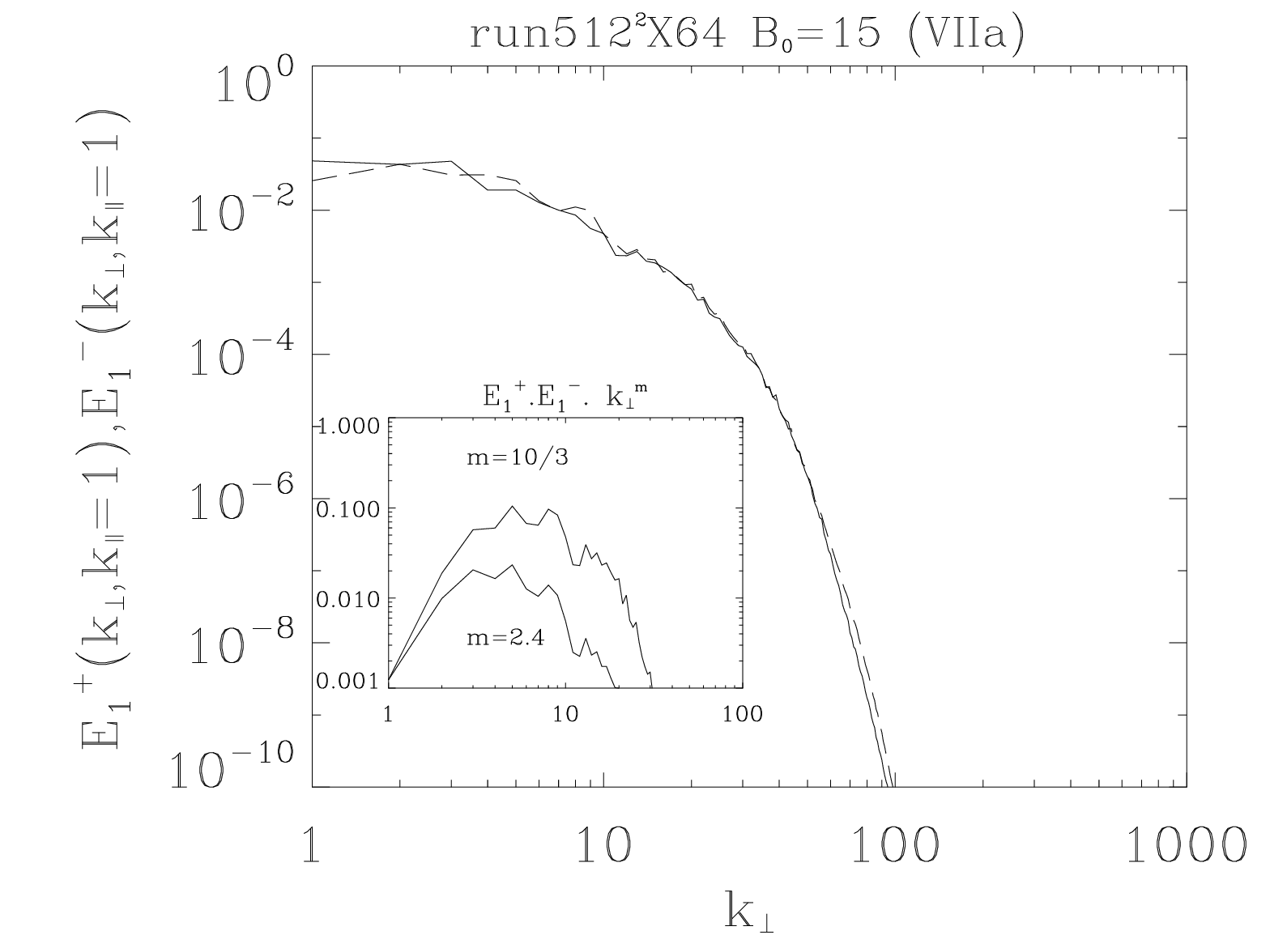}}
\end{tabular}
\caption{Energy spectra of shear-Alfv\'en waves $E_1^+$ (solid line) and $E_1^-$ (dashed line) \vs 
the perpendicular wavenumbers $\kpe$ after integration over all parallel wavenumbers (top), for 
$\kpa=0$ (middle) and for $\kpa=1$ (bottom). Each of these three columns presents, from left to right 
respectively, runs ${\bf Va}$ and ${\bf VIa}$ ($\nu=10^{-3}$ with $B_0=15$ and $30$), and the 
hyperviscous run ${\bf VIIa}$ ($\nu=10^{-6}$ with $B_0=15$). Inset: compensated product of energy 
spectra, $E^+ E^- k^m$ for a given $m$. 
\label{figSpec2Dk}}
\end{figure*}

Figure \ref{figSpec2Dk} shows anisotropic spectra for shear-Alfv\'en waves (polarity $+$) at times for 
which turbulence is fully developed ($t \sim 4$). First, we see spectra $E_1^\pm(\kpe)$ (top) 
which are defined as
\be 
E_1^\pm(\kpe)=\int E_1^\pm(\kpe,\kpa) d\kpa \, .
\ee
Then two other sets of spectra are given: $E_1^\pm(\kpe,\kpa=0)$ and $E_1^\pm(\kpe,\kpa=1)$ 
(middle and bottom panels respectively). The most interesting case seems to be the middle panel, \ie 
the spectra of the two-dimensional (2D) state, from which we see a clear inertial 
range where a power law may be extracted. An attempt is made to find this power law by computing the 
compensated spectra $E^+ E^- k^m$. Different values are proposed in the insets. We see that the 
2D state is characterized by approximately $m=14/3$ which means on average a spectrum steeper than 
the Kolmogorov one. This scaling is clearly different from the value found for $E_1^\pm(\kpe)$ where 
the Kolmogorov value $m=10/3$ is better fitted. The last case $E_1^\pm(\kpe,\kpa=1)$ is the most 
difficult one to analyze and no clear scaling appears. 
Note that the hyperviscous runs do not exhibit significant differences with, for example, a wider 
inertial range. In fact, the latter effect is easier seen for spectra plotted at fixed, but large, $k_\parallel$
($k_\parallel>1$).
Finally note the difference between these spectra and those found in Figure \ref{figSpecEP1D} (left) 
with an inertial range easier to determine in Figure \ref{figSpec2Dk} which may be attributed to the 
choice of the representation (anisotropic spectra instead of reduced spectra).

\subsection{Anisotropic scaling laws}

\begin{figure*}[ht]
\begin{tabular}{cc}
\resizebox{75mm}{!}{\includegraphics{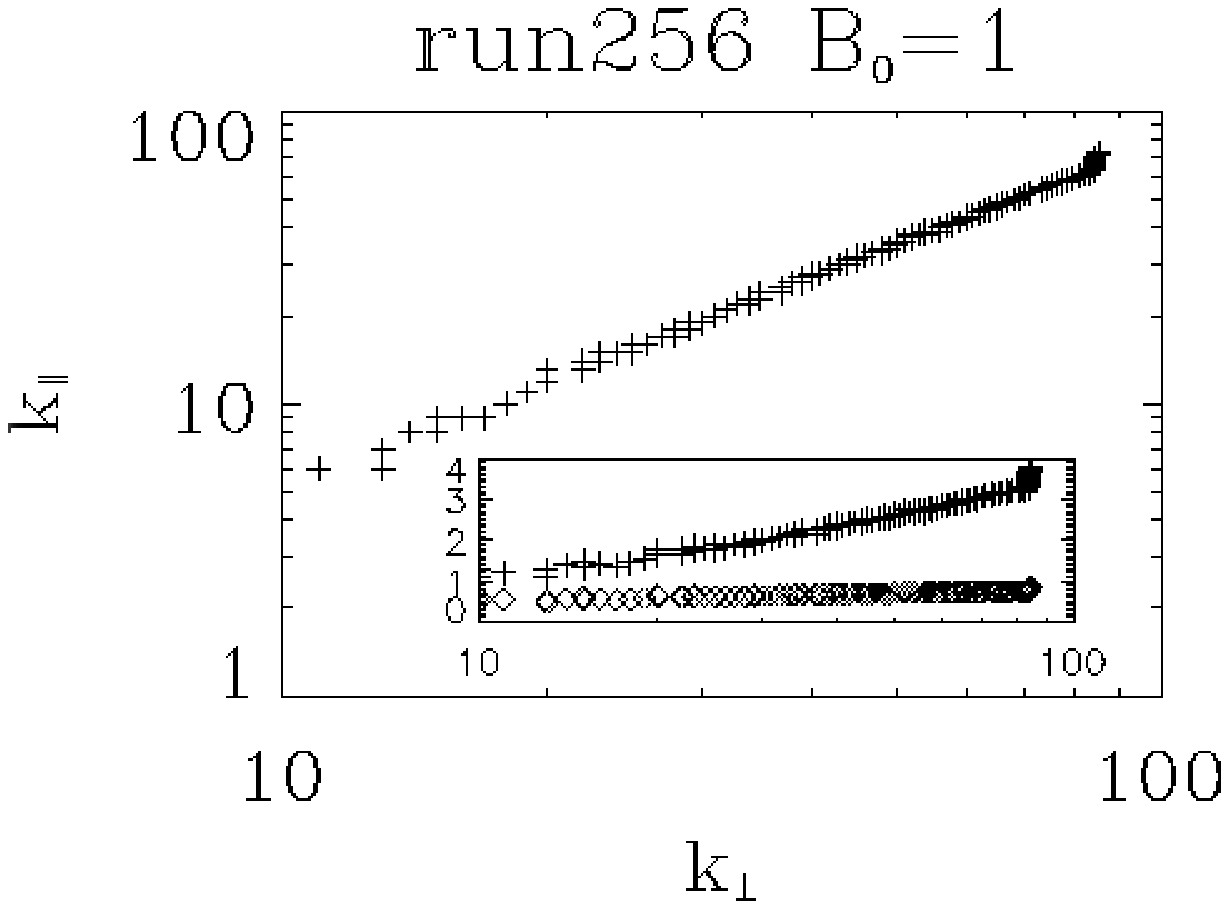}}
\resizebox{75mm}{!}{\includegraphics{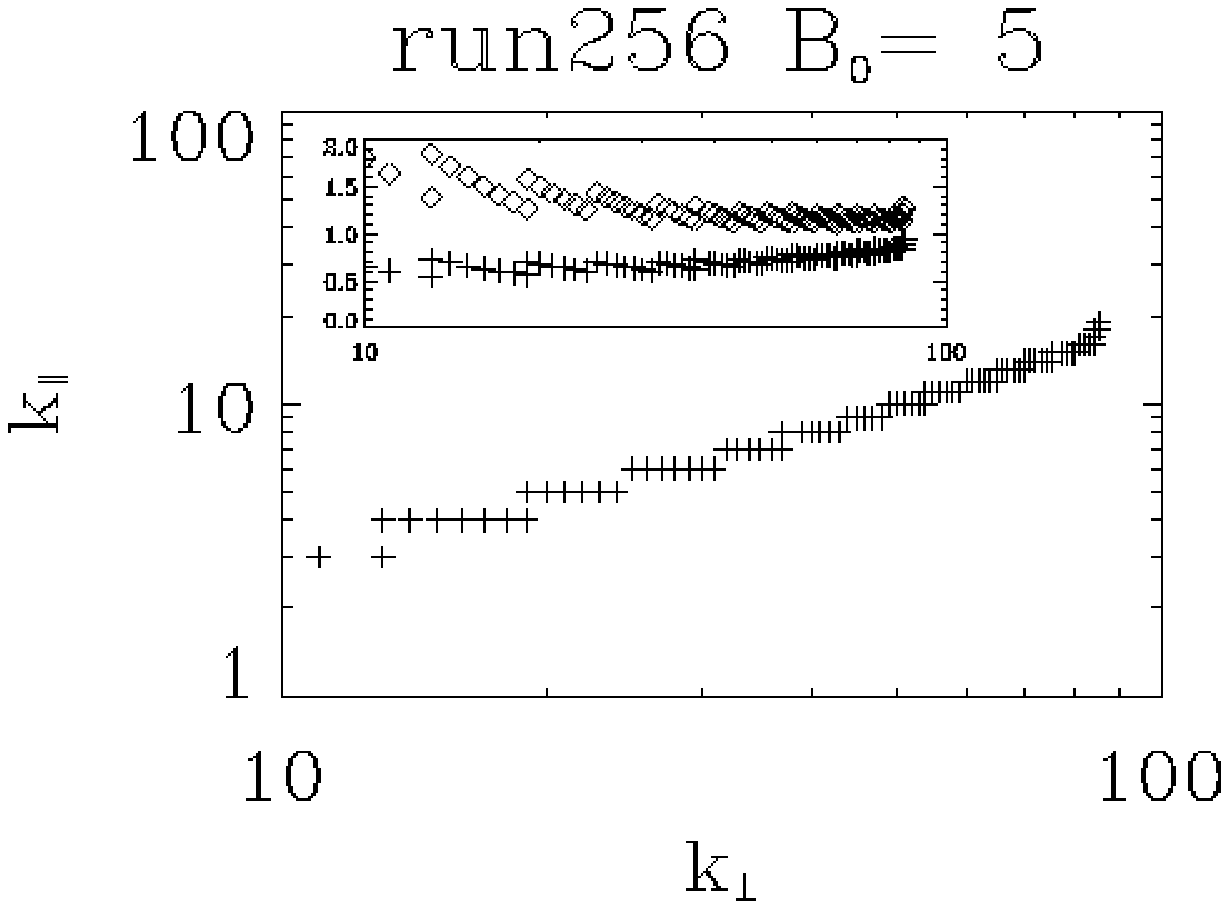}}\\
\resizebox{75mm}{!}{\includegraphics{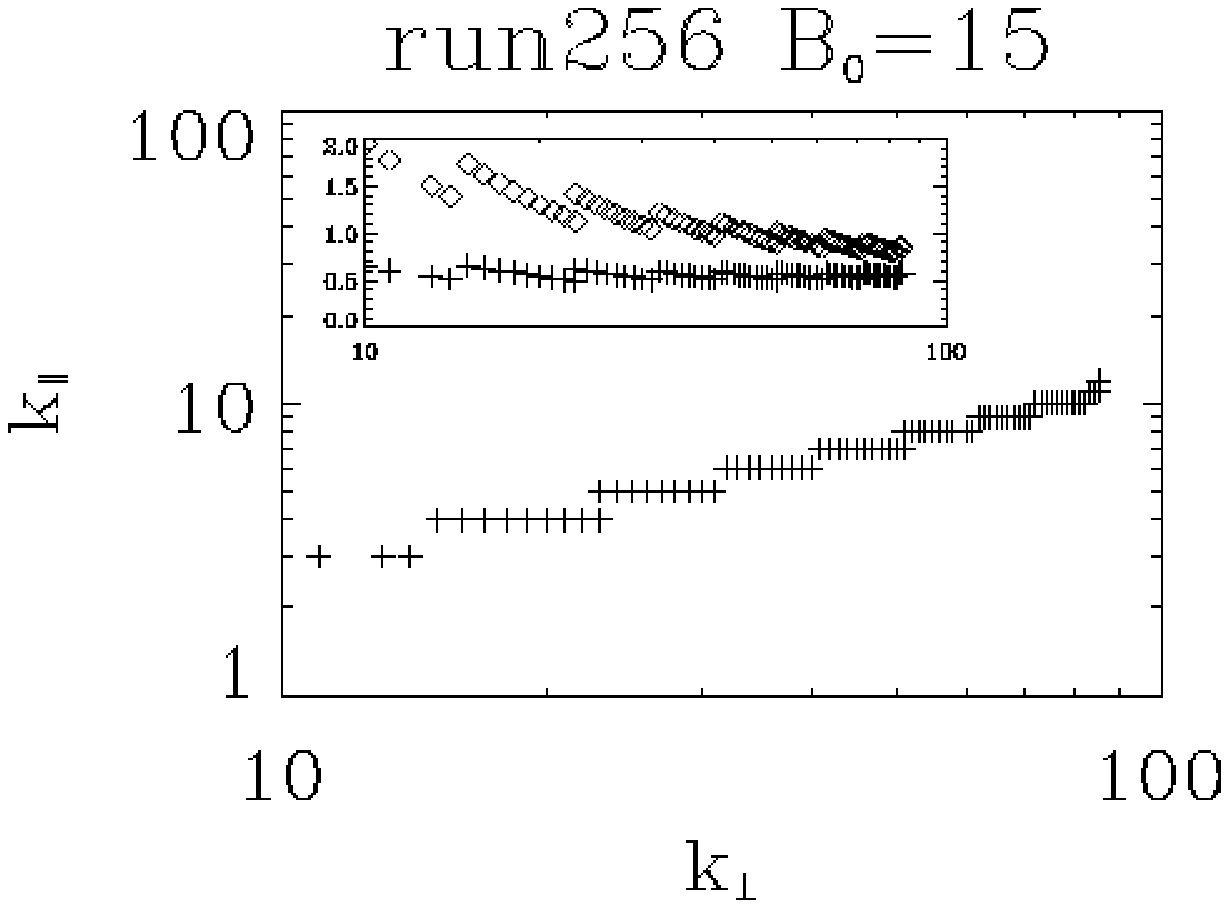}}
\resizebox{75mm}{!}{\includegraphics{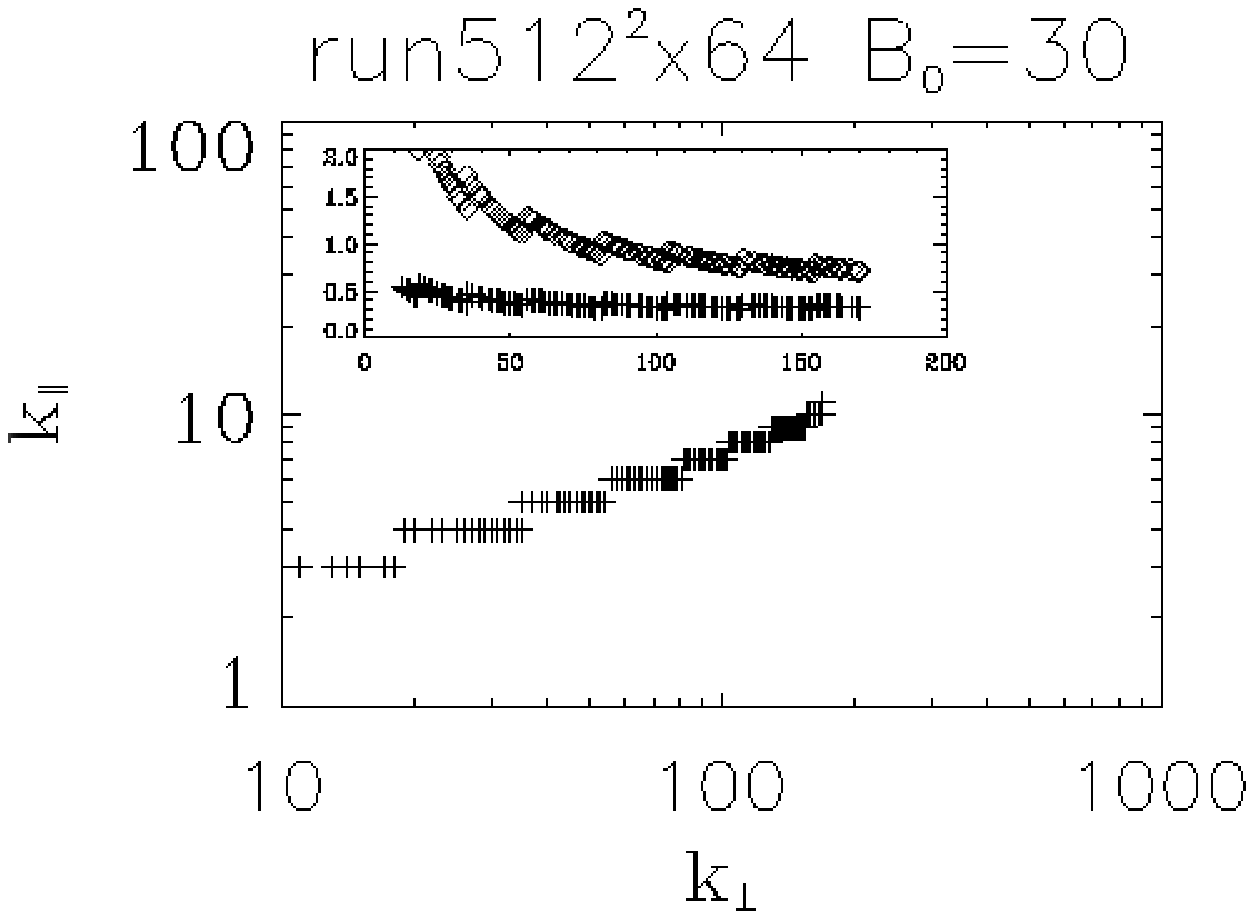}}
\end{tabular}
\caption{Anisotropic scaling laws between wavenumbers $\kpe$ and $\kpa$ (see text) for runs 
${\bf IIa}$ to ${\bf IVa}$ and run ${\bf VIa}$. The inset displays two compensated scalings: 
$\kpa(\kpe) \, \kpe^{-1}$ (diamonds) and $\kpa(\kpe) \, \kpe^{-2/3}$ (crosses).
\label{k}}
\end{figure*}
In order to extract a scaling law between parallel and perpendicular wavenumbers, we plot the modes 
($\kpe$,$\kpa$) corresponding to the equality $E_1^+(\kpe)=E_1^+(\kpa)$ with 
\be
E_1^+(k_\parallel)=\int E_1^+(k_\perp,k_\parallel) dk_\perp \, .
\ee
The result is given in Figure \ref{k} for runs ${\bf IIa}$ ($t\sim2$), ${\bf IIIa}$-${\bf IVa}$ ($t\sim3$) and 
${\bf VIa}$ ($t\sim4$). We clearly see different slopes for different magnitudes of $B_0$, with an 
isotropic law $\kpa \sim \kpe$ for $B_0=1$ and an anisotropic law around $\kpa \sim \kpe^{2/3}$ for 
$B_0 \ge 15$ (see insets). For all cases, we see that the scaling law extends to the dissipative range. 
The same behavior is found for the pseudo-Alfv\'en waves (not shown here). 
Note that with this method the scaling law extracted suffers from an average effect since each 
spectrum is obtained after summation over the parallel or the perpendicular direction. 
Nevertheless, the anisotropic prediction proposed by \cite{GS95} is often recovered, but as it was 
explained above for a sub-critical balance between the Alfv\'en and nonlinear times, which may be 
understood in a wider context \cite{Galtier2005} as discussed in the introduction.

%%%%%%%%%%%%%%%%%%%%%%%%%%%%%%%%%%%%%%%%%%%%%%%%%
%\section{Flow geometry}
\section{Visualizations}

\subsection{Spectral space}

\begin{figure*}[ht]
\begin{tabular}{cc}
\resizebox{58mm}{!}{\includegraphics{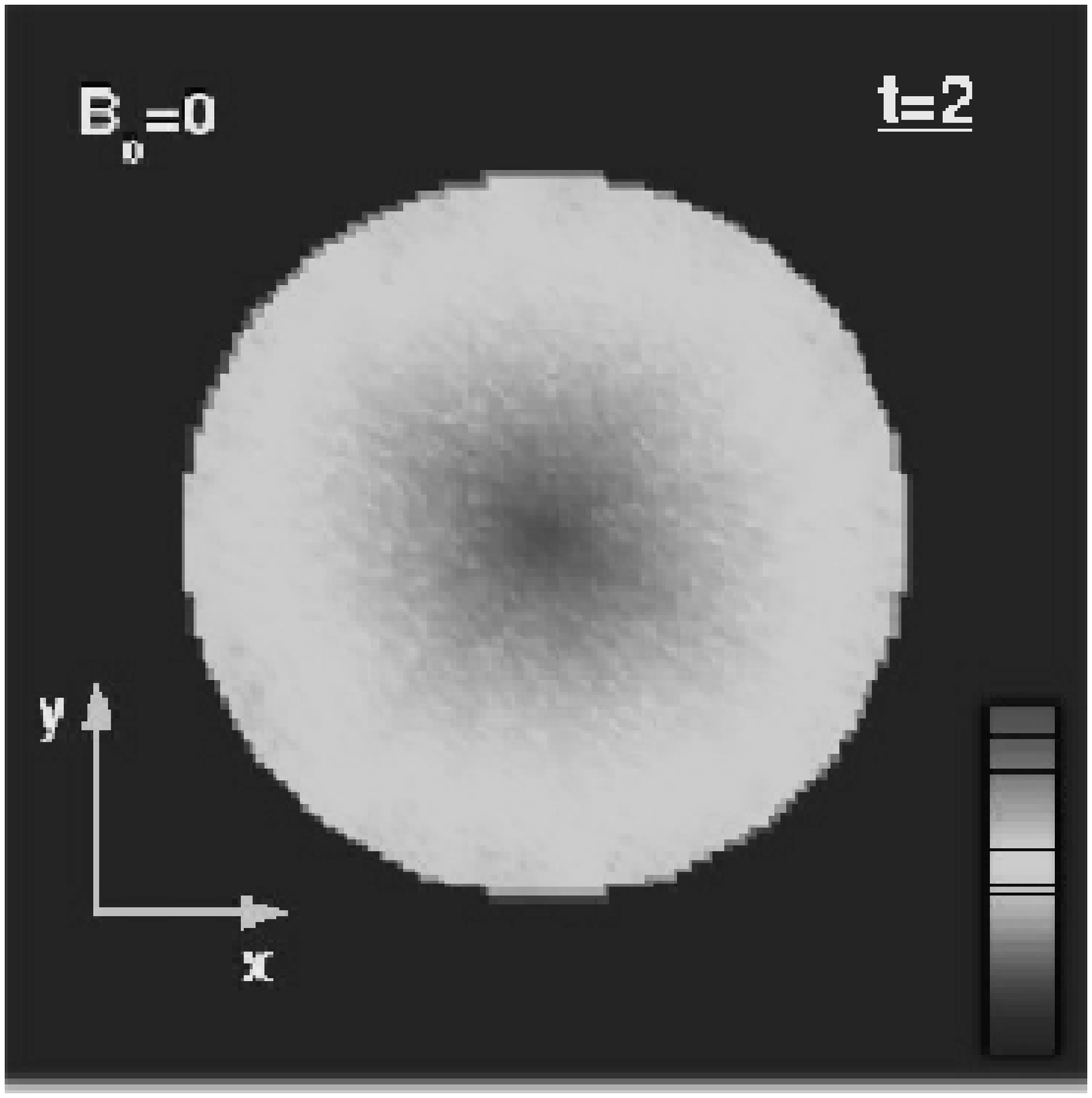}}
\resizebox{59mm}{!}{\includegraphics{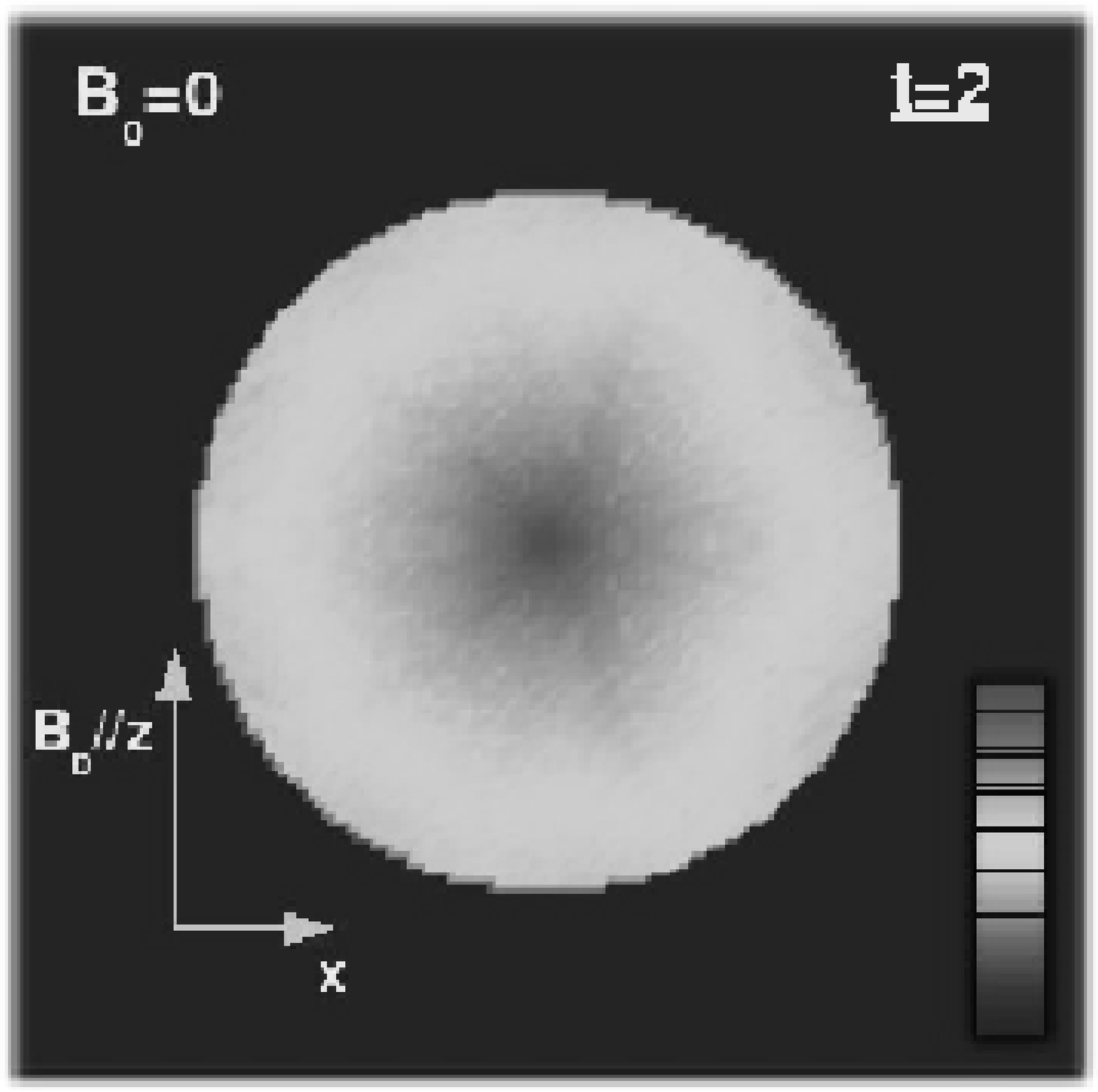}}\\
\resizebox{58mm}{!}{\includegraphics{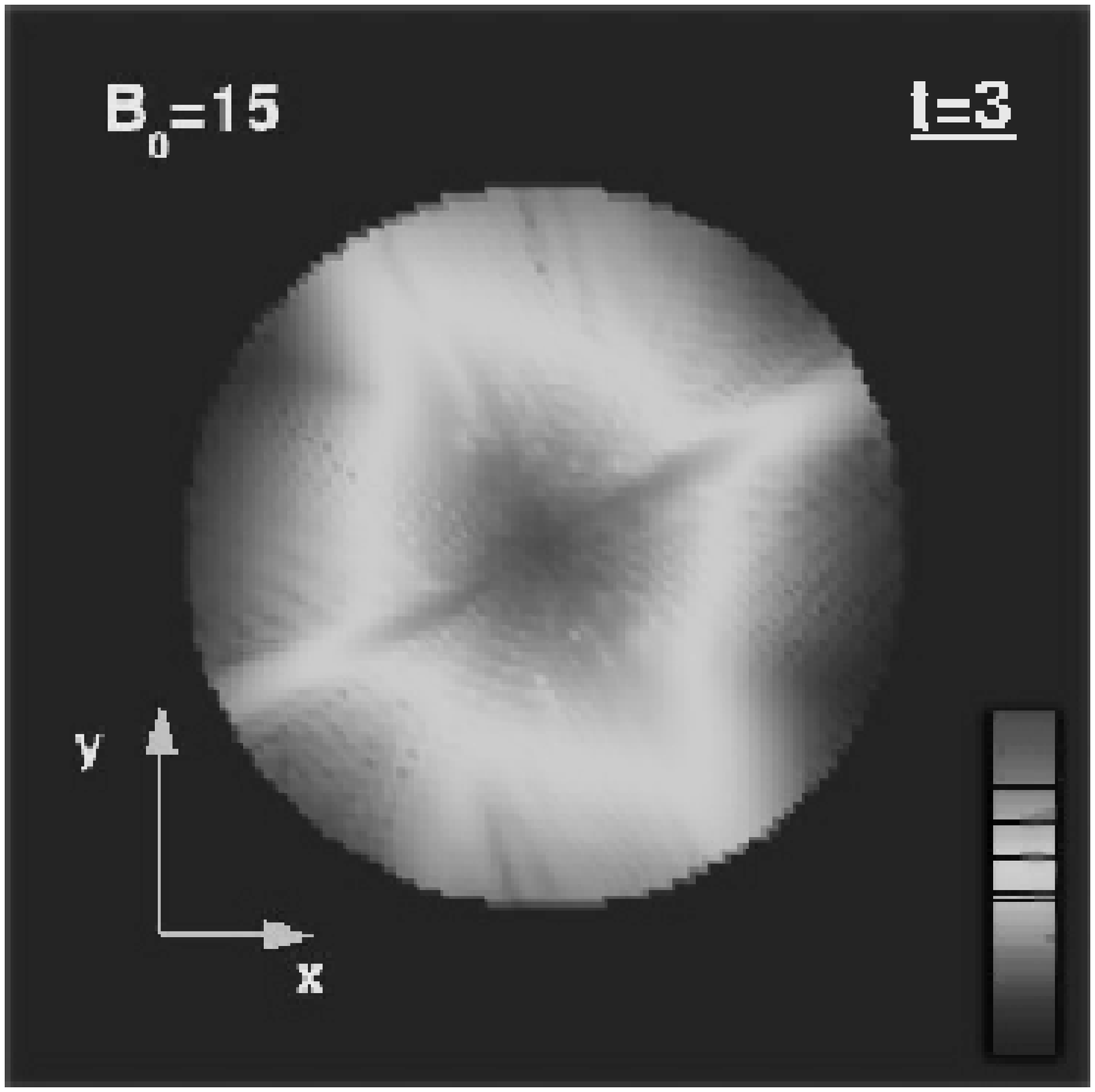}}
\resizebox{58mm}{!}{\includegraphics{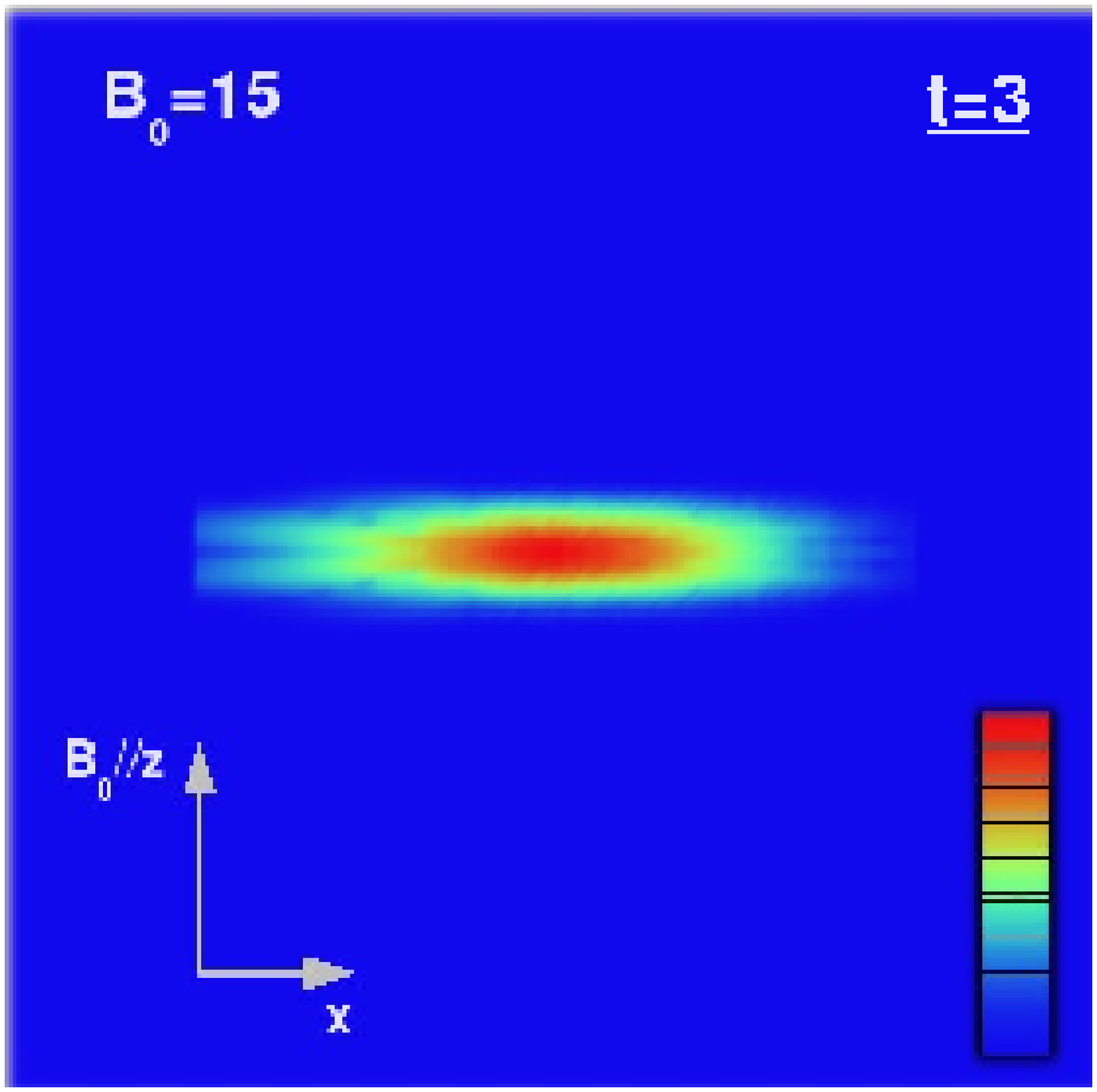}}
\end{tabular}
\caption{$E^+(\bf k)$ cuts in Fourier space at $k_{\parallel}=0$ (left) and $k_y=0$ (right) for
flows at $B_0=0$ (top; run {\bf Ia}) at $t=2$, and $B_0=15$ (bottom; run {\bf IVa}) at $t=3$.
Color bars normalized to 1 for the maximum intensity (red) and to 0 for the minimum (blue) one.}
\label{VisuSpec1}
\end{figure*}

At a time at which energy spectra are fully developped, Figure \ref{VisuSpec1} displays perpendicular, 
at $k_{\parallel}=0$, and parallel, at $k_y=0$, cuts in Fourier space for $E^+(\bf k)$ in flows at 
$\nu=\eta=4.10^{-3}$ without ($B_0=0$, $t=2$; run {\bf Ia}) or with ($B_0=15$, $t=3$; run {\bf IVa}) an 
applied magnetic field. Initially, for both flows, the isotropic energy injection corresponds to spherical 
shells with maximum radius $k=8$. 
The spectra then evolves depending on the level of the flow magnetization.
Indeed, at $B_0=0$, the maximum spectral radius increases in all directions, meaning
an isotropic energy transfer towards small scales, while at $B_0=15$, the 
three-dimensional energy spectrum collapses into ellipsoidal shapes with ratio $1/6$th,
corresponding to an anisotropic transfer, strongly inhibited in the ${\bf B_0}$ parallel
direction. In this case, in ${\bf B_0}$ perpendicular planes (shown here at $k_{\parallel}=0$), 
one can observe a loss of excitation at higher modes together with a loss of axisymmetry, with two 
preferred directions, compared to non-magnetized flows. 
\begin{figure*}[ht]
\begin{tabular}{cc}
\resizebox{58mm}{!}{\includegraphics{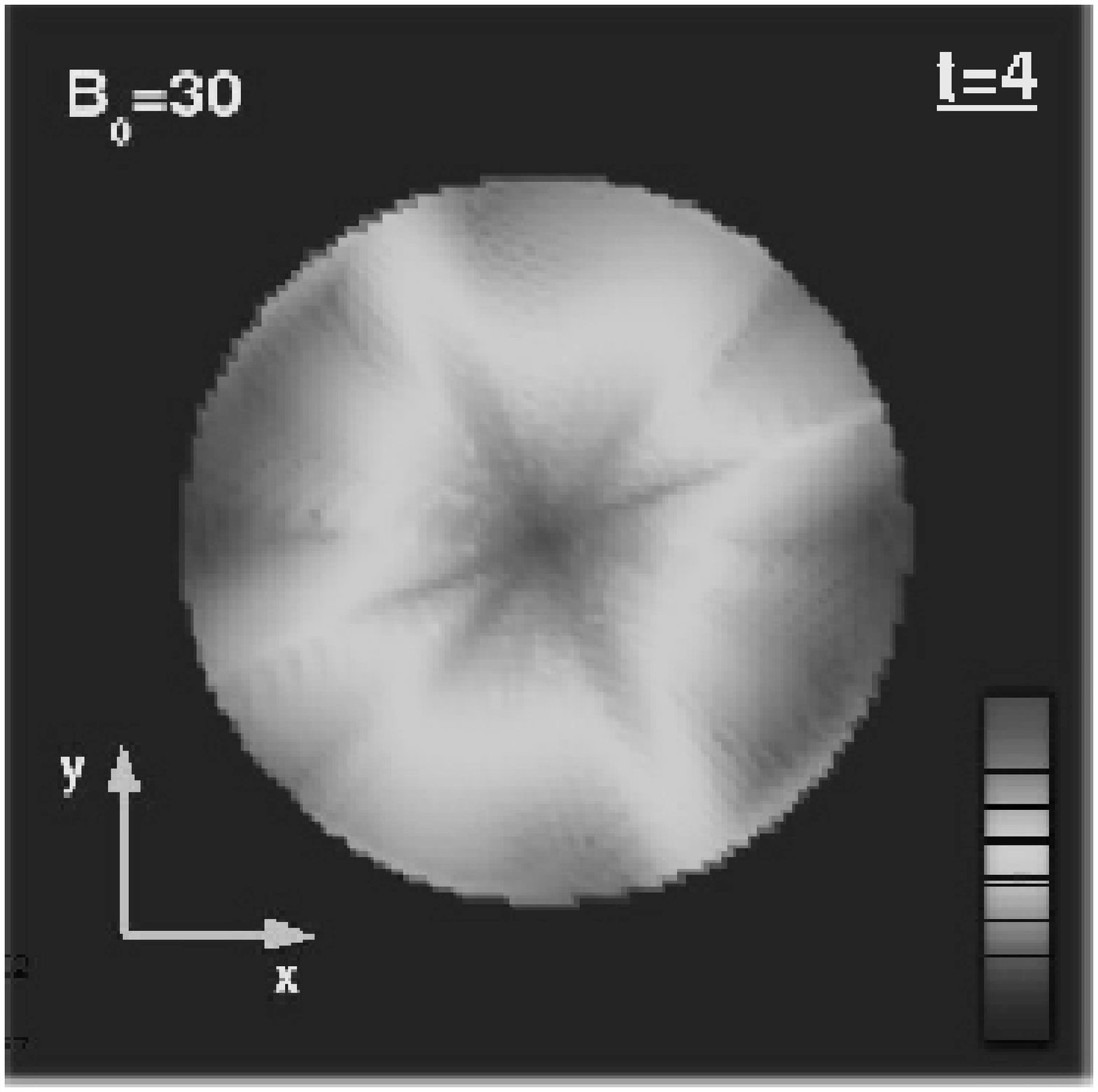}}
\resizebox{58mm}{!}{\includegraphics{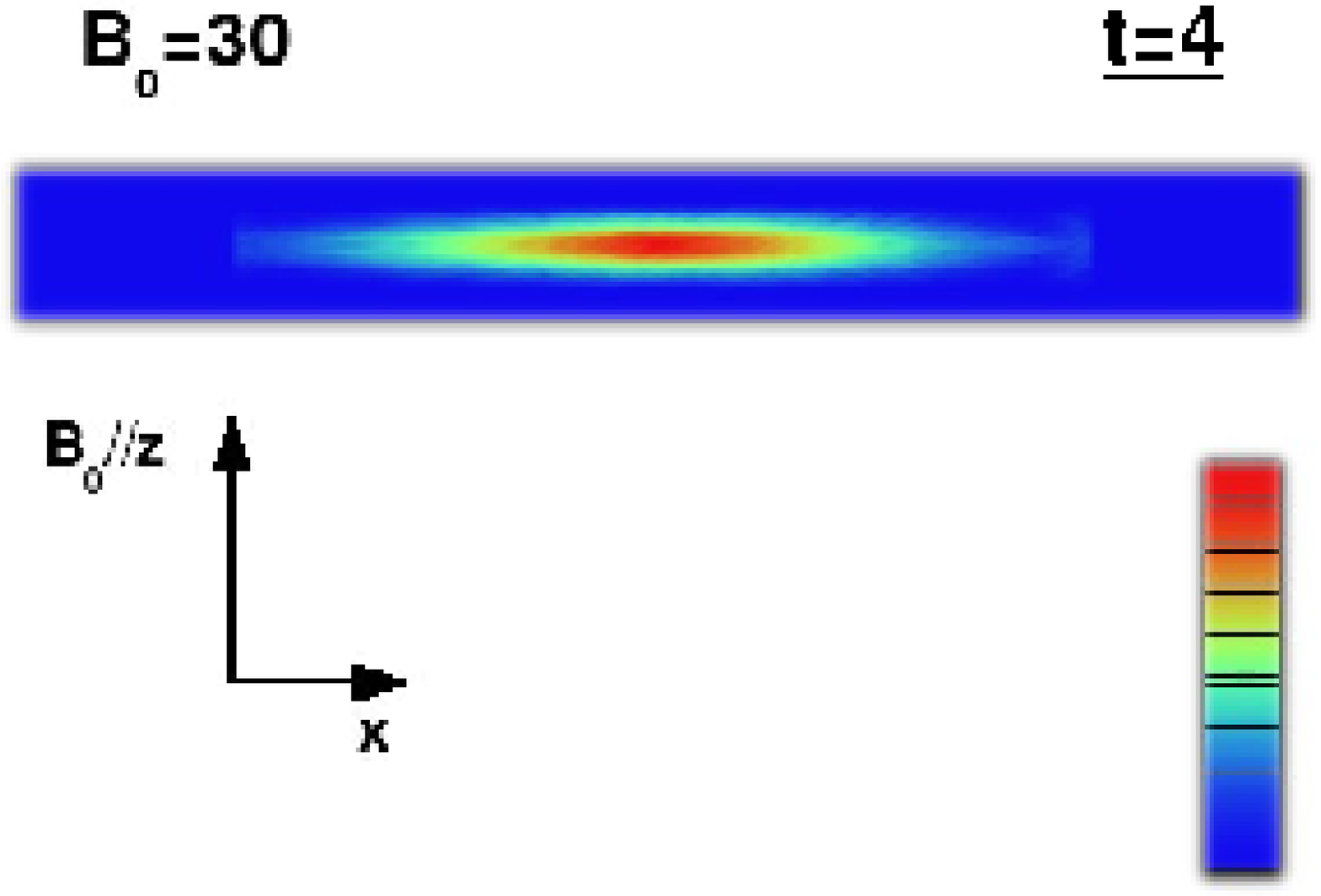}}
\end{tabular}
\caption{ $E^+(\bf k)$ cuts in Fourier space at $k_{\parallel}=0$ (left) and $k_y=0$ (right) for
flows at $B_0=30$ (run {\bf VIa} using $512^2 \times 64$ grid points) at $t=4$.
Color bars normalized to 1 for the maximum intensity (red) and to 0 for the minimum (blue) one.}
\label{VisuSpec2}
\end{figure*}

Figure \ref{VisuSpec2} shows the case of strongly magnetized flows, $B_0=30$, at lower 
viscosity $\nu=10^{-3}$, and resolved with $512^2 \times 64$ grid points ($t=4$; run {\bf VIa}). 
The aspect ratio of the spectral ellipsoidal shape decreases up to $1/10$th and 
in transverse planes, a star shape with several "jets" appears. As time evolves (not shown), 
the number of these jets increases leading to an enhanced isotropy in tranverse planes (at 
$k_\parallel \geq 0$). In all flows, similar observations stand for $E^-(\bf k)$ spectra.

\subsection{Physical space}

\begin{figure*}[ht]
\begin{tabular}{cc}
\resizebox{72mm}{!}{\includegraphics{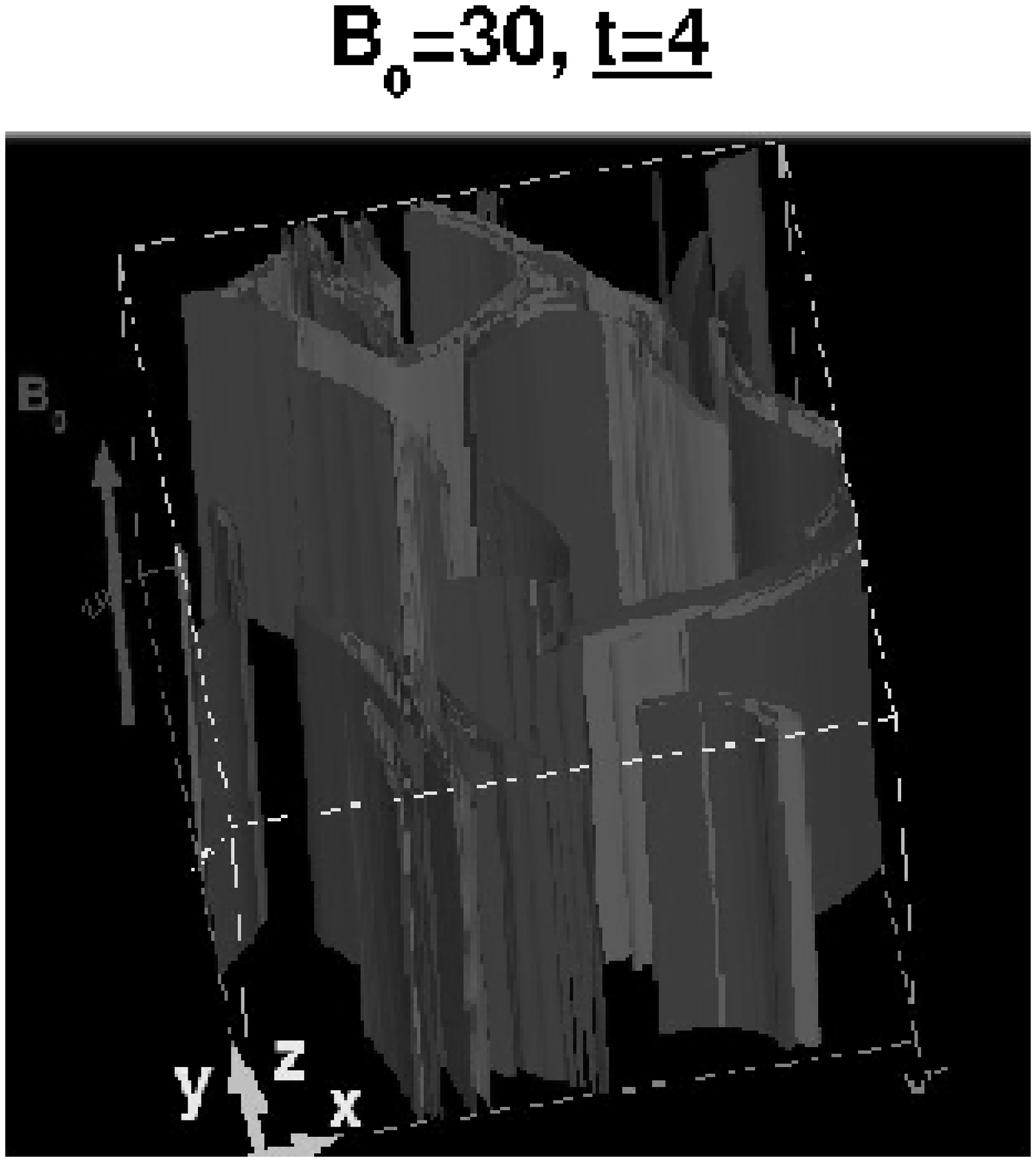}}
\resizebox{72mm}{!}{\includegraphics{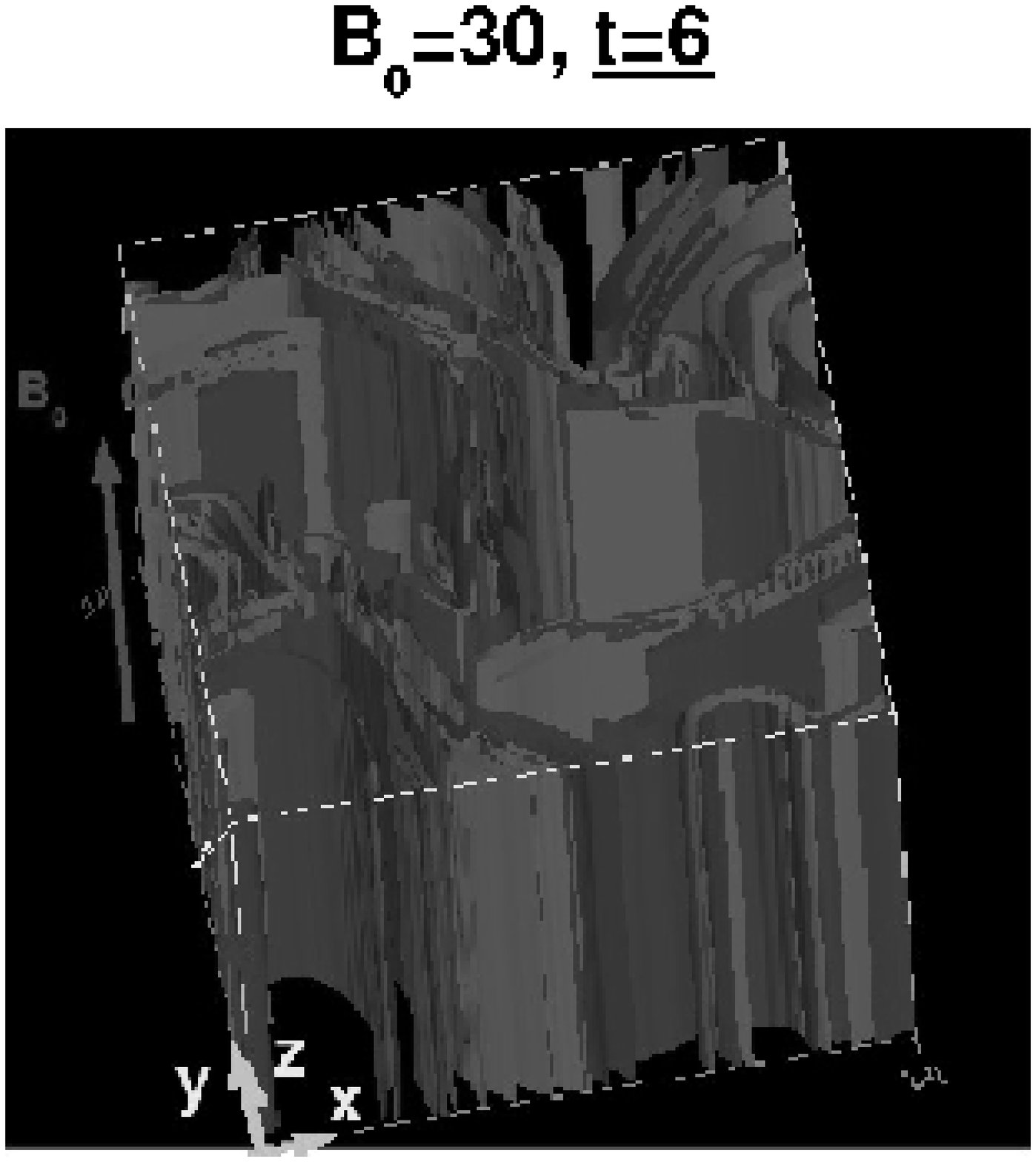}}
\end{tabular}
\caption{Isosurfaces of vorticity (blue) and current (orange) intensities for the 2D state $k_{\parallel}=0$
(see text) for a flow with $B_0=30$ (run {\bf VIa}), drawn at $20\%$ of their respective maxima.
At $t=4$ (left), $|{\bf w}|_{max}=8.8$ and $|{\bf j}|_{max}=11.3$, and at $t=6$ (right) 
$|{\bf w}|_{max}=6.6$ and $|{\bf j}|_{max}=8.1$.}
\label{VisuFilter1}
\end{figure*}
\begin{figure*}[ht]
\begin{tabular}{cc}
\resizebox{72mm}{!}{\includegraphics{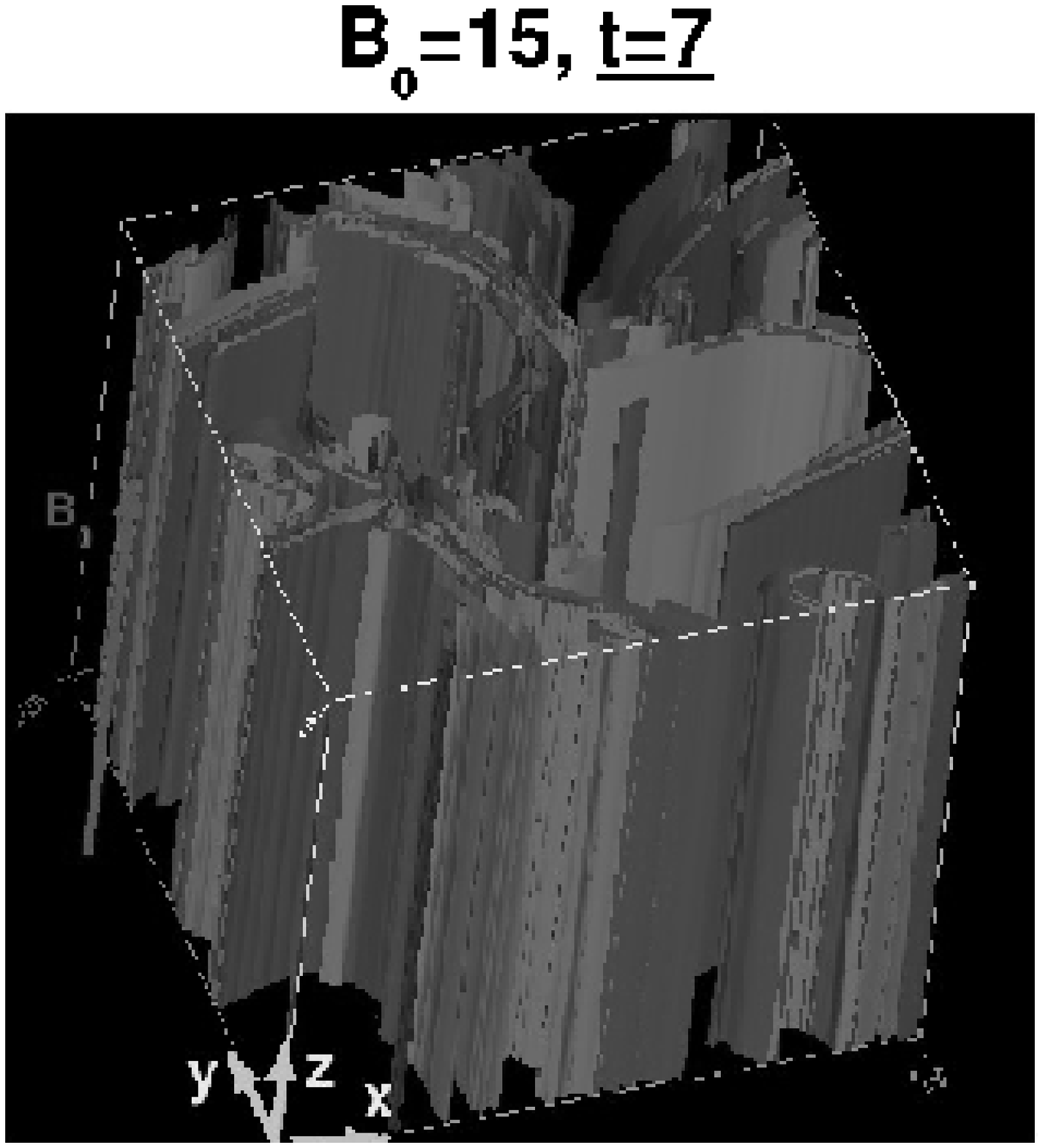}}
\resizebox{72mm}{!}{\includegraphics{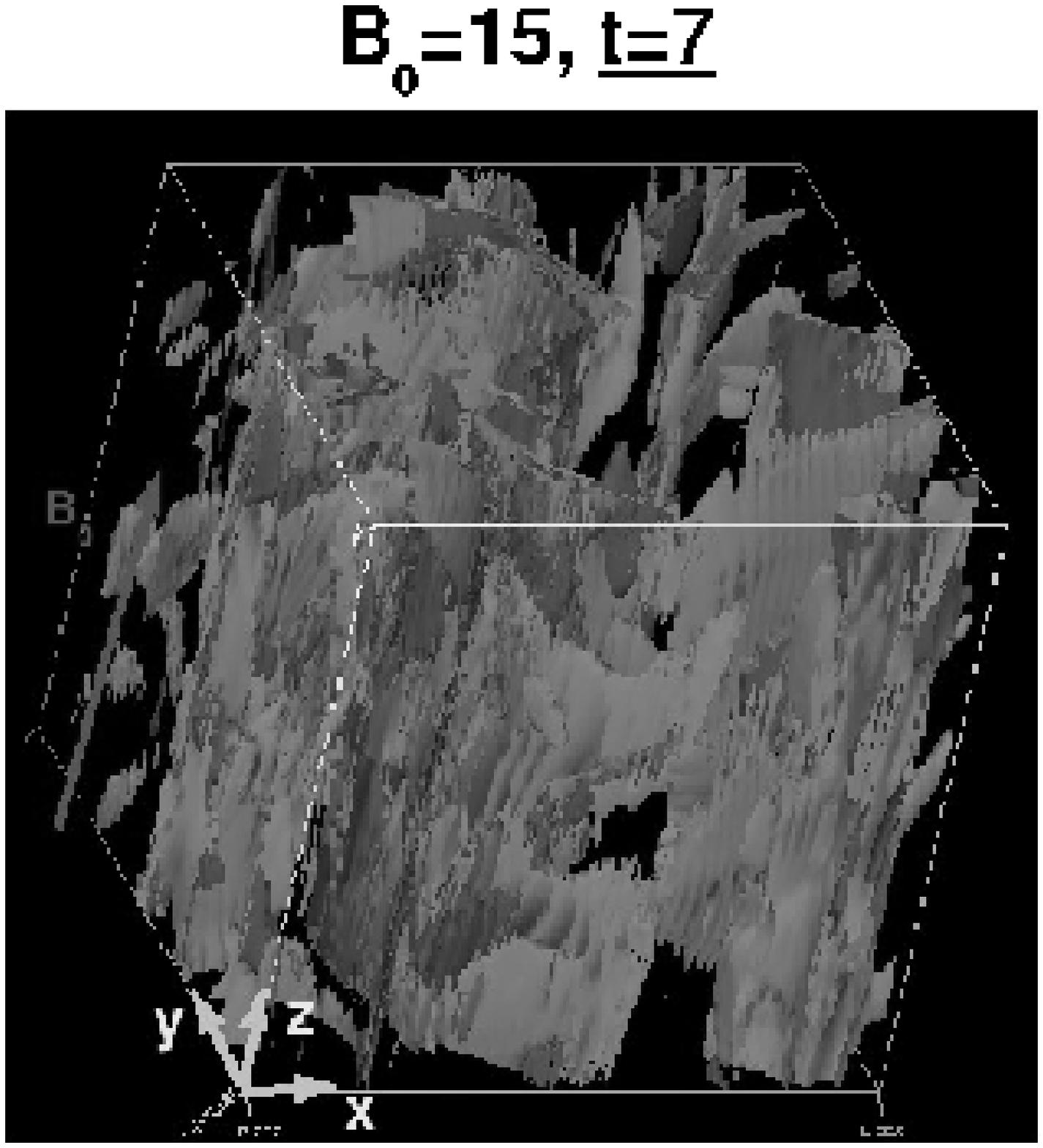}}
\end{tabular}
\caption{Filtered vorticity (blue) and current (orange) intensities with 
$k_{\parallel}=0$ (2D state, left) and $k_{\parallel}>0$ (right) for $B_0=15$ (run {\bf IVa}) :
isosurfaces are drawn at $27\%$ and $20\%$ of their respective maxima ($k_{\parallel}=0$;
$|{\bf w}|_{max}=5.7$ and $|{\bf j}|_{max}=9.$, and $k_{\parallel}>0$; 
$|{\bf w}|_{max}=18.5$ and $|{\bf j}|_{max}=19.2$).}
\label{VisuFilter2}
\end{figure*}
In order to understand the observed spectral structures in transverse planes for
magnetized flows, we first visualize their spatial counterparts,
once some Fourier amplitudes at wavevectors $(k_x,k_y,k_z=k_{\parallel})$ are filtered  
for a given field. Hence structures only corresponding to the 2D state are obtained with 
$(k_x,k_y,k_\parallel > 0)$ modes filtered, and structures for 3D modes ($k_\parallel > 0)$ 
are obtained with $(k_x,k_y,k_\parallel = 0)$ modes filtered. 
Figure~\ref{VisuFilter1}, for a flow with $B_0=30$ (run {\bf VIa}), displays vorticity and
current isosurfaces for the 2D state at the same time as Figure~\ref{VisuSpec2}, $t=4$,
and at a later time $t=6$. The transverse spectral star shape is related to the spatial
distribution of the vorticity and current sheets, perpendicularly to two peculiar directions
at $t=4$, and more irregularly distributed at $t =6$ for which higher number of jets is observed
in spectral transverse planes (not shown). 

\begin{figure*}[ht]
\begin{tabular}{cc}
\resizebox{72mm}{!}{\includegraphics{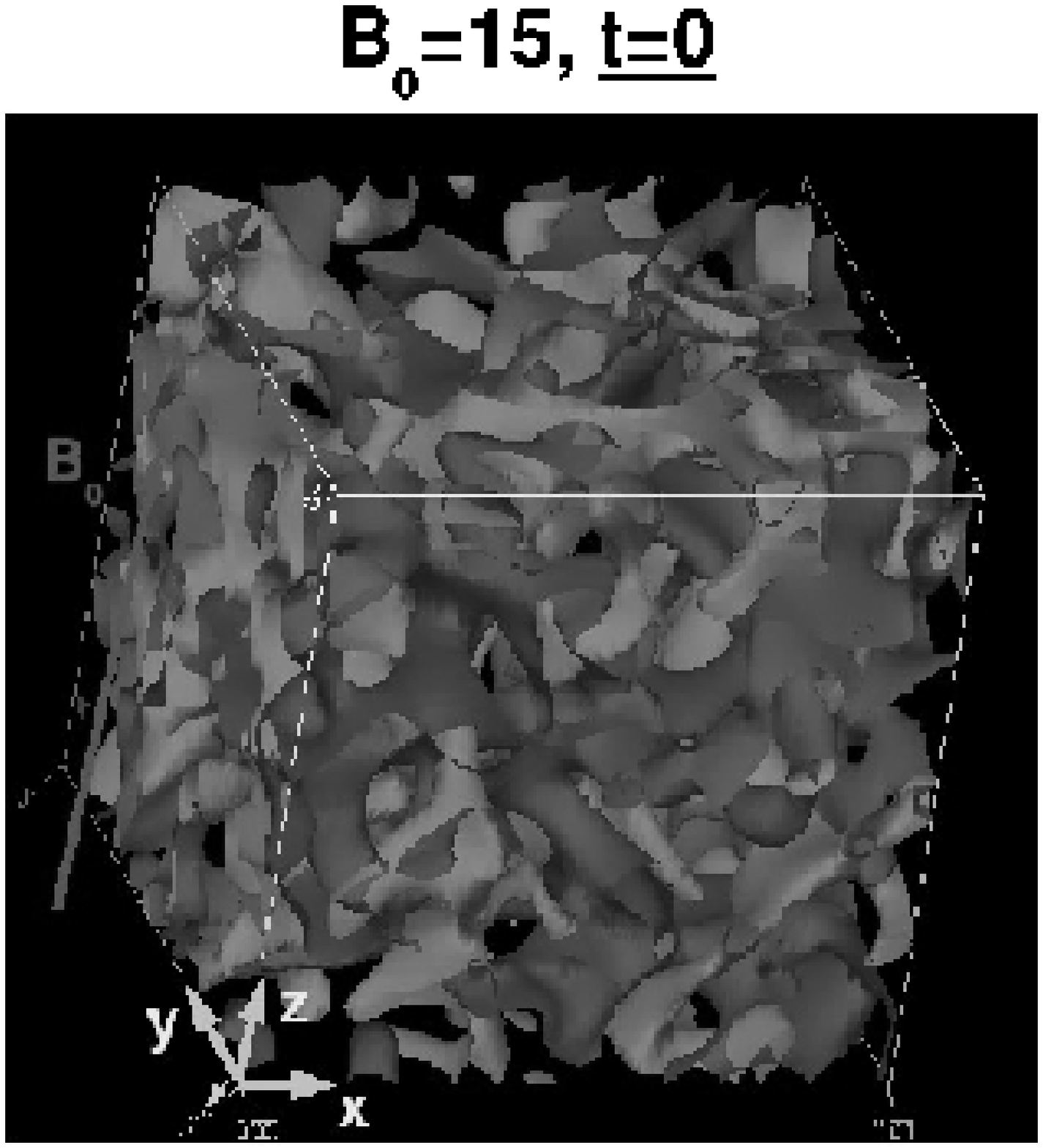}}
\resizebox{72mm}{!}{\includegraphics{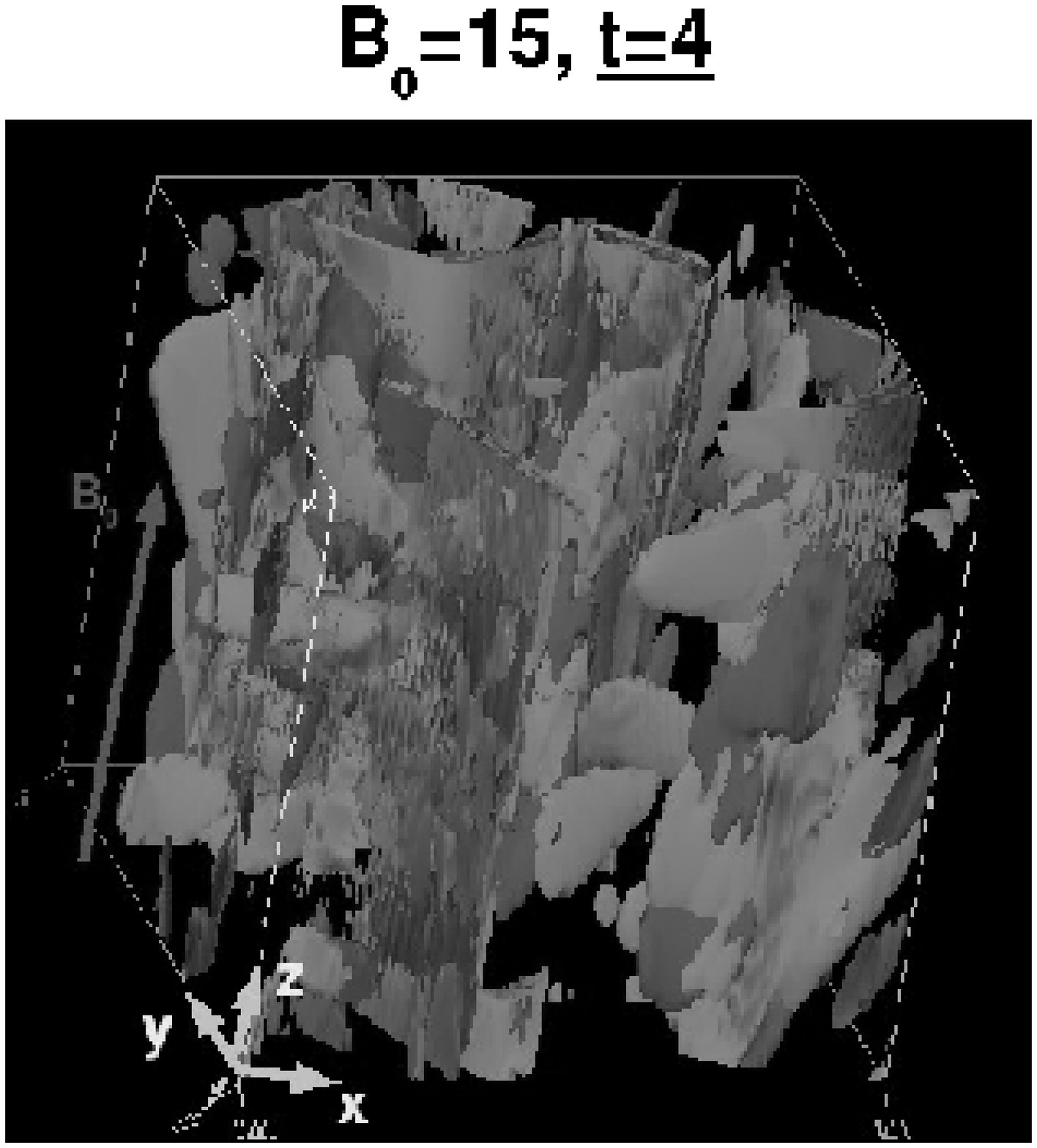}}\\
\resizebox{72mm}{!}{\includegraphics{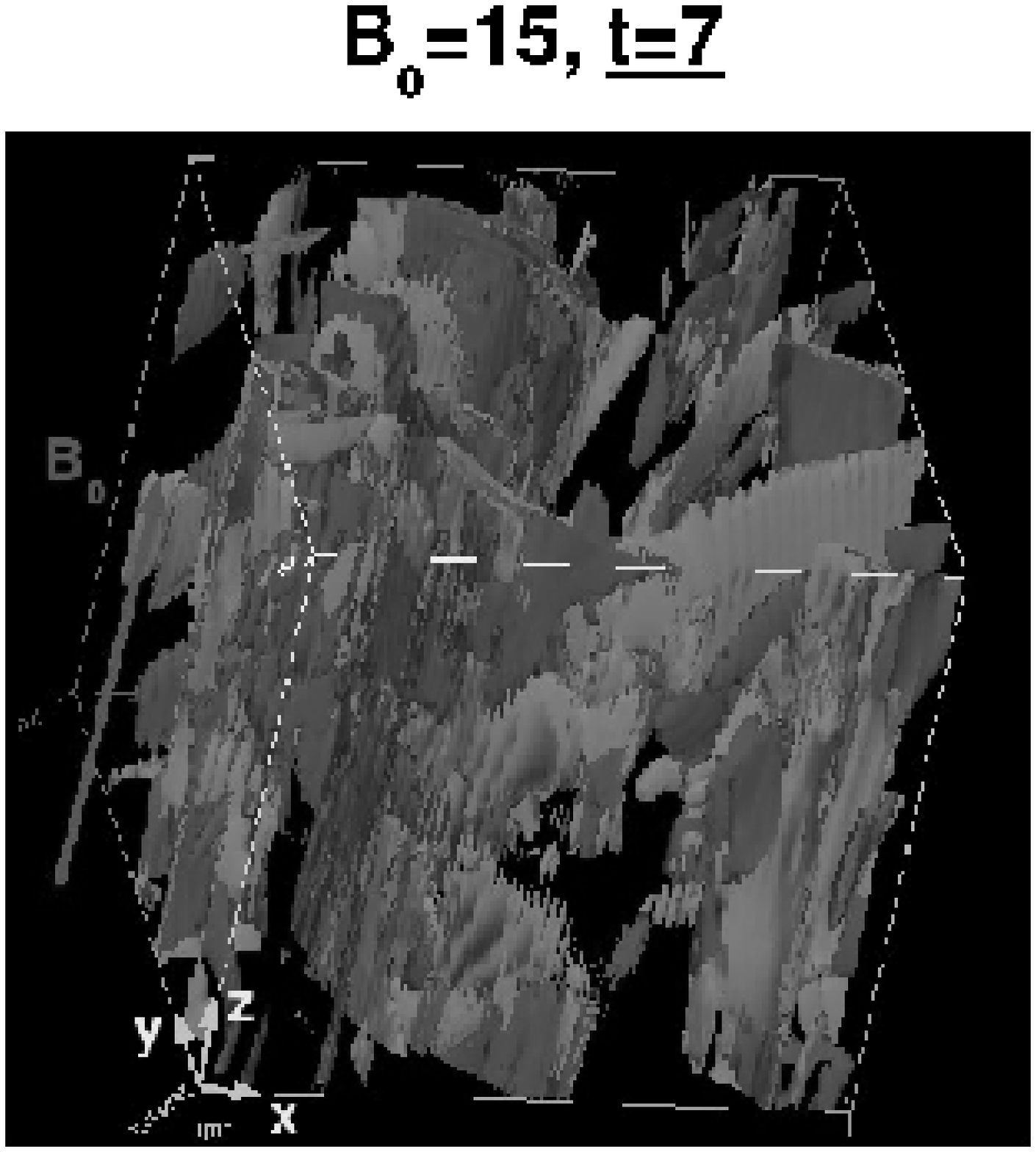}}
\resizebox{72mm}{!}{\includegraphics{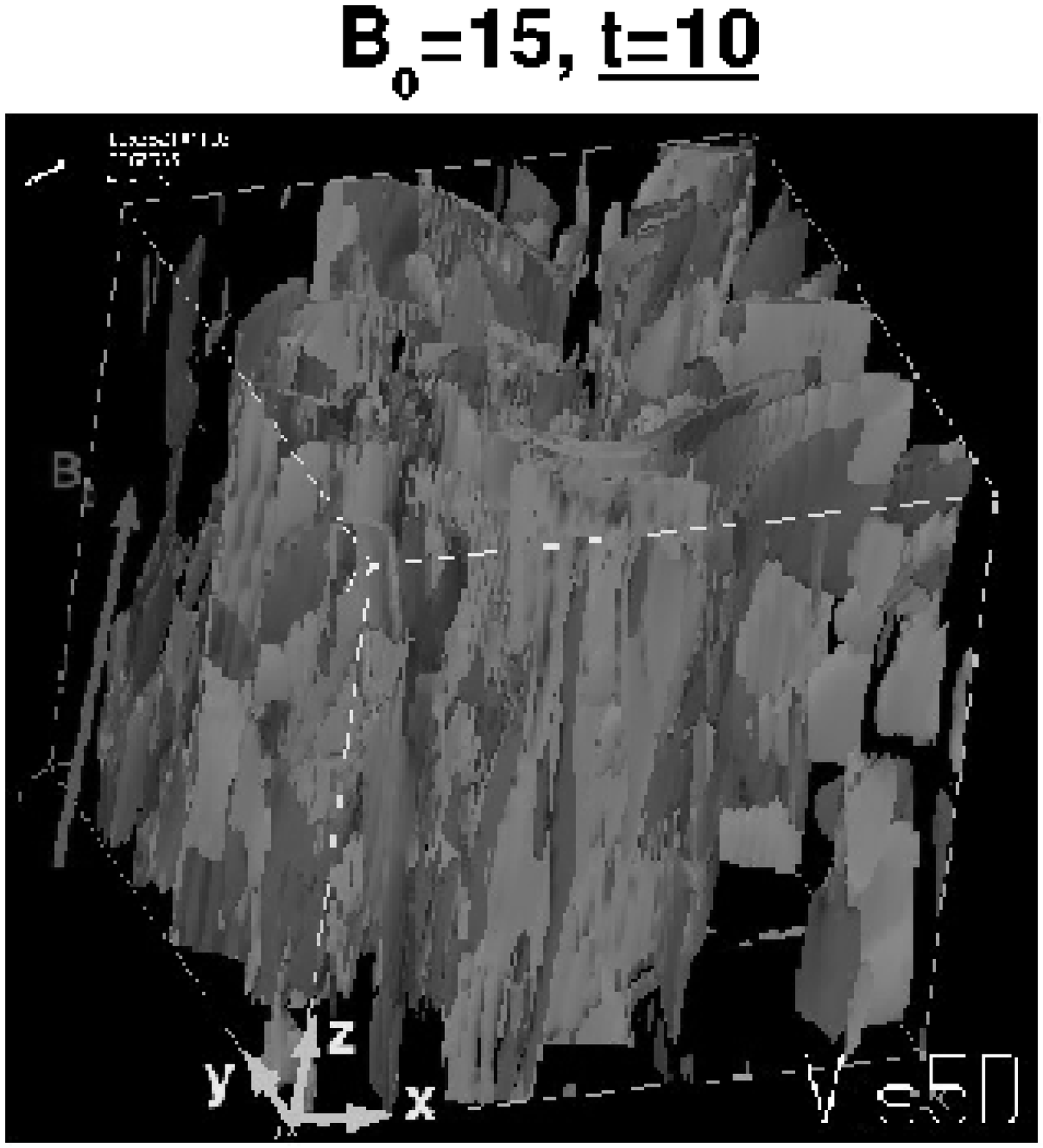}}
\end{tabular}
\caption{Temporal evolution of vorticity (blue) and current (orange) intensities 
for the same run as in Figure~\ref{VisuFilter2}. Isosurfaces drawn at $27\%$ and $20\%$ of
$|{\bf w}|_{max}$ and  $|{\bf j}|_{max}$ instantaneous maxima. 
At $t=0$; $|{\bf w}|_{max}=2.7$ and $|{\bf j}|_{max}= 3.$, 
at $t=4$; $|{\bf w}|_{max}=24.3$ and  $|{\bf j}|_{max}=27.6$, 
at $t=7$, $|{\bf w}|_{max}=17.9$ and  $|{\bf j}|_{max}= 25.6$, and
at $t=10$, $|{\bf w}|_{max}=9.2$ and  $|{\bf j}|_{max}=15.4$.}
\label{Visu3D}
\end{figure*}
Similarly, for a flow with $B_0=15$ (run {\bf IVa}), vorticity and current isosurfaces for states with  
$k_{\parallel}=0$ and $k_{\parallel}>0$ are shown in Figure~\ref{VisuFilter2} at $t=7$, when
the total energy loss is about $40\%$. The 2D state structures are again related to the
star shape in transverse spectral planes (see Figure \ref{VisuSpec1}), while the vorticity and 
current sheets with $k_{\parallel}>0$ present filamentary structures. From our knowledge it is 
the first time that such filaments are reported for (strongly) magnetized flows. 

When looking at the dynamics in physical space (without filtering), shown in Figure~\ref{Visu3D}, 
the vorticity and current intensities are superimposed sheet-like structures aligned along the 
ambiant magnetic field. At $t=7$, a filament formation is observed within the sheets. This can 
be related to the filamentary structures with $k_{\parallel}>0$ (see Figure~\ref{VisuFilter2})
that do not exist in the 2D state, meaning that this sheet filamentation is mainly due
to the wave components. At later times, $t=10$, with a total energy loss of about $55\%$, the 
vorticity and current sheets are disruped by dissipation effects. 

\begin{figure*}[ht]
\begin{tabular}{cc}
\resizebox{85mm}{!}{\includegraphics{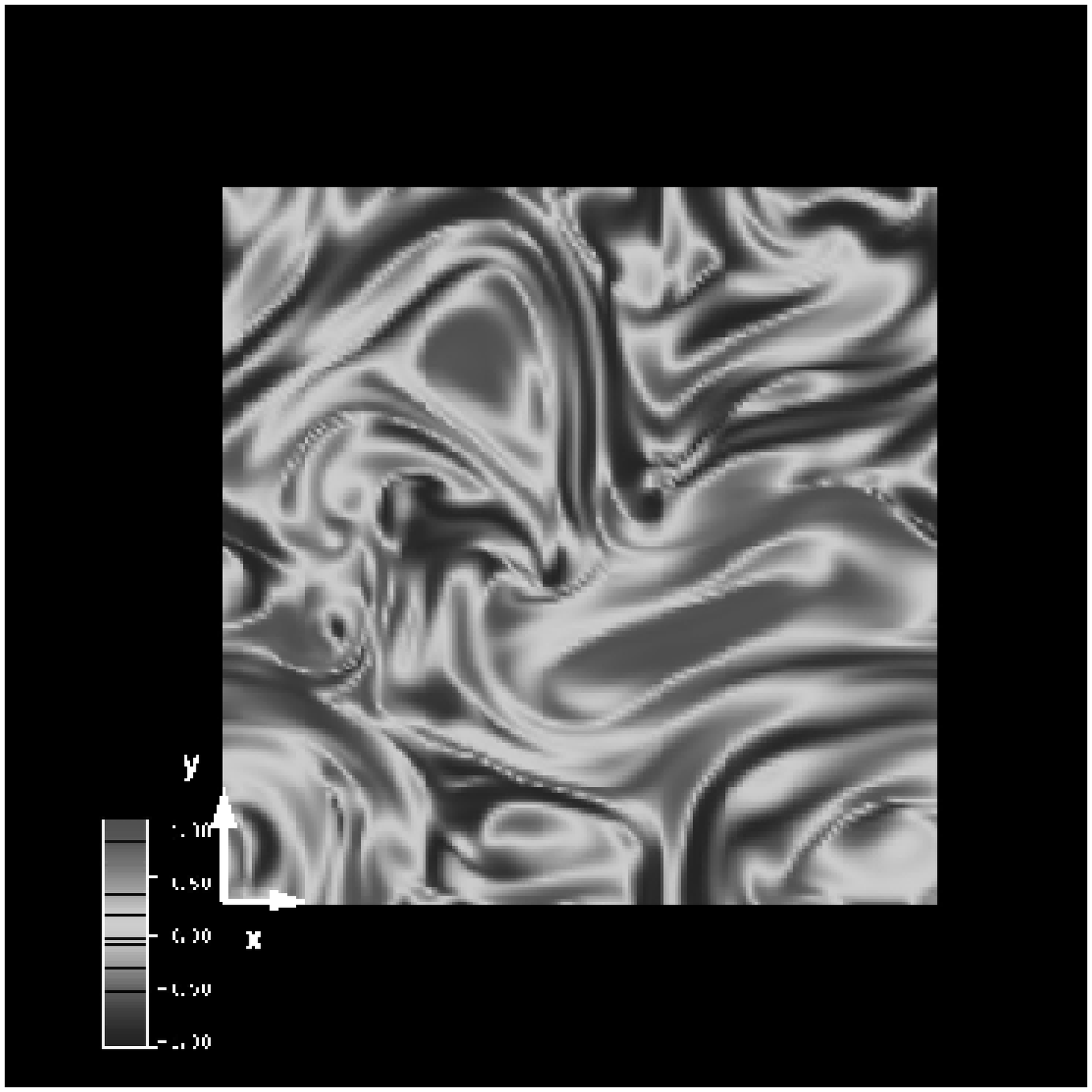}}
\resizebox{85mm}{!}{\includegraphics{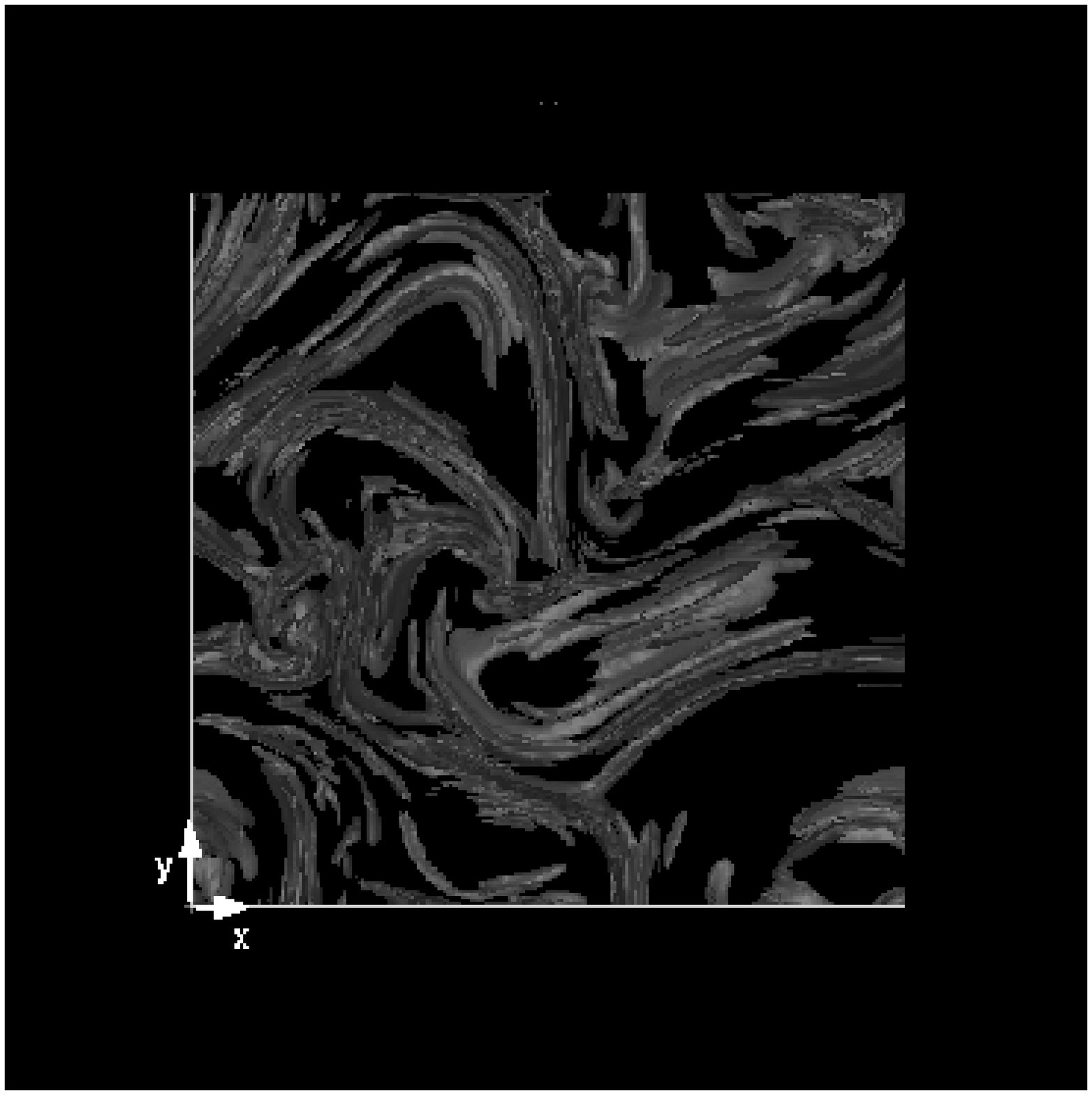}}\\
\resizebox{85mm}{!}{\includegraphics{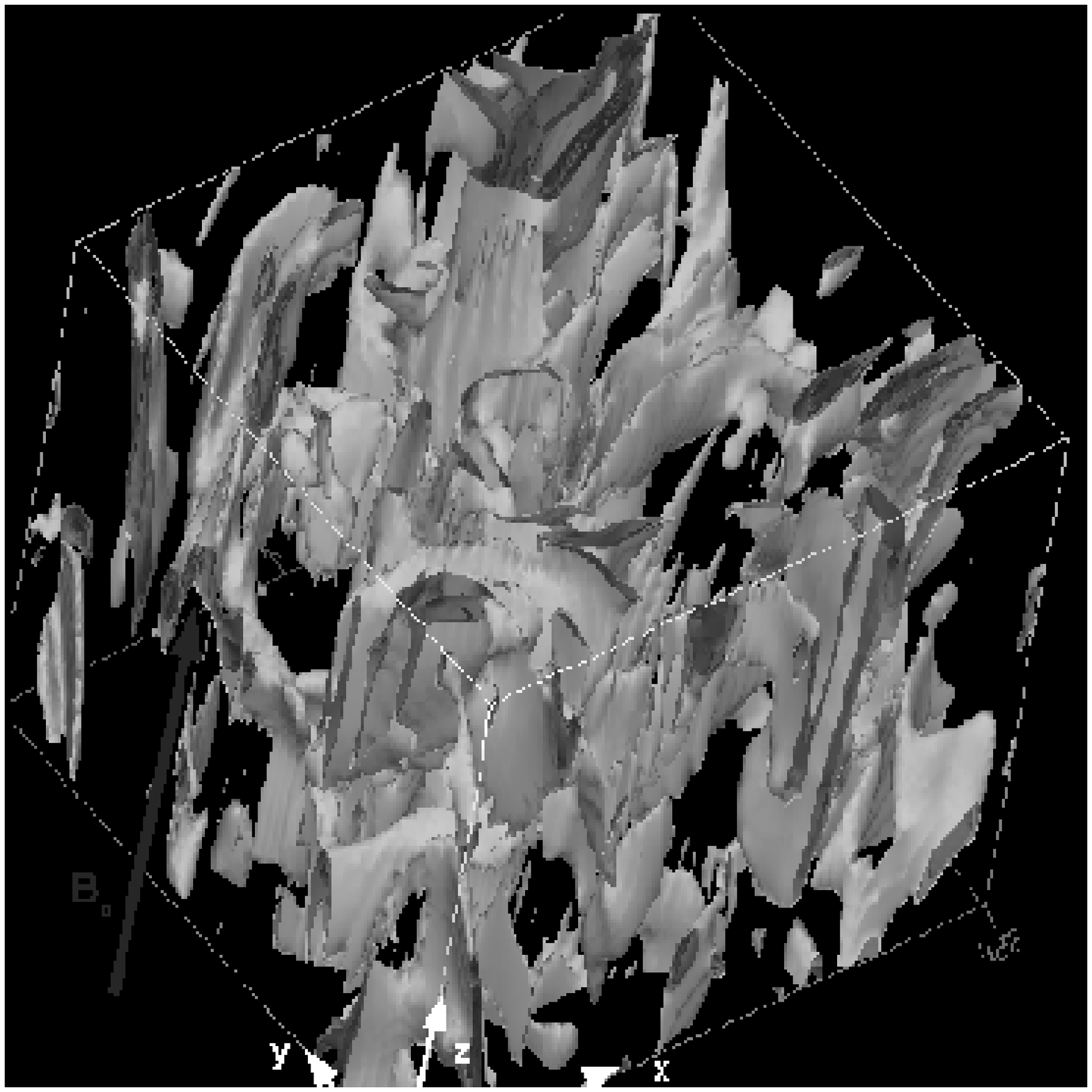}}
\resizebox{85mm}{!}{\includegraphics{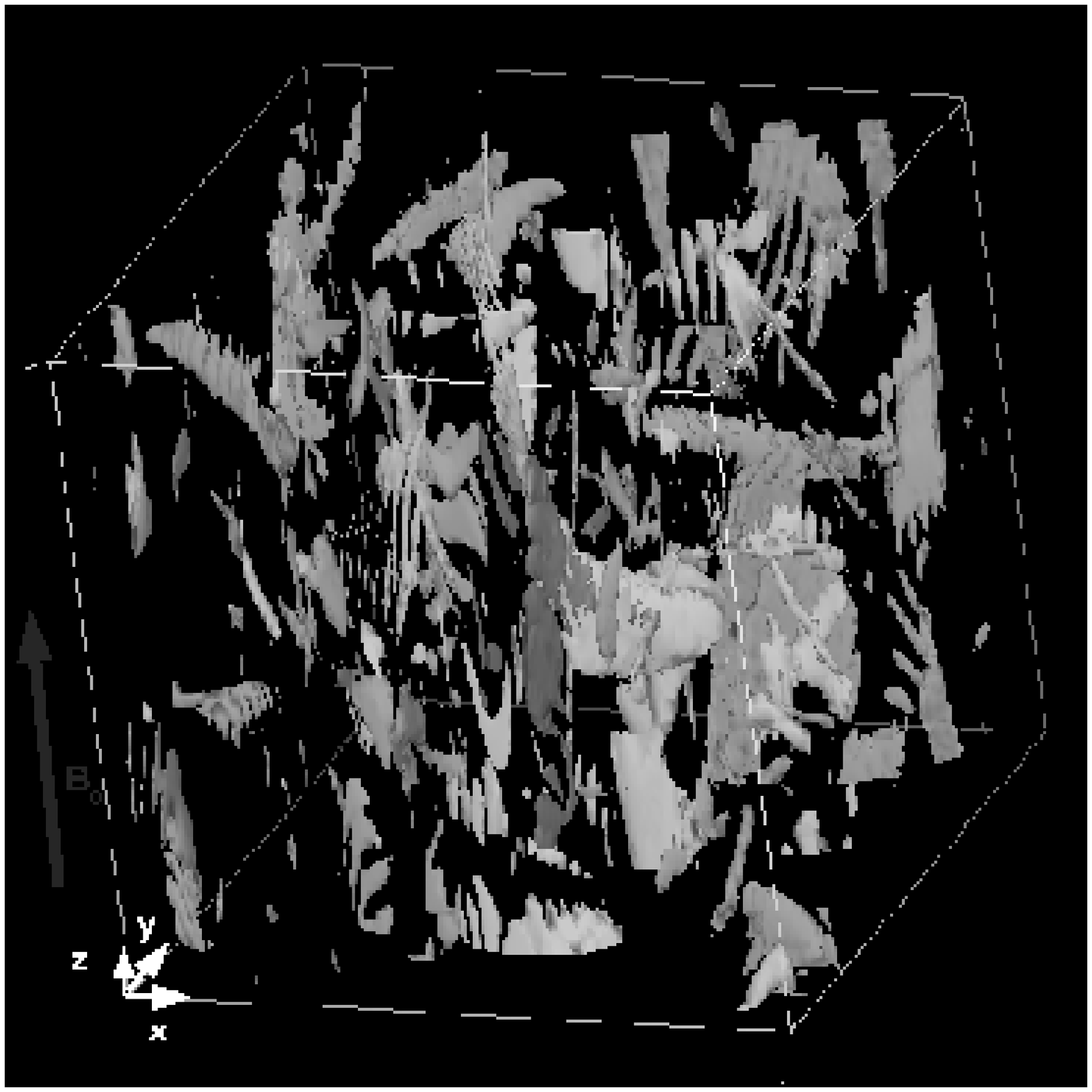}}
\end{tabular}
\caption{Top-left: isosurfaces (at $z=\pi$) of the cross-correlation coefficient; Top-right: the 
corresponding isosurfaces of the current (red) and vorticity (blue) drawn at $21\%$ of their 
respective maxima; Bottom-left: isosurfaces of the cross-correlation coefficient at $-0.70$ (yellow), 
$-0.75$ (blue) and $-0.90$ (red); Bottom-right: isosurfaces of the cross-correlation coefficient at 
$0.96$ (yellow) and $-0.96$ (pink). (Run {\bf IVa} at time $t=10$.)}
\label{Visu3Dcor}
\end{figure*}
Figure \ref{Visu3Dcor} shows the distribution of the cross-correlation values at $z=\pi$ (top) and 
in the entire numerical box (bottom). A comparison with the current and vorticity distribution shows 
that the high (absolute) value of the cross-correlation coincide with the position of dissipative 
structures which means that the velocity and magnetic field fluctuations tend to be aligned at small
scales. This result corroborates recent works on the dynamic alignment in MHD \cite{Boldyrev06} 
where a statistical model is proposed.

%%%%%%%%%%%%%%%%%%%%%%%%%%%%%%%%%%%%%%%%%%%%%%%%%
\section{Runs $Ib$ and $IIb$}

In a last set of simulations we change the initial condition, as explained in Section \ref{bb}, to evaluate 
in particular their influence on the dynamics. It is thought that this new initial condition is more 
appropriate to turbulent flows with a modal spectrum at large-scales (larger than the integral length-scale) 
in $\kpe^3$ which is in agreement with the phenomenology for freely decaying turbulence \cite{Bigot07a}. 
These runs correspond to a strong magnetized flow ($B_0=15$) and high resolution ($512^2 \times 64$). 
We will not focus on the temporal decay which has been analyzed recently \cite{Bigot07a} and we will 
only look at the spectral behavior. 

\begin{figure}[ht]
\begin{tabular}{cc}
\resizebox{88mm}{!}{\includegraphics{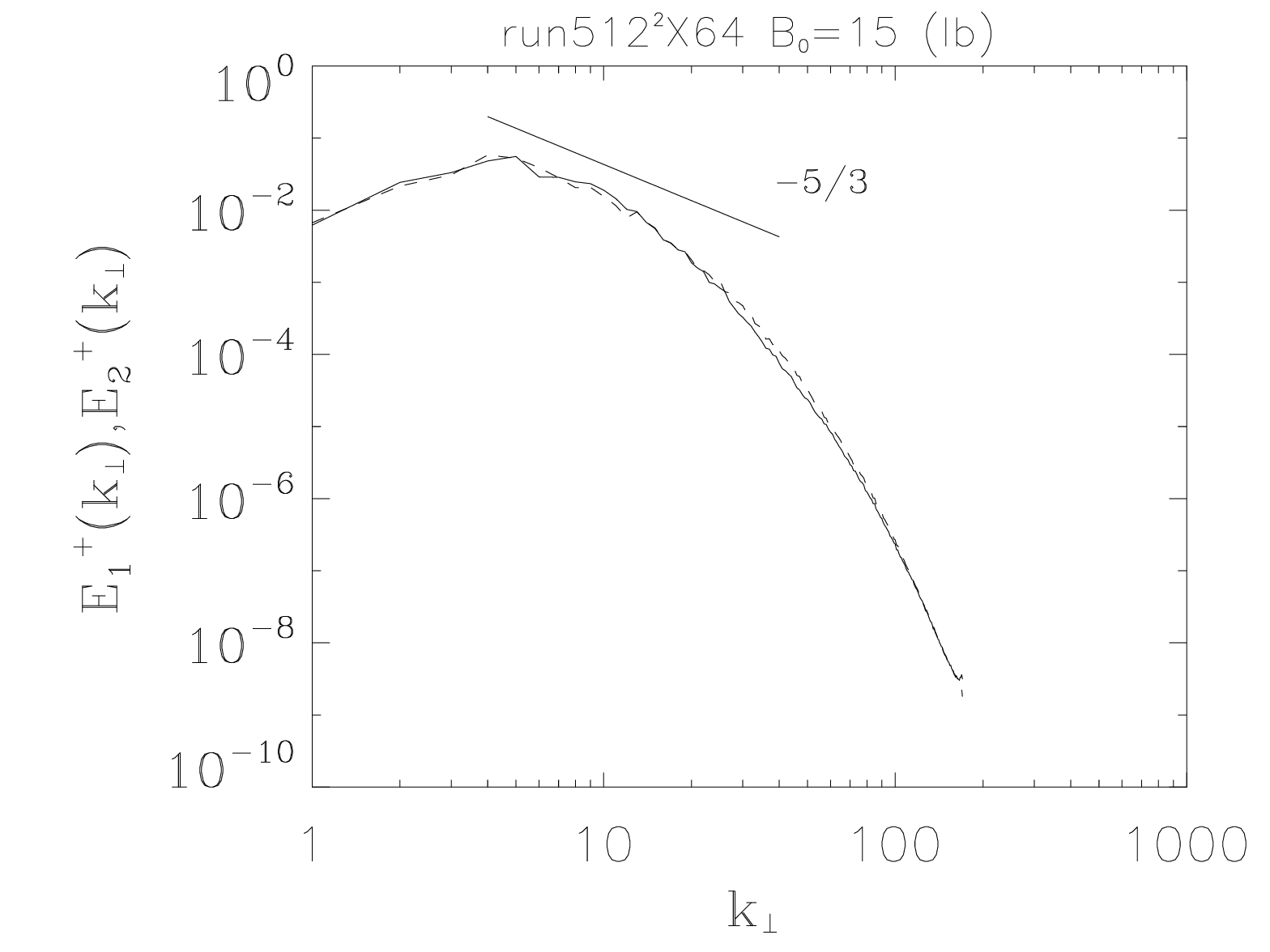}}
%\resizebox{88mm}{!}{\includegraphics{FigDeclinHyper_kt.eps}}
\end{tabular}
\caption{Energy spectra $E^+_{1,2}(\kpe)$ for shear- (solid) and pseudo-Alfv\'en (dashed line) waves
integrated over all parallel wavenumbers in the viscous case ({\bf Ib}).
%(top) and the hyperviscous case ({\bf IIb}) (bottom). 
The straight line follows a $\kpe^{-5/3}$ law. 
\label{figSpecDec_kt}}
\end{figure}

In Figure \ref{figSpecDec_kt} we show the 1D spectra $E^+_{1,2}(\kpe)$ for shear- and pseudo-Alfv\'en 
waves integrated over all parallel wavenumbers. The time chosen is the one for which we have a fully 
developed turbulence. Despite the high resolution 
%and the use of hyperviscosity, 
no clear inertial range appears. The Kolmogorov scaling is given as a reference that is roughly followed. 

The next Figure \ref{figSpecDec_k0} gives at the same time the energy spectra $E_{1,2}^+$ and 
$E_{1,2}^-$ of shear- and pseudo-Alfv\'en waves for the 2D state ($\kpa=0$). The most remarkable 
result is the presence of a relatively extended inertial range characterized by a compensated energy 
spectrum on average around the IK prediction, \ie in $\kpe^{-3/2}$ (note however the presence of a 
bottleneck effect for the viscous run (left)). We conclude that the integration over the parallel 
wavenumbers tends to hide the true scaling by an average effect. 
\begin{figure*}
\begin{tabular}{cc}
\resizebox{90mm}{!}{\includegraphics{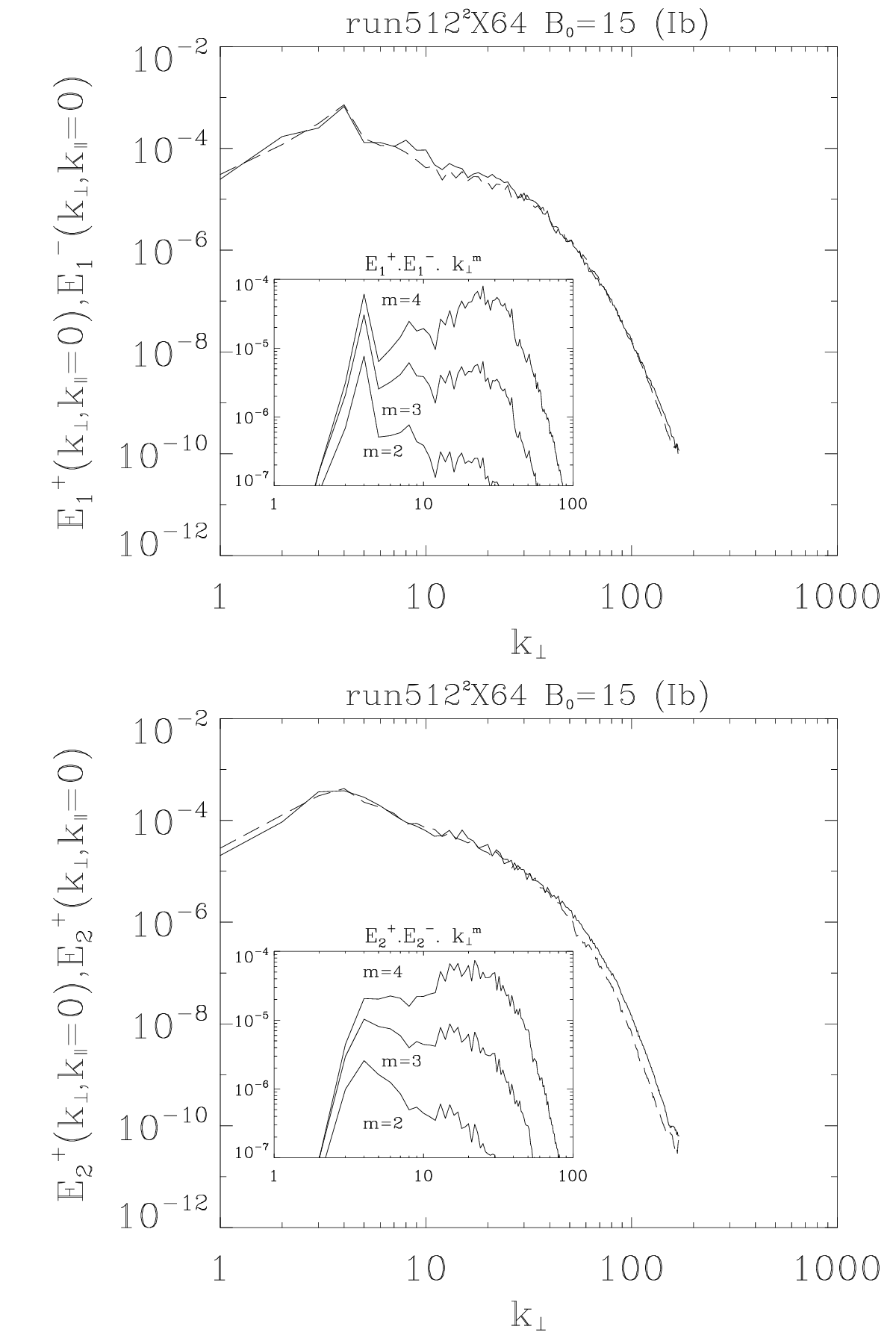}}
\resizebox{90mm}{!}{\includegraphics{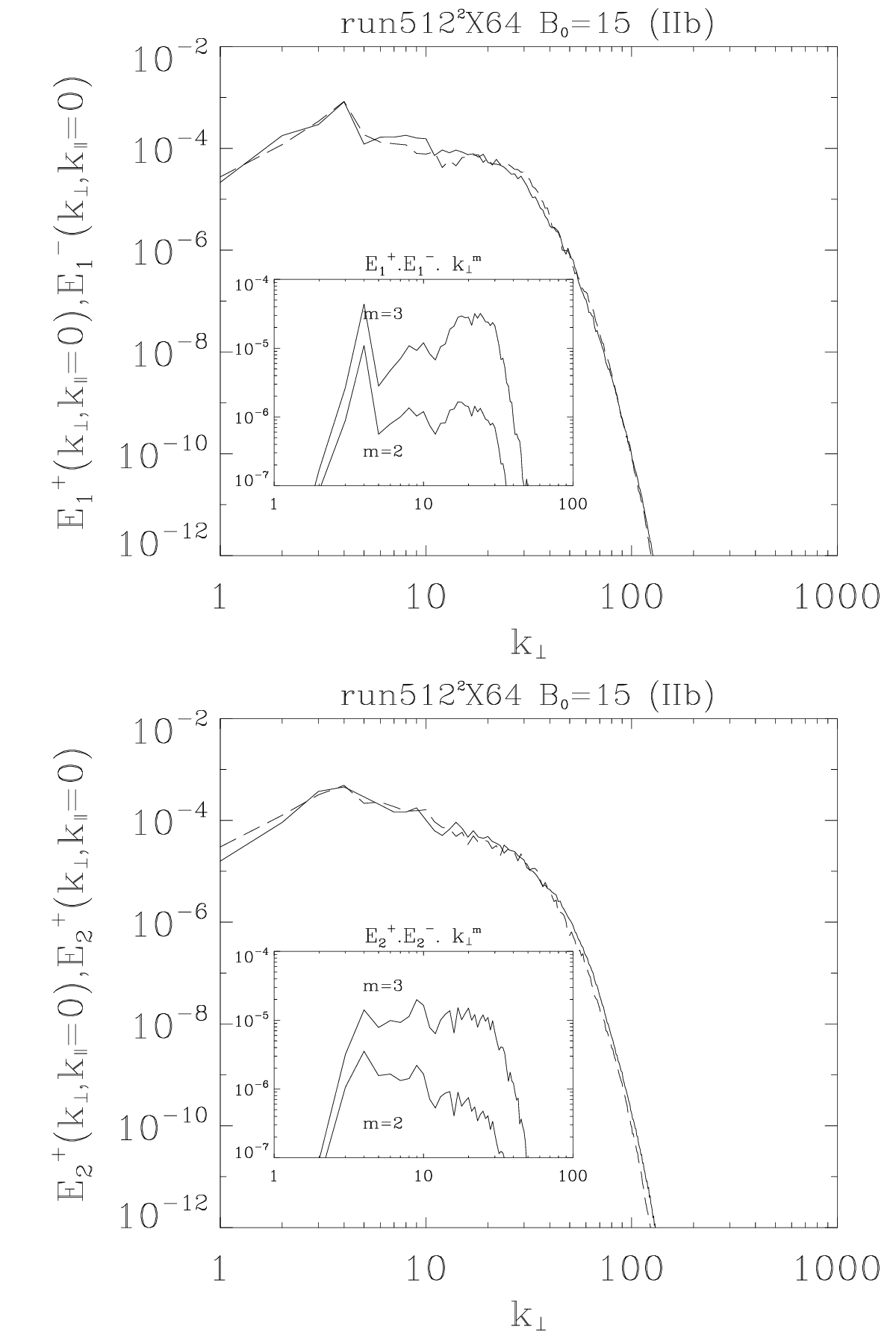}}
\end{tabular}
\caption{Energy spectra $E_{1,2}^+$ (solid) and $E_{1,2}^-$ (dashed) of shear- (top) and pseudo- 
(bottom) Alfv\'en waves for the 2D state ($\kpa=0$). The viscous ({\bf Ib}) (left) and the hyperviscous 
case ({\bf IIb}) (right) are shown. Inset: compensated energy spectra 
$E_{1,2}^+(\kpe,0) E_{1,2}^-(\kpe,0) k^m$.
\label{figSpecDec_k0}}
\end{figure*}

\begin{figure*}
\begin{tabular}{cc}
\resizebox{90mm}{!}{\includegraphics{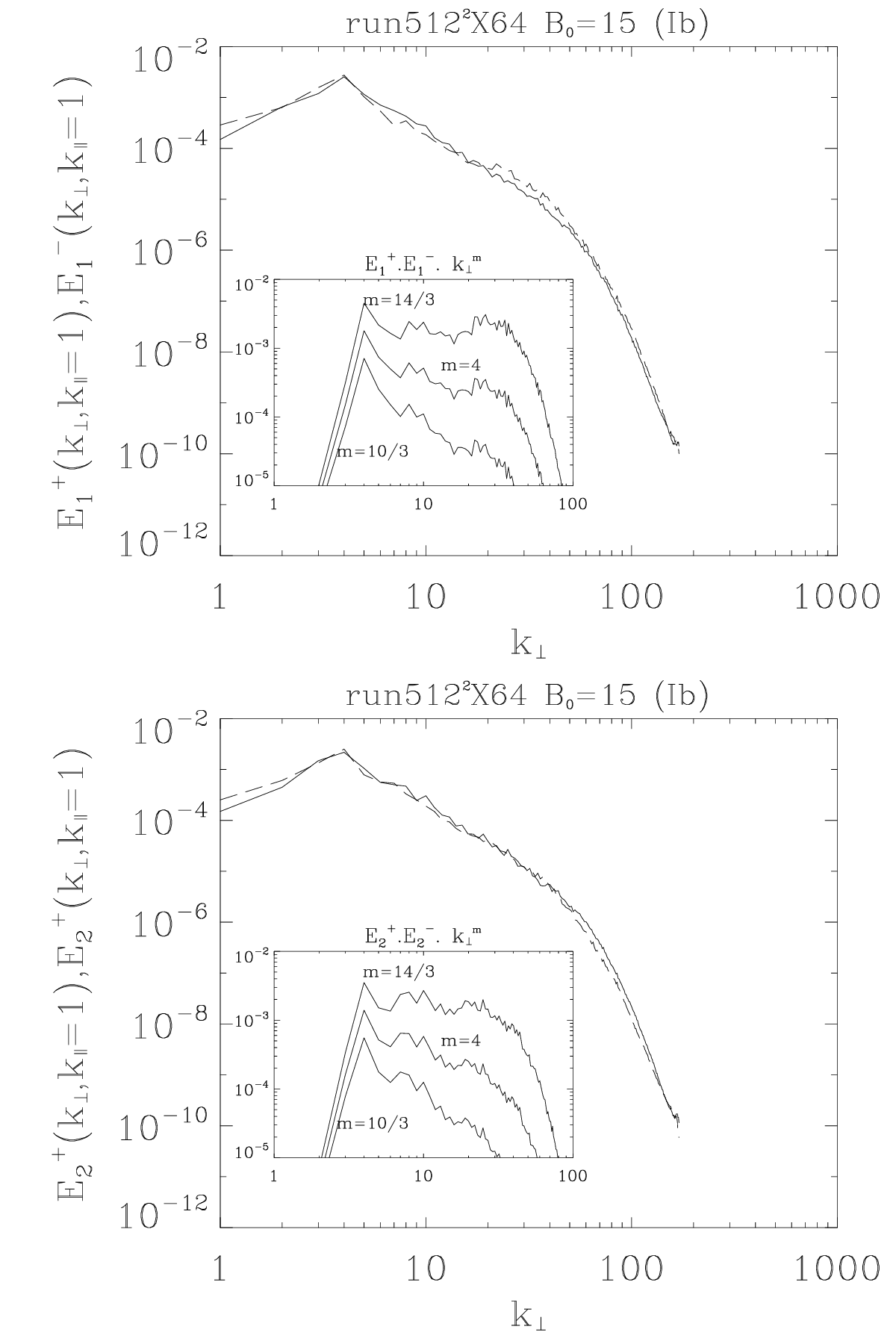}}
\resizebox{90mm}{!}{\includegraphics{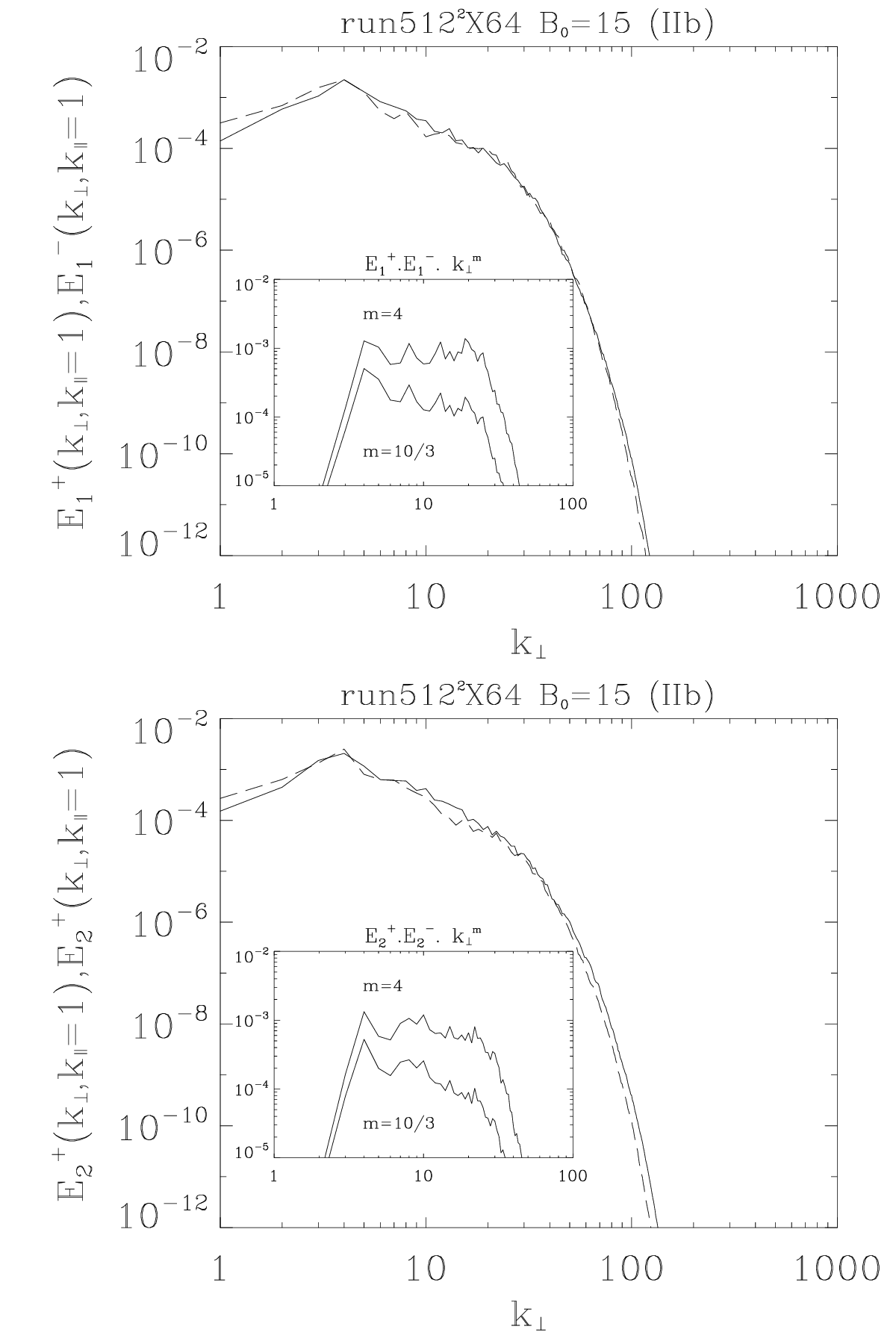}}
\end{tabular}
\caption{Energy spectra $E_{1,2}^+$ (solid) and $E_{1,2}^-$ (dashed) of shear- (top) and pseudo- 
(bottom) Alfv\'en waves for $\kpa=1$. The viscous ({\bf Ib}) (left) and the hyperviscous 
case ({\bf IIb}) (right) are shown. Inset: compensated energy spectra 
$E_{1,2}^+(\kpe,1) E_{1,2}^-(\kpe,1) k^m$.
\label{figSpecDec_k1}}
\end{figure*}

The last Figure \ref{figSpecDec_k1} gives at the same time the energy spectra $E_{1,2}^+$ and 
$E_{1,2}^-$ of shear- and pseudo-Alfv\'en waves at a fixed parallel wavenumber ($\kpa=1$). 
Once again a relatively extended inertial range is found. It is characterized by a compensated energy 
spectrum steeper than the previous one with an index around $\kpe^{-2}$ and $\kpe^{-7/3}$ for 
respectively the hyperviscous and viscous case. Other spectra at higher fixed parallel wavenumbers 
are not shown because there are characterized by a smaller inertial range from which it is difficult to 
find a power law scaling. 

%%%%%%%%%%%%%%%%%%%%%%%%%%%%%%%%%%%%%%%%%%%%%%%%%%%
\section{Summary and conclusion}

In this paper, we present a set of 3D direct numerical simulations of incompressible decaying MHD 
turbulence in which the influence of an external uniform magnetic field ${\bf B_0}$ is investigated. 
A parametric study in terms of $B_0$ intensity is made to show the development of anisotropy. 
In general, the temporal evolutions show oscillations that are associated with the presence of Alfv\'en 
waves. The dynamics is slower for strongly magnetized flows with, in particular, a cross-correlation 
between the velocity and the magnetic field fluctuations frozen on average around its initial (small) 
value but with, locally, a wide range of possible values. 
For all temporal results, one can see that the flows with the highest values of $B_0$ ($\ge 5$) behave 
quite similarly while for $B_0=1$, the flow presents a transient regime between the case without 
background magnetic field and the other cases. 
We also discuss the presence of a sub-critical balance between the Alfv\'en and nonlinear times with 
both a global and a spectral definition. This regime is still associated with the anisotropic scaling laws
(\ref{aniso1}) between the perpendicular and the parallel wavenumbers. 
The nonlinear dynamics of strongly magnetized flows is characterized by a different $\kpe$-spectrum 
if it is plotted at a fixed $\kpa$ (2D spectrum) or if it is integrated (averaged) over all $\kpa$ (1D 
spectrum). In the former case a much wider inertial range is found with a steep power law, closer to the 
wave turbulence prediction than the Kolmogorov one like in the latter case. 
Note that the inertial range of these spectra is better seen for the shear- and pseudo-Alfv\'en 
waves rather than for the cartesian fields.

One of the most important results of this paper is the difference found between the $\kpe$--spectra 
plotted after integration over $\kpa$ and those at a given $\kpa$. This point is generally not discussed 
in numerical works whereas it appears to be a fundamental aspect of this problem. Direct numerical 
simulations of the Alfv\'en wave turbulence regime seems to be still out of the current numerical 
capacity \cite{Naza07} and only the detection of the transition towards such a regime seems possible. 
In such a study it is crucial to avoid any noisy effect linked, for example, to the initial condition (or 
forcing) that could favored one particular type of spectrum. But the other effect that could hide the true 
dynamics of strongly magnetized flows is the averaging effect as we have clearly seen in the second 
set of simulations: the presence of an inertial range was not obvious from a first global analysis 
(Figure \ref{figSpecDec_kt}) whereas it was clear from 2D spectra (Figure \ref{figSpecDec_k0}). 
This averaging effect may be due to the moderate spatial resolution 
used but also to the regime which is in a transition phase towards the wave turbulence regime. If we 
extrapolate such a result to natural plasmas like the one found in the interplanetary medium (inner solar 
wind) then we may interpret the current spectra as averaging spectra (since we are not able to report 
spectra at a given parallel wavenumber with only one spacecraft). Then it is not surprising that we 
observe both an anisotropic flow with approximately a Kolmogorov scaling. 

In a recent numerical analysis \cite{Boldyrev08} dedicated to the development of anisotropy and wave turbulence in forced incompressible reduced MHD flows, a change of spectral slope was reported for the 
$\kpe$--energy spectrum when the forcing is applied on a larger range of parallel wavenumbers with no 
driving of the $\kpa=0$ modes. In the light of the present paper this finding may be interpreted as a way 
to decrease the averaging effect which is mainly due to the dissipative scales. Indeed, when a larger 
parallel wavenumbers is excited the spectrum integrated over all $\kpa$ is more sensitive to the non
dissipative parallel wavenumbers and tends therefore to reveal the true scaling. Another way to avoid 
this noisy effect would have been to plot the spectra at a given but low parallel wavenumber in order 
to avoid the dissipative range. 

The question of the power law index predicted by the wave turbulence theory has not been addressed  
so far. A $\kpe^{-2}$--spectrum is expected for strongly magnetized flows in the regime of wave 
turbulence (even without assuming a restriction on $\kpe$ and $\kpa$ \cite{Galtier06}). 
In our simulations a scaling close to this value is found when hyperviscosity is used 
(Figure \ref{figSpecDec_k1}) in the second set of simulations at $\kpa=1$ whereas the 2D state 
(spectrum at $\kpa=0$) scales on average around $\kpe^{-3/2}$. This latter result is the same as 
the one found generally in 2D isotropic MHD turbulence \cite{politano98,biskamp01}. The steep power 
law reported in $\kpe^{-\alpha}$ with $\alpha \in [2,7/3]$ may be attributed to the very first sign of the 
wave turbulence regime that should be confirmed nevertheless at higher resolution. The $\alpha=7/3$ 
case is {\it a priori} unexpected although it was seen as a transient regime before the finite energy flux 
solution settles down \cite{Galtier2000}. However, no change of slope is observed in our simulation 
because, in particular, for 
larger times the reduction of the inertial range does not allow to conclude about the inertial scaling law.  
The case $\alpha=7/3$ is also a solution predicted by a heuristic model based on a sub-critical balance 
between the Alfv\'en and the nonlinear times \cite{Galtier2005}. In this case, the total energy spectrum 
satisfies the relations  $E(\kpe,\kpa) \sim \kpe^{-\alpha} \kpa^{-\beta}$, with $3\alpha + 2 \beta =7$. 
Thus the $\alpha=7/3$ solution implies no $\kpa$--scale dependence which could be linked to the 
weakness of parallel transfers. 

Our analysis in the physical space has revealed an important new information about structures. 
A filament formation is observed within the current and vorticity sheets. This important property may 
be explained by the specific condition of our simulation (large $B_0$ and large Reynolds number) and 
has to be confirmed at higher Reynolds numbers. 
The classical picture of current sheets in MHD turbulence may be wrong in the strongly anisotropic 
case and filaments are may be the right picture. This result may be compared with astrophysical 
plasmas like in the solar corona where extremely thin (dissipative) coronal loops (filaments or 
"strands") are observed. Although their presence is well accepted, the origin of these filaments is 
still not well explained. Turbulence and Alfv\'en wave could be the main ingredients \cite{bigotheat}.

Other questions about scaling laws for structure functions and intermittency for strongly magnetized 
flows are not discussed here. Forcing numerical simulations are then necessary which is out of the 
scoop of this paper. The unbalance case has not been addressed in this paper. It is also an important 
issue not only from a theoretical point of view but also from an observational point of view since the 
most analyzed astrophysical plasma, the inner solar wind, is mainly made of outwards propagating 
Alfv\'en waves. This point is left for future works.

%%%%%%%%%%%%%%%%%%%%%%%%%%%%%%%%%%%%%%%%%%%%%%%%%%%
\begin{acknowledgments}
We thank A. Alexakis for useful discussions.
This work is supported by INSU/PNST-PCMI Programs and CNRS/GdR Dynamo.
This work was supported by the ANR project no. 06-BLAN-0363-01 ``HiSpeedPIV''.
Computations time was provided by IDRIS (CNRS) Grant No. 070597. 
\end{acknowledgments}

%%%%%%%%%%%%%%%%%%%%%%%%%%%%%%%%%%%%%%%%%%%%%%%%%%%


\begin{thebibliography}{99}
\bibitem{Tajima}
T. Tajima, and K. Shibata, {\it Plasma Astrophysics} (Westview Press, Boulder, USA 2002).
\bibitem{goldstein99}
M.L. Goldstein and D.A. Roberts, Phys. Plasmas {\bf 6}, 4154 (1999). 
\bibitem{galtier06}
S. Galtier, J. Low Temp. Phys. {\bf 145}, 59 (2006).
\bibitem{iro}
P.S. Iroshnikov, Soviet Astron. {\bf 7}, 566 (1964).
\bibitem{Kraichnan65}
R.H. Kraichnan, Phys. Fluids {\bf 8}, 1385 (1965).
\bibitem{MontgoTurner}
D. Montgomery, and L. Turner, Phys. Fluids {\bf 24}, 825 (1981).
\bibitem{Shebalin}
J.V. Shebalin, W.H. Matthaeus and D. Montgomery, J. Plasma Phys. {\bf 29}, 525 (1983).
\bibitem{Higdon}
J.-C. Higdon, Astrophys. J. {\bf 285}, 109 (1984).
\bibitem{Oughton94}
S. Oughton, E.R. Priest and W.H. Mattaheus, J. Fluid Mech. {\bf 280}, 95 (1994).
\bibitem{GS95}
P. Goldreich, and S. Sridhar, Astrophys. J. {\bf 438}, 763 (1995).
\bibitem{ng96}
C.S. Ng and A. Bhattacharjee, Astrophys. J. {\bf 465}, 845 (1996).
\bibitem{Kinney}
R.M. Kinney and J.C. McWilliams, Phys. Rev. E {\bf 57}, 7111 (1998). 
\bibitem{Matthaeus98}
W.H. Matthaeus, S. Oughton, S. Ghosh and M. Hossain, Phys. Rev. Lett. {\bf 81}, 2056 (1998). 
\bibitem{Galtier2000}
S. Galtier, S.V. Nazarenko, A.C. Newell and A. Pouquet, J. Plasma Phys. {\bf 63}, 447 (2000).
\bibitem{Galtier2002}
S. Galtier, S.V. Nazarenko, A.C Newell and A.  Pouquet, Astrophys. J. {\bf 564}, L49 (2002).
\bibitem{milano2001}
L.J. Milano, W.H. Matthaeus, P. Dmitruk and D.C. Montgomery, Phys. Plasmas, {\bf 8}, 2673 (2001).
\bibitem{Muller03}
W.-C. M\"uller, D. Biskamp and R. Grappin, Phys. Rev. E {\bf 67}, 066303 (2003).
\bibitem{verma}
M.K. Verma, Phys. Rep. {\bf 401}, 229 (2004).
\bibitem{chandran}
B.D.G. Chandran, Phys. Rev. Lett. {\bf 95}, 265004 (2005).
\bibitem{Muller}
W.-C. M\"uller and R. Grappin, Phys. Rev. Lett. {\bf 95}, 114502 (2005).
\bibitem{Boldyrev06}
S. Boldyrev, Phys. Rev. Lett. {\bf 96}, 115002 (2006).
\bibitem{Bigot07a}
B. Bigot, S. Galtier and H. Politano, Phys. Rev. Lett. {\bf 100}, 074502 (2008). 
\bibitem{solarobs}
T.S. Horbury, in {\it Plasma Turbulence and Energetic Particles in Astrophysics},
edited by M. Ostrowski and R. Schlickeiser (U. Jagiellonski, Cracow, 1999) p. 115.
\bibitem{Dasso05}
S. Dasso, L.J. Milano, W.H. Matthaeus and C.W. Smith, Astrophys. J. {\bf 635}, L181 (2005).
\bibitem{elmegreen}
B.G. Elmegreen and J. Scalo, Annu. Rev. Astron. Astrophys. {\bf 42}, 211 (2004); 
J. Scalo and B.G. Elmegreen, Annu. Rev. Astron. Astrophys. {\bf 42}, 275 (2004).
\bibitem{Galtier2005}
S. Galtier, A. Pouquet and A. Mangeney, Phys. Plasmas {\bf 12}, 092310 (2005).
\bibitem{Cho00}
J. Cho and E. T. Vishniac, Astrophys. J. {\bf 539}, 273 (2000).
\bibitem{MaronGoldreich01}
J. Maron and P. Goldreich, Astrophys. J. {\bf 554}, 1175 (2001).
\bibitem{shaikh}
D. Shaikh and G. Zank, Astrophys. J. {\bf 656}, L17 (2007).
\bibitem{ZLF}
V.E. Zakharov, V. L'vov and G.E. Falkovich,  
{\it Kolmogorov Spectra of Turbulence I: Wave Turbulence.} (Springer-Verlag, Berlin, Germany 1992). 
\bibitem{Newell01}
A.C. Newell, S.V. Nazarenko and L. Biven, Physica D {\bf 152-153}, 520 (2001). 
\bibitem{Naza07}
S.V. Nazarenko, New J. Phys. {\bf 9}, 307 (2007). 
\bibitem{Boldyrev08}
J.-C. Perez and S. Boldyrev, Astrophys. J. {\bf 672}, L61 (2008).
\bibitem{Galtier99}
S. Galtier, H. Politano and A. Pouquet, J. Plasma Phys. {\bf 61}, 507 (1999). 
\bibitem{Oughton04}
S. Oughton, P. Dmitruk, and W.H. Matthaeus, Phys. Plasmas {\bf 11}, 2214 (2004).
\bibitem{smith99}
L. Smith, and F. Waleffe, Phys. Fluids {\bf 11}, 1608 (1999).
\bibitem{bruno05}
R. Bruno, and V. Carbone, Living Rev. Solar Phys. {\bf 2}, 1 (2005). 
\bibitem{Alexakis07b}
A. Alexakis, B. Bigot, H. Politano and S. Galtier, Phys. Rev. E {\bf 76}, 056313 (2007). 
\bibitem{Alexakis07a}
A. Alexakis, Astrophys. J. {\bf 667}, L93 (2007). 
\bibitem{Boldyrev06}
J. Mason, F. Cattaneo and S. Boldyrev, Phys. Rev. Lett. {\bf 97}, 255002 (2006).
\bibitem{Galtier06}
S. Galtier and B.D.G. Chandran, Phys. Plasmas {\bf 13}, 114505 (2006).
\bibitem{politano98}
H. Politano, A. Pouquet and V. Carbone, Europhys. Lett. {\bf 43}, 516 (1998). 
\bibitem{biskamp01}
D. Biskamp and E. Schwarz, Phys. Plasmas {\bf 8}, 3282 (2001). 
\bibitem{bigotheat}
B. Bigot, S. Galtier and H. Politano, accepted for Astron. Astrophys.
\end{thebibliography}
\end{document}